\documentclass[12pt,a4]{article}
\usepackage{relsize}
\usepackage{array}
\usepackage{a4wide}
\usepackage[hidelinks]{hyperref}
\usepackage{latexsym}
\usepackage{cite}
\usepackage{comment}
\usepackage{caption}
\usepackage{mathrsfs}
\usepackage{tikz}
\usetikzlibrary{arrows.meta,calc,decorations.markings,math,arrows.meta,shapes.misc, decorations.pathreplacing,decorations.pathmorphing}
\usepackage{amssymb}
\usepackage{amsmath}
\usepackage{graphicx}
\usepackage{enumerate}
\usepackage{todonotes}
\usepackage{a4wide}
\usepackage{tensor}
\usepackage{comment}

\newcommand{\be}{\begin{equation}\label}
\newcommand{\ee}{\end{equation}}
\newcommand{\bea}{\begin{eqnarray}\label}
\newcommand{\eea}{\end{eqnarray}}

\makeatletter
\newcommand*{\textoverline}[1]{$\overline{\hbox{#1}}\m@th$}
\newcommand*\bigcdot{\mathpalette\bigcdot@{.65}}
\newcommand*\bigcdot@[2]{\mathbin{\vcenter{\hbox{\scalebox{#2}{$\m@th#1\bullet$}}}}}
\makeatother

\newcommand{\n}{n}
\renewcommand{\k}{k}
\newcommand{\Npt}{N}
\newcommand{\s}{s}

\tikzset{cross/.style={cross out, draw=black, fill=none, minimum size=2*(#1-\pgflinewidth), inner sep=0pt, outer sep=0pt}, cross/.default={2pt}}
\tikzset{
  on each segment/.style={
    decorate,
    decoration={
      show path construction,
      moveto code={},
      lineto code={
        \path [#1]
        (\tikzinputsegmentfirst) -- (\tikzinputsegmentlast);
      },
      curveto code={
        \path [#1] (\tikzinputsegmentfirst)
        .. controls
        (\tikzinputsegmentsupporta) and (\tikzinputsegmentsupportb)
        ..
        (\tikzinputsegmentlast);
      },
      closepath code={
        \path [#1]
        (\tikzinputsegmentfirst) -- (\tikzinputsegmentlast);
      },
    },
  },
  mid arrow/.style={postaction={decorate,decoration={
        markings,
        mark=at position .5 with {\arrow[#1]{stealth}}
      }}},
}
\tikzset{snake it/.style={decorate, decoration=snake}}

\newcommand{\nn}{\nonumber}

\numberwithin{equation}{section}       % equation numbers in each section

\date{}
\begin{document}

   \vspace{1.8truecm}

 \centerline{\LARGE \bf {\sc Five-Dimensional Non-Lorentzian Conformal Field }}
 \vskip6pt

 \centerline{\LARGE \bf {\sc  Theories and their Relation to Six-Dimensions}}

 %\centerline{\LARGE \bf {\sc Five-Dimensional  $SU(1,3)$ Correlators}}
 %\vskip6pt
 
 %\centerline{\LARGE \bf {\sc  and Six-dimensional Conformal Field Theories.}}
%  \vskip 12pt
 
 %\centerline{\LARGE \bf {\sc  and their Si Origin } }

% \centerline{\LARGE \bf {\sc  Five-Dimensional non-Lorentzian CFT's }}
  %\vskip 12pt
 
 %\centerline{\LARGE \bf {\sc  And Their Relation to Six-Dimensions } }
   
\vskip 2cm
  \centerline{
   {\large {\bf  {\sc N.~Lambert,${}^{\,a}$}}\footnote{E-mail address: \href{neil.lambert@kcl.ac.uk}{\tt neil.lambert@kcl.ac.uk}}     \,{\sc A.~Lipstein,$^{\,b}$}\footnote{E-mail address: \href{mailto:arthur.lipstein@durham.ac.uk}{\tt arthur.lipstein@durham.ac.uk}}\, {\sc R.~Mouland${}^{\,a}$}\footnote{E-mail address: \href{rishi.mouland@kcl.ac.uk}{\tt rishi.mouland@kcl.ac.uk}}   {\sc    and P.~Richmond${}^{\,a}$}}\footnote{E-mail address: \href{mailto:paul.richmond@kcl.ac.uk}{\tt paul.richmond@kcl.ac.uk}}  }  
     
\vspace{1cm}
\centerline{${}^a${\it Department of Mathematics}}
\centerline{{\it King's College London }} 
\centerline{{\it The Strand, WC2R 2LS, UK}} 
  
\vspace{1cm}
\centerline{${}^b${\it Department of Mathematical Sciences}}
\centerline{{\it Durham University}} 
\centerline{{\it  Durham, DH1 3LE, UK}}

\begin{abstract}

\noindent

We study correlation functions   in five-dimensional non-Lorentzian theories with an $SU(1,3)$ conformal symmetry. Examples of such theories have recently been obtained as $\Omega$-deformed Yang-Mills Lagrangians arising from a null reduction of six-dimensional superconformal field theories on a conformally compactified Minkowski space. The correlators exhibit a rich structure with many novel properties compared to conventional correlators in Lorentzian conformal field theories. Moreover, identifying the instanton number with the Fourier mode number of the dimensional reduction offers a hope to formulate six-dimensional conformal field theories in terms of  five-dimensional Lagrangian theories. To this end we show that the Fourier decompositions of six-dimensional correlation functions solve the Ward identities of the  the $SU(1,3)$ symmetry, although more general solutions are possible.  Conversely we illustrate how one can reconstruct  six-dimensional  correlation functions from those of a five-dimensional theory, and do so explicitly at 2- and 3-points. We also show that,  in a suitable decompactification limit $\Omega\to 0$, the correlation functions become those of the DLCQ description.

%original: We discuss non-Lorentzian five-dimensional  field theories  with an $SU(1,3)$ conformal symmetry. In particular we determine the form of the two and three point functions. Such theories can be obtained by conformally compactifying a six-dimensional conformal field theory along a null direction. Performing a Fourier expansion in the compact direction leads to a five-dimensional description where the Fourier mode is interpreted as an instanton number over the four spatial dimensions. Unlike familiar DLCQ constructions this approach does not require that the operators are periodic in the compact direction. We show how the Ward identities of the five-dimensional theory agree with those of the six-dimensional theory. This offers a hope for a formulation of six-dimensional CFT's in terms of five-dimensional Lagrangian field theories with an $SU(1,3)$ conformal symmetry.

\end{abstract}

\pagebreak

\tableofcontents

%\allowdisplaybreaks

\section{Introduction}\label{sect: Introduction}

Understanding higher-dimensional  conformal field theories (CFT's) is one of the major challenges of Theoretical Physics. Progress in these theories is hampered by the fact that they are not expected to admit a Lagrangian description, at least in a traditional sense. Although CFT's have many applications in physics, six-dimensional examples are especially important in M-theory where they describe the worldvolume dynamics of M5-branes and hence, {\it via} the AdS/CFT correspondence they can be used to define M-theory in asymptotically AdS$_7\times S^4$ spacetimes. 

M-theory also admits M2-branes described by strongly coupled three-dimensional CFT's. 
The first construction of  these CFT's was given in \cite{Bagger:2006sk,Gustavsson:2007vu,Bagger:2007jr}. This theory has a Lagrangian with the required $SO(2,3)$ conformal symmetry and $SO(5)$ R-symmetry along with 16 supersymmetries and 16 superconformal symmetries. However it is only capable of describing a small number of interacting M2-branes. It turns out that to find a description for an arbitrary number of interacting M2-branes requires one to give up some symmetries of the Lagrangian. In particular the ABJM models \cite{Aharony:2008ug} have $SO(2,3)$ conformal symmetry but only an $SU(4)$ R-symmetry along with 12 supersymmetries and 12 superconformal symmetries.  However there is a topological conserved current   $J_M\sim \star{\rm tr} (F)$ which gives an additional $U(1)$ symmetry. The missing symmetries arise   in the quantum theory through the use of monopole operators at special, order one,  values of the coupling, and enhance $U(1)\times SU(4)$   to $SO(8)$, as well as supplying the missing  4 super and 4 superconformal symmetries. Thus we have, {\it via} the AdS/CFT correspondence, a complete description of M-theory in asymptotically AdS$_4\times S^7$ spacetimes.

To date there is no satisfactory construction of the six-dimensional CFT for M5-branes. 
In \cite{Lambert:2019jwi,Lambert:2019fne} we constructed Lagrangian field theories for  an arbitrary number of M5-branes reduced on a null direction in conformally compactified Minkowski space. The resulting five-dimensional theories have a non-Lorentzian Lagrangian description with an $SU(1,3)$  conformal symmetry, $SO(5)$ R-symmetry, 8 supersymmetries and 16 superconformal symmetries.    Furthermore   these theories retain a knowledge of the sixth spacetime dimension through a topological $U(1)$ current generated by $J_I\sim \star{\rm tr}(F\wedge F)$. Thus they offer an M5-brane analogue of the ABJM description of M2-branes: see Table 1 for a comparison. In particular one sees that the M2-brane $SO(2,3)$ conformal symmetry is swapped for an $SO(5)$ M5-brane R-symmetry, whereas the M2-brane $SU(4)$ R-symmetry is swapped for an M5-brane $SU(1,3)$ conformal symmetry.\footnote{We note that a six-dimensional Lagrangian, with $SO(1,5)\times SO(5)$ Lorentz and R-symmetries and 16 supersymmetries can be constructed for two M5-branes \cite{Lambert:2019diy} (although in this case the translational and conformal symmetries are spontaneously broken), offering an M5-brane analogue of the BLG M2-brane Lagrangian. }
 The conjecture of \cite{Lambert:2019jwi} is that at strong coupling  the topological $J_I$ current will enhance the $SU(1,3)\times U(1)$  symmetry to $SO(2,6)$, add the missing 8 supersymmetries   and thereby describe the six-dimensional M5-brane CFT. The total symmetries of the M2 and M5-brane CFT's expected from these enhancements by topological currents are listed in table 2. We see that the key enhancement that is required is $SU(4)\times U(1)\to SO(8)$ for M2-branes and $SU(1,3)\times U(1)\to SO(2,6)$ for M5-branes. Indeed  from a group theoretic point of view these are the same, just with a different choice of signature for the inner-product.   If this enhancement occurs then the additional supersymmetries and superconformal symmetries must also arise and {\it vice-versa}. In terms of the field theory, the role of monopole operators is presumably replaced by so-called instanton operators \cite{Lambert:2014jna, Tachikawa:2015mha}.
 Part of the motivation of this paper is to explore how the enhancement of conformal symmetry could happen at the level of correlation functions. 
 
\begin{table}
\begin{center}
	\begin{tabular}{|l|l|l|}
	\hline Symmetry &	M2 & M5\\
		\hline \hline
	Conformal &$SO(2,3)$ &  $SU(1,3)$  \\
		\hline
	R-symmetry &$ SU(4)$ &  $  SO(5)$   \\
	\hline
	Topological & U(1) & U(1)\\ \hline
	Supersymmetry & 12 & 8\\ \hline
	Superconformal &12 & 16\\ \hline
	\end{tabular}
\caption{Symmetries of M-brane Lagrangians}
\end{center}
	\end{table}
	
	\begin{table}
\begin{center}
	\begin{tabular}{|l|l|l|}
	\hline Symmetry &	M2 & M5\\
		\hline \hline
	Conformal &$SO(2,3)$ &  $SO(2,6)$  \\
		\hline
	R-symmetry &$ SO(8)$ &  $  SO(5)$   \\
	\hline
	%Topological & U(1) & U(1)\\ \hline
	Supersymmetry & 16 & 16\\ \hline
	Superconformal &16 & 16\\ \hline
	\end{tabular}
\caption{Symmetries of M-brane CFT's}
\end{center}
	\end{table}

 It has been previously argued that the M5-brane CFT can arise from a discrete light cone quantization (DLCQ) description where a null direction of six-dimensional Minkowski space is periodically identified with period $R_+$ \cite{Aharony:1997th,Aharony:1997an}. Those authors proposed that the dynamics of $M$ M5-branes at fixed light cone momentum $P_+ = \n/R_+$, $\n\in {\mathbb Z}$ is described by quantum mechanics, with `time' played by $x^-$, on the moduli space of charge $\n$, $SU(M)$ instantons on ${\mathbb R}^4$. In principle one can then recover the full six-dimensional theory by taking the limit $R_+\to \infty$. 
 
 It has also been argued that five-dimensional maximally supersymmetric Yang-Mills is in fact non-perturbatively well-defined despite being naively power counting non-renormalizable \cite{Douglas:2010iu,Lambert:2010iw}. This corresponds to the M5-brane CFT compactified on a circle of radius $R_5 = g^2_{YM}/4\pi^2$. In this way the full six-dimensional CFT should arise as a strong coupling limit $R_5\to \infty$. 
Another approach conformally maps the M5-brane CFT to ${\mathbb R}\times S^5$ of arbitrary radius $R$ and then performs a Hopf reduction of $S^5$ to $\mathbb{C}P^2$ \cite{Kim:2012tr}. In the resulting five-dimensional Lagrangian on ${\mathbb R}\times \mathbb{C}P^2$  the coupling constant is controlled by an integer $k$ which plays the role of a Chern-Simons level. The full six-dimensional theory corresponds to the strong coupling point $k=1$. Other constructions include deconstruction from four-dimensional superconformal field theories \cite{ArkaniHamed:2001ie} as well as variety of proposals to capture at least a part of the dynamics through various novel  constructions.

The approach adopted in this paper contains elements of many of these previous constructions but we hope with additional benefits.  In particular we perform a conformal compactification  that maps a null direction in six-dimensions to $x^+\in [-\pi R,\pi R]$, where $R$ is an arbitrary scale introduced by the conformal compactification, while still preserving $\partial/\partial x^+$ as an isometry. This construction naturally arises from holographic considerations analogous to those of the ABJM theory, as shown in \cite{Lambert:2019jwi}. Thus we can form a natural expansion in Fourier modes labelled by $\n\in {\mathbb Z}$ with period $2\pi R$. As such, the map from six-dimensional fields to five-dimensional ones is invertible, at least for a large class of operators. However we can also restrict $\n\in \k{\mathbb Z}$ for an integer $\k$, thereby imposing an analogue of a ${\mathbb Z}_\k$ orbifold, leading to coupling constant $g^2_{YM}=4\pi^2 R/k$ in the five-dimensional gauge theory that can be made small. We claim that the enhancement of symmetries required to describe the full six-dimensional theory occurs at $\k=1$, for any value of $R$.  On the other hand we will see that the degenerate limit $\k,R\to\infty$ with $R_+=R/k$  fixed reproduces the   DLCQ description. Indeed taking such a limit of our correlation functions leads to those of the DLCQ prescription.

Furthermore other constructions in M-theory and  massive type IIA string theory lead to six-dimensional  CFT's with $(1,0)$ supersymmetry. 
In \cite{Lambert:2019fne} a large class of five-dimensional Lagrangians with $SU(1,3)$ conformal symmetry but half as many super and superconformal symmetries were constructed, corresponding to  a similar conformally compactified null reduction of  $(1,0)$ CFT's.  Independently of their application to six-dimensional CFT's, string  and M-theory, these Lagrangians form a new class of interacting five-dimensional field theories with novel features which are of interest in their own right. Thus it is of interest to further explore the features of  non-Lorentzian  field theories with $SU(1,3)$ conformal symmetry.

As explained above, the main motivation of this paper is to get a better handle on the worldvolume theory for M5-branes and other six-dimensional CFT's. Being a six-dimensional superconformal field theory, the main observables which can be used to probe the dynamics   are its correlation functions. Although it is very challenging to compute anything dynamical ({\it i.e.}\ unprotected) in this theory from first principles, there has recently been a great deal of progress in computing correlations functions using a combination of conformal bootstrap and holographic methods. In particular 3-point correlators were first computed in \cite{Bastianelli:1999ab,Bastianelli:1999vm,Eden:2001wg} and 4-point correlators were computed in the supergravity approximation in \cite{Arutyunov:2002ff,Heslop:2004du,Beem:2015aoa,Rastelli:2017ymc}. The study of 4-point correlators beyond the supergravity approximation was initiated in \cite{Heslop:2017sco} and further developed in \cite{Chester:2018dga}, which fixed coefficients of higher-derivative corrections using a powerful chiral algebra conjecture formulated in \cite{Beem:2014kka}. Higher derivative corrections to eleven-dimensional supergravity were also studied from various other points of view in \cite{Chester:2018lbz,Chester:2018aca,Alday:2020tgi}.

\subsection{Plan and Summary of the Paper}

In this paper we will consider five-dimensional correlators in theories with an $SU(1,3)$ symmetry. Note that this is not a standard conformal group of the form $O(2,d)$ and is best thought of as a non-relativistic conformal symmetry group since it contains a Lifshitz scaling. As a result, far less is known about correlators with this symmetry. To our knowledge, non-relativistic conformal correlators were first studied in \cite{Henkel:1993sg} in the context of condensed matter physics. In that paper, the Ward identities were solved at 2- and 3-points, showing that 3-point correlators are not completely fixed by symmetry, in contrast to conformal correlators in Lorentzian theories. The same structure was later found when computing 2-point functions of protected operators of the M5-brane theory using the  DLCQ proposal \cite{Aharony:1997an}. Supersymmetric extensions of the conformal Ward identities were formulated and solved at 2-points in \cite{Henkel:2005dj}. More recently, the prospect of carrying out the bootstrap for two-dimensional Galilean conformal theories was explored in \cite{Chen:2020vvn}.

As described above, our motivation for this paper was a certain class of non-Lorentzian $\Omega$-deformed Yang-Mills gauge theories which admit an $SU(1,3)$ conformal symmetry and provide the Lagrangians for six-dimensional CFT's conformally compactified along a null direction \cite{Lambert:2019jwi,Lambert:2020jjm}. On the other hand, our results do not rely on this Lagrangian and can potentially apply to any theory with this symmetry. We first solve the conformal Ward identities, revealing an intricate mathematical structure. Perhaps the most notable feature is that the correlators naturally decompose into a perturbative part with power-law decay and an oscillating non-perturbative part. In addition the coordinate dependence can be expressed using complex structure which is reminiscent of that found in two-dimensional CFT's. Another notable feature is that 3-point correlators are once again not fully fixed by the symmetry while 4-point correlators can depend on five conformal cross-ratios in contrast to the standard result in relativistic CFT where there are only two cross-ratios. As a result, crossing symmetry (which forms the backbone of the conformal bootstrap) is more intricate in theories with non-relativistic conformal symmetry.

We then ask the question how are these general solutions constrained by the condition that they arise from conformal compactification of Lorentzian six-dimensional CFT correlators? In this case, five-dimensional correlators are labelled by Fourier modes along the null direction and dimensional reduction implies many additional constraints. For example, it fixes the normalisation of five-dimensional 2-point functions in terms of combinatoric factors which vanish if the mode number is less than half of the scaling dimension of each operator, implying that zero modes cannot contribute. This resolves a long-standing ambiguity in the role of zero modes in DLCQ theories \cite{Nakanishi:1976vf,Fitzpatrick:2018ttk}. Moreover, if we identify Fourier modes with instantons in the five-dimensional theory, this implies that only anti-instantons can propagate, which is consistent with the observation that the five-dimensional Lagrangians derived in \cite{Lambert:2019jwi,Lambert:2020jjm} localise onto anti-self-dual field configurations. At three points, dimensional reduction also fixes the functional form of the correlators and implies various sum rules on the Fourier modes which essentially enforce that the sum over Fourier modes can never go positive. In the case of the gauge theories of \cite{Lambert:2019jwi,Lambert:2020jjm} this implies that instantons cannot be produced. At four points, the functional form in six-dimensions is not fixed and hence for simplicity we consider dimensional reduction of disconnected free correlators and show that the corresponding five-dimensional correlators nontrivially decompose into products of 2-point correlators. This also serves to illustrate how the conformal cross-ratios derived earlier in the paper appear in practice. 

Lastly we consider the limit $R \rightarrow \infty$ where the conformal compactification is removed. However we can get a non-trivial limit by keeping modes with  $\n/R$ finite (so we can write $\n = \n_+\k$, $R=\k R_+$ with $\k\in {\mathbb Z}$ and take the limit $\k\to\infty$). This effectively leads to a situation where the operators in flat six-dimensional Minkowski space are constrained to be periodic in $x^+$ with period $2\pi R_+$.  In this limit the 2-point correlators we obtain reduce to the DLCQ correlators found in \cite{Aharony:1997an}. Therefore our results can be used to extend their analysis up to 4-points. It should be noted that various divergences arise when computing DLCQ correlators from dimensional reduction, which are regulated in the $SU(1,3)$ theory. Hence, the $\Omega$-deformed Lagrangian for conformally compactified M5-branes appears to be more fundamental than the original DLCQ proposal. However we note that \cite{Aharony:1997an} discusses a resolution of the instanton moduli space that is reminiscent of an $\Omega$-like deformation.  Ultimately, identifying the Fourier modes with  instanton-solitons in five-dimensional gauge theories offers the tantalising possibility of computing correlators using Lagrangian methods and then re-constructing six-dimensional correlators {\it via} a Fourier series.

The rest of this paper is organised as follows. In section \ref{sect: Ward Identities in Five-Dimensions} we consider the Ward identities of 2, 3 and higher-point functions that arise in five-dimensional theories with an $SU(1,3)$ conformal symmetry. This symmetry fixes the 2-point functions (up to a constant) and determines the 3-point functions up to a single function. Furthermore one sees that there is a natural complex structure that appears where the correlation functions factorise into a product of holomorphic and anti-holomorphic components. In section \ref{sect: Null Conformal Compactification} we discuss a null conformal compactification of six-dimensional Minkowski space. Performing a Fourier expansion leads to a five-dimensional theory with $SU(1,3)$ conformal symmetry but no Lorentz invariance and a Kaluza-Klein-like tower of operators. For a wide class of operators this mapping from six to five dimensions is invertible. We also show how the correlation functions of the six-dimensional theory reduce to those of the five-dimensional theory with specific expressions for the otherwise undetermined function.   In section \ref{sect: Recovering the DLCQ description} we take a limit of our construction where the conformal mapping degenerates into a null compactification of Minkowski space. In this limit only operators which are periodic in the null coordinate can be mapped to the five-dimensional theory. We show that the correlation functions reduce to those of a DLCQ construction. In section \ref{sect: Conclusion} we give our comments and conclusions. In the appendices we discuss some details of the evaluation of various integrals we encounter in the main analysis and in particular the role that the six-dimensional causal structure plays in the form of an $i\varepsilon$ prescription to regulate the integrals. 

\section{Constraining Correlators in Five Dimensions}\label{sect: Ward Identities in Five-Dimensions} 

\subsection{The Symmetry Algebra}

The geometric symmetry algebra of the five-dimensional theories we are considering is given by  central extension to $\frak{su}(1,3)$. The symmetry generators of $SU(1,3)$ are\linebreak $\{P_-,P_i,B, C^I,T,M_{i+},K_+\}$ with $i=1,\dots,4$, $I=1,2,3$ and correspond respectively to what were called Types I-VII in \cite{Lambert:2019fne}. In addition there is the central element $P_+$. A subset of the commutation relations of the algebra is
%\begin{align}
%  [T,P_-] 	&= -2P_- \, ,					&		[M_{i+},P_j] 	&= -\delta_{ij} P_+ - \tfrac{1}{2}\Omega_{ij} T - \tfrac{2}{R} \delta_{ij} B + \Omega_{ik}\eta^I_{jk} C^I \, ,\nn\\
%  [T,K_+] 	&= 2K_+ \, ,					&		[P_-,P_i] 				&= 0		  				\, ,\nn\\
%  [K_+,P_-] &= -2T	\, ,					&		[P_-,M_{i+}] 			&= \tilde{P}_i   			\, ,\nn\\
%&										& 		[K_+,P_i] 				&= -2M_{i+} 		\, ,\nn\\
%  [T,P_i] 	&= -P_i \, ,		&		[K_+,M_{i+}] 			&= 0  						\, ,\nn\\
%  [T,M_{i+}]&= M_{i+} \, ,	&		[P_i,P_j] 		&= -\Omega_{ij} P_-  						\, , \nn\\
%&										& 		[M_{i+},M_{j+}] &= -\tfrac{1}{2} \Omega_{ij} K_+ \, .
%\label{eq: extended su(1,3) algebra}
%\end{align}
\begin{align}
[M_{i+},P_j] \ &= \  -\delta_{ij} P_+ - \tfrac{1}{2}\Omega_{ij} T - \tfrac{2}{R} \delta_{ij} B + \Omega_{ik}\eta^I_{jk} C^I    \, ,	&	[T,P_-] \ 	&= \ -2P_-					\, ,\nn\\
  [T,K_+] \	 &= \ 2K_+ \, ,					&		[P_-,P_i] 				\ &= \ 0		  				\, ,\nn\\
  [K_+,P_-] \ &= \ -2T	\, ,					&		[P_-,M_{i+}] 		\ &= \ P_i   			\, ,\nn\\
[M_{i+},M_{j+}] \ &= \ -\tfrac{1}{2} \Omega_{ij} K_+ \, , &										 		[K_+,P_i] 				\ &= \ -2M_{i+} 		\, ,\nn\\
  [T,P_i] 	\ &= \ -P_i \, ,		&		[K_+,M_{i+}] 			\ &= \  0  						\, ,\nn\\
  [T,M_{i+}] \ &= \ M_{i+} \, ,	&		[P_i,P_j] 		\ &= \ -\Omega_{ij} P_-  						\, ,
\label{eq: extended su(1,3) algebra}
\end{align}
where $\Omega_{ij}$ is anti-symmetric, anti-self-dual and satisfies $\Omega_{ij}\Omega_{jk} = -R^{-2}\delta_{ik}$.
Here $R$ is a constant with dimensions of length. 
The rotations $B,C^I$ form an $\frak{u}(1)\oplus \frak{su}(2)$ subalgebra;
\begin{align}
  [B,C^I]	\ = \ 0\, ,\qquad [C^I, C^J] \ = \ -\varepsilon^{IJK}C^K \, .
\end{align}
In particular these generate all rotations in the four-dimensional plane that leave $\Omega_{ij}$ invariant. 
The remaining brackets are neatly summarised by noting that the `scalar' generators $S=P_-,T, K_+$ are inert under the rotation subgroup, i.e.\ $[S,B]=[S,C^I]=0$, while the `one-form' generators $W_i=P_i,M_{i+}$ transform as
\begin{align}
  	[W_i,B] 	\ &= \ -\tfrac{1}{2} R\,\Omega_{ij} W_j 	\, ,	\qquad [W_i,C^I] \	= \ \tfrac{1}{2}\eta^I_{ij} W_j \, .
\end{align}

As with any geometric symmetry, this algebra then admits a representation in terms of vector fields under commutation. Writing $(x^-, x^i)$ for the coordinates on our five-dimensional space, and $(\partial_-,\partial_i)$ for their derivatives, the vector field representation is then  \begin{align}
	\left( P_+ \right)_\partial 	\ &= \ 0 \, , 	\nn\\
	\left( P_- \right)_\partial \ 	&= \ \partial_-  \, ,  \nn\\
	\left( P_i \right)_\partial \ 	&= \ \tfrac{1}{2}\Omega_{ij} x^j \partial_- + \partial_i	\, , \nn\\
	\left( B \right)_\partial 		\ &= \ -\tfrac{1}{2}R\,\Omega_{ij}x^i\partial_j 		\, , \nn\\
	\left( C^I \right)_\partial 	\ &= \ \eta^I_{ij}x^i\partial_j 								\, , \nn\\
	\left( T \right)_\partial 		\ &= \ 2x^- \partial_- + x^i \partial_i					\, , \nn\\
	\left( M_{i+} \right)_\partial 	\ &= \ \left( \tfrac{1}{2}\Omega_{ij} x^- x^j - \tfrac{1}{8}R^{-2} x^j x^j x^i \right)\partial_- + x^- \partial_i  +\tfrac{1}{4}( 2\Omega_{ik}x^k x^j + 2\Omega_{jk}x^k x^i - \Omega_{ij}x^k x^k )\partial_j	\, , \nn\\
	\left( K_{+} \right)_\partial 	 \ &= \ ( 2 ( x^- )^2 - \tfrac{1}{8} R^{-2} ( x^i x^i )^2 )\partial_- +( \tfrac{1}{2} \Omega_{ij} x^j x^k x^k + 2 x^- x^i )\partial_i	\, .
	\label{eq: 5d algebra vector field rep}
\end{align}
In particular, since $\left( P_+ \right)_\partial=0$, this is just a representation of $\frak{su}(1,3)$.\\

We can finally build representations of the algebra (\ref{eq: extended su(1,3) algebra}). Although the algebra is not a conventional conformal algebra, it shares many properties with one. We still have a five-dimensional subalgebra of translations generated by $\{P_-, P_i\}$, although it is not Abelian, and a Lifshitz scaling $T$ which plays the role of the usual dilatation. Further, we still have pairs of ladder operators that raise and lower an operator or state's eigenvalues under $T$, except unlike usual conformal algebras there are two different gradations. The pair $(P_i, M_{i+})$  of raise and lower $T$ by one unit, while $(P_-, K_+)$ raise and lower by two units.

We therefore proceed in analogy with the familiar construction of conformal algebra representations. We first consider representations of the subalgebra that stabilises the origin $\{x^-=0, x^i=0\}$, which is generated by $\{B,C^I,T,M_{i+},K_+\}$. Then, an operator $\mathcal{O}(0)$ at the origin transforms as
\begin{align}
  	[\mathcal{O}(0),B] \ 	&= \ R_\mathcal{O}[B]\,\mathcal{O}(0)	\, ,\nn\\
  	[\mathcal{O}(0),C^I] 	\ &= \ R_\mathcal{O}[C^I]\,\mathcal{O}(0)\, .
\end{align}
for some representations $R_\mathcal{O}[B]$, $R_\mathcal{O}[C^I]$, while we also have $[\mathcal{O}(0),P_+]=-ip_+\mathcal{O}(0)$.

Further supposing that $R_\mathcal{O}[C^I]$ is irreducible, by Schur's lemma we must have
\begin{align}
	[\mathcal{O}(0),T] 				\ 	&= \ \Delta \mathcal{O}(0)	 \, ,	\nn\\
	[ \mathcal{O}(0),M_{i+}]		\ &= \ 0		\, ,					\nn\\
	[ \mathcal{O}(0),K_+]				\ &= \ 0		 \, ,					\nn\\
	[\mathcal{O}(0),P_+]					\ &= \ -ip_+\mathcal{O}(0) \, ,
	\label{eq: 5d alg at origin}
\end{align}
for some $p_+,\Delta\in\mathbb{C}$.

Then, an operator $\mathcal{O}(x)$ at a generic point $(x^-,x^i)$ is defined by
\begin{align}
  \mathcal{O}(x) \ = \ \exp\left( -x^- P_- - x^i P_i \right) \mathcal{O}(0) \exp\left( x^- P_- + x^i P_i \right) \, .
\end{align}
The action of $P_-$ and $P_i$ at generic points is then determined by requiring
\begin{align}
  \mathcal{O}(x+\epsilon) - \mathcal{O}(x) \ = \ \epsilon^- \partial_- \mathcal{O}(x) + \epsilon^i \partial_i\mathcal{O}(x) \, ,
\end{align}
to leading order in $\epsilon^-,\epsilon^i$. Making use of the relation
\begin{align}
  \exp\left( \left( x^-+\epsilon^- \right)P_- + \left( x^i + \epsilon^i \right)P_i \right) &= \exp\left( x^- P_- + x^i P_i \right)\exp\left( \epsilon^- P_- + \epsilon^j P_j \right)\exp\left( -\tfrac{1}{2}\Omega_{kl}\epsilon^k x^l P_- \right)	\nn\\
  &=\exp\left( \tfrac{1}{2}\Omega_{kl}\epsilon^k x^l P_- \right)\exp\left( \epsilon^- P_- + \epsilon^j P_j \right)\exp\left( x^- P_- + x^i P_i \right)\, ,
\end{align}
we find
\begin{align}
  	[\mathcal{O}(x),P_-]		\	 &= \ \partial_- \mathcal{O}(x)	\ = \ \left( P_- \right)_\partial \mathcal{O}(x)	\, , \nn\\
  	[\mathcal{O}(x),P_i]	\ &= \ \left( \partial_i + \tfrac{1}{2}\Omega_{ij} x^j \partial_- \right) \mathcal{O}(x) 	\ = \ \left( P_i \right)_\partial \mathcal{O}(x) \, ,
\end{align}
as expected. Using (\ref{eq: extended su(1,3) algebra}), we determine the action of the whole algebra on $\mathcal{O}(x)$ to be
\begin{align}
  	[\mathcal{O}(x),P_+]		\ &=	 \ -ip_+ \mathcal{O}(x)		\, ,												\nn\\
  	[\mathcal{O}(x),P_-]		\ &=	 \ \left( P_- \right)_\partial \mathcal{O}(x)	\, ,							\nn\\
  	[\mathcal{O}(x),P_i]		\ &= \ 	\left( P_i \right)_\partial \mathcal{O}(x)	\, ,							\nn\\
  	[\mathcal{O}(x),B]		\ &=	 \ \left( B \right)_\partial \mathcal{O}(x) + R_\Phi[B]   \mathcal{O}(x)\, ,	\nn\\
  	[\mathcal{O}(x),C^I]		\ &=	 \ \left( C^I \right)_\partial \mathcal{O}(x) + R_\Phi[C^I] \mathcal{O}(x)	\, ,\nn\\
  	[\mathcal{O}(x),T]		\ &= \	\left( T \right)_\partial \mathcal{O}(x) +  \Delta \mathcal{O}(x)	\, ,\nn\\
  	[\mathcal{O}(x),M_{i+}]	\ &= \	\left( M_{i+} \right)_\partial \mathcal{O}(x) + \left(\tfrac{1}{2}\Delta \Omega_{ij} x^j - ip_+ x^i	 +\tfrac{2}{R}x^i R_\Phi[B] - \Omega_{ik} \eta^I_{jk} x^j R_\Phi[C^I]\right) \mathcal{O}(x)			\, ,\nn\\
  	[\mathcal{O}(x),K_{+}]			\ &=	 \ \left( K_+ \right)_\partial \mathcal{O}(x)  + \left(2\Delta\, x^- - ip_+ x^i x^i +\tfrac{2}{R}x^i x^i R_\Phi[B] - x^i x^j \Omega_{ik}\eta^I_{jk} R_\Phi[C^I]   \right) \mathcal{O}(x) \, .
  	\label{eq: 5d algebra action on fields}
\end{align}
Then, under an infinitesimal conformal transformation generated by generator $G$, we have that the coordinates and a generic local operator $\mathcal{O}(x)$ transform as
\begin{align}
  	x^\mu \ &\to \ x'^\mu(x) \ = \ x^\mu + G_\partial \,x^\mu \, ,\nn\\
	\mathcal{O}(x) \ &\to \ \mathcal{O}' (x') \ = \ \mathcal{O}(x') - [\mathcal{O}(x'),G] \ = \ \mathcal{O}(x) + G_\partial \mathcal{O}(x) - [\mathcal{O}(x),G] \, .
\end{align}

\subsection{Ward Identities}

The general Ward identity associated with symmetry generator $G$ acting on an $N$-point correlation function is
\begin{align}
	0 \ =& \ \sum_{a=1}^N \langle \mathcal{O}^{(1)}(x^-_1,x^i_1) \cdots [\mathcal{O}^{(a)}(x^-_a,x^i_a),G] \cdots \mathcal{O}^{(N)}(x^-_N,x^i_N) \rangle \, ,
\end{align}
where each $\mathcal{O}^{(a)}(x^-_a,x^i_a)$ denotes a generic operator with associated eigenvalues\footnote{We have suppressed the $+$ subscript on $p_+$ here and in the remaining subsections to improve readability. Moreover, $a,b$ indices are never subject to the Einstein summation convention.} $\Delta_a,p_a$ inserted at spacetime point $(x^-_a,x^i_a)$. Focusing on scalar operators for which $R_\mathcal{O}[B]=0=R_\mathcal{O}[C^I]$, the Ward identities for the generators $\{P_+,P_-,P_i,B, C^I,T,M_{i+},K_+\}$ read %
{\allowdisplaybreaks
\begin{align}
	0 \ =& \ \sum_{a=1}^N \Bigg( - i p_a \Bigg) \langle \mathcal{O}^{(1)}(x^-_1,x^l_1) \cdots \mathcal{O}^{(N)}(x^-_N,x^l_N) \rangle \, , \label{WIpplus} \\
	0 \ =& \ \sum_{a=1}^N \Bigg(  \frac{\partial}{\partial x^-_a} \Bigg) \langle \mathcal{O}^{(1)}(x^-_1,x^l_1) \cdots \mathcal{O}^{(N)}(x^-_N,x^l_N) \rangle \, , \label{WIpminus} \\
	0 \ =& \ \sum_{a=1}^N \Bigg(  \frac{\partial}{\partial x^i_a} + \tfrac{1}{2}\Omega_{ij} x^j_a \frac{\partial}{\partial x^-_a} \Bigg) \langle \mathcal{O}^{(1)}(x^-_1,x^l_1) \cdots \mathcal{O}^{(N)}(x^-_N,x^l_N) \rangle \, , \label{WIpi} \\
	0 \ =& \ \sum_{a=1}^N  \Bigg(   - \tfrac{1}{2} R\,\Omega_{ij} x^i_a \frac{\partial}{\partial x^j_a} \Bigg) \langle \mathcal{O}^{(1)}(x^-_1,x^l_1) \cdots \mathcal{O}^{(N)}(x^-_N,x^l_N) \rangle \, , \label{WIB} \\
	0 \ =& \ \sum_{a=1}^N \Bigg(   \eta^I_{ij} x^i_a \frac{\partial}{\partial x^j_a} \Bigg) \langle \mathcal{O}^{(1)}(x^-_1,x^l_1) \cdots \mathcal{O}^{(N)}(x^-_N,x^l_N) \rangle \, , \label{WIC} \\
	0 \ =& \ \sum_{a=1}^N \Bigg(   2 x^-_a \frac{\partial}{\partial x^-_a} + x^i_a \frac{\partial}{\partial x^i_a} + \Delta_a \Bigg) \langle \mathcal{O}^{(1)}(x^-_1,x^l_1) \cdots \mathcal{O}^{(N)}(x^-_N,x^l_N) \rangle \, , \label{WIT} \\
	0 \ =& \ \sum_{a=1}^N \Bigg(   \big( \tfrac{1}{2}\Omega_{ij} x^-_a x^j_a - \tfrac{1}{8}R^{-2} x^j_a x^j_a x^i_a \big) \frac{\partial}{\partial x^-_a} + x^-_a \frac{\partial}{\partial x^i_a} + \tfrac{1}{2} \Delta_a \Omega_{ij} x^j_a - i p_a x^i_a \nn \\
	&\quad + \tfrac{1}{4} \big( 2\Omega_{ik}x^k_a x^j_a + 2\Omega_{jk}x^k_a x^i_a - \Omega_{ij}x^k_a x^k_a \big) \frac{\partial}{\partial x^j_a} \Bigg) \langle \mathcal{O}^{(1)}(x^-_1,x^l_1) \cdots \mathcal{O}^{(N)}(x^-_N,x^l_N) \rangle \, , \label{WIMiplus}\\
	0 \ =& \ \sum_{a=1}^N \Bigg( \big( 2 ( x^-_a )^2 - \tfrac{1}{8} R^{-2} ( x^i_a x^i_a )^2 \big) \frac{\partial}{\partial x^-_a} + 2\Delta x^-_a - i p_a x^i_a x^i_a \nn \\
	&\quad \qquad + \big( \tfrac{1}{2} \Omega_{ij} x^j_a x^k_a x^k_a + 2 x^-_a x^i_a \big) \frac{\partial}{\partial x^i_a} \Bigg) \langle \mathcal{O}^{(1)}(x^-_1,x^l_1) \cdots \mathcal{O}^{(N)}(x^-_N,x^l_N) \rangle \, .	 \label{WIKplus}
\end{align}
}
There are some consequences of these equations which hold for any $N$-point function; the first equation immediately imposes
\begin{align}
	0 \ =& \ \sum_{a=1}^N  p_a \, , \label{momemtumconserv}
\end{align} 
whilst equations \eqref{WIpminus} through \eqref{WIC} force the correlation function to be a function of the variables
\begin{align}
	\tilde{x}_{ab} \ \equiv & \ x^-_a - x^-_b + \tfrac{1}{2} \Omega_{ij} x^i_a x^j_b \, , \qquad  x_{ab}^i x_{cd}^i \ \equiv \ ( x_a^i - x_b^i ) ( x_c^i - x_d^i ) \, .
\end{align}
It will also be convenient to define:
\begin{equation}
\xi_{ab} \ \equiv \ \frac{x_{ab}^{i} x_{ab}^{i}}{\tilde{x}_{ab}} \, \ ,
\end{equation}
as well as  the complex combination
\begin{align}
z_{ab} \ &= \ \tilde{x}_{ab} + \frac{i}{4R} |x_{ab}|^2  \, . \label{Defzonetwo}
\end{align}

The  further constraints that the remaining equations \eqref{WIT} - \eqref{WIKplus},  corresponding to the generators $T,M_{i+}$ and $K_+$, place on correlation functions have to be considered on a case-by-case basis and we turn to that question now.

\subsection{2-point Functions}\label{Sec2ptFn}

We begin with the simplest non-trivial correlation function of scalar operators and define
\begin{align}
	\langle \mathcal{O}^{(1)}(x^-_1,x^i_1) \mathcal{O}^{(2)}(x^-_2,x^i_2) \rangle \ = \ F( \tilde{x}_{12},|x_{12}|^2 ) \, ,
\end{align}
with $|x_{ab}|^2=x_{ab}^i x_{ab}^i$. We take as an ansatz
\begin{align}
	F \ = \ ( \tilde{x}_{12} )^{-\alpha} G( \tilde{x}_{12} , |x_{12}|^2 ) \, , \label{2ptFnAnsatz}
\end{align}
for some constant $\alpha$.
Then the Ward identity \eqref{WIT} associated with Lifshitz scaling is solved if
\begin{align}
	0 \ =& \ \alpha - \frac{1}{2} ( \Delta_1 + \Delta_2 ) \, ,\nonumber \\
	0 \ =& \ \bigg( \tilde{x}_{12} \frac{\partial}{\partial \tilde{x}_{12}} + |x_{12}|^2 \frac{\partial}{\partial |x_{12}|^2} \bigg) G( \tilde{x}_{12} , |x_{12}|^2 ) \, .
\end{align}
The second condition forces the functional form of $G$ to be
\begin{align}
	G( \tilde{x}_{12} , |x_{12}|^2 ) \ = \ G\bigg( \frac{|x_{12}|^2}{\tilde{x}_{12}} \bigg) \ = \ G( \xi_{12} ) \, .
\end{align}
Moving onto the Ward identity associated with the operator $M_{i+}$ we find using $F=( \tilde{x}_{12} )^{-\frac{1}{2} ( \Delta_1 + \Delta_2 )} G(\xi_{12})$ leads to
\begin{align}
	0 \ =& \ \frac{1}{4} ( \Delta_1 - \Delta_2 ) \Omega_{ij} x^j_{12} G(\xi_{12}) + x_{12}^i \bigg[ \Big( \frac{1}{8R^2} \xi_{12}^2 + 2 \Big) \frac{\mathrm{d}}{\mathrm{d} \xi_{12}} + \frac{1}{16R^2} ( \Delta_1 + \Delta_2 ) \xi_{12} - i p_1 \bigg] G(\xi_{12}) \, .
\end{align}
Since $x_{12}^i$ and $\Omega_{ij} x^j_{12}$ are independent for each value of $i$ and assuming a non-trivial $G(\xi_{12})$, we have two equations for this Ward identity to hold: 
\begin{align}
	0 \ =& \ \frac{1}{4} ( \Delta_1 - \Delta_2 ) \, ,\nonumber \\
	0 \ =& \ \bigg[ \Big( \frac{1}{8R^2} \xi_{12}^2 + 2 \Big) \frac{\mathrm{d}}{\mathrm{d} \xi_{12}} + \frac{1}{16R^2} ( \Delta_1 + \Delta_2 ) \xi_{12} - i p_1 \bigg] G(\xi_{12}) \, .
\end{align}
Solving we find $\Delta_1=\Delta_2$ and
\begin{align}
	G(\xi_{12}) \ = \ C_{\Delta_1,p_1} ( \xi_{12}^2 + 16 R^2 )^{-\frac{1}{2} \Delta_1} e^{\left( 2i p_1 R \arctan\big( \tfrac{\xi_{12}}{4R} \big) \right)} \, ,
\end{align}  
where $C_{\Delta_1,p_1}$ is a constant which may depend on $\Delta_1$ and $p_1$. The above formula can be written more explicitly using the identity
\begin{equation}
e^{2i\arctan (x)}=\frac{i-x}{i+x}.
\end{equation}
The final Ward identity \eqref{WIKplus} places no further constraints on $G(\xi_{12})$.
Therefore the 2-point correlation function is fully determined up to a constant by the symmetries, in direct analogy with the more familiar case of the $SO(p,q)$ conformal group. It is given by
\begin{align}
	\langle \mathcal{O}^{(1)}(x^-_1,x^i_1) \mathcal{O}^{(2)}(x^-_2,x^i_2) \rangle \ = \ \delta_{0,p_1+p_2} \delta_{\Delta_1,\Delta_2} \frac{C_{\Delta_1,p_1}}{ [ \tilde{x}_{12} ( \xi_{12}^2 + 16 R^2 )^{\frac{1}{2}} ]^{\Delta_1}} e^{ 2i R p_1  \arctan\big( \tfrac{\xi_{12}}{4R} \big) } \, .
\end{align}

Using the definition (\ref{Defzonetwo}) may now factorise the 2-point function into holomorphic and anti-holomorphic parts;
\begin{align}
\langle \mathcal{O}^{(1)}(x^-_1,x^i_1) \mathcal{O}^{(2)}(x^-_2,x^i_2) \rangle \ =& \ \delta_{0,p_1+p_2}\delta_{\Delta_1,\Delta_2}C_{\Delta_1,p_1}
\left(\frac{1}{z_{12}}\right)^{\Delta_1/2 -p_1 R} \left(\frac{1}{\bar z_{12}}\right)^{ \Delta_1/2 +p_1 R}  \, .
\end{align}
This resembles a 2-point function in a two dimensional CFT. From the Lagrangian gauge theory perspective it can be decomposed into a perturbative piece with power-law decay times an oscillating non-perturbative piece associated  to a non-vanishing instanton number.

\subsection{3-point Functions}\label{subsec: 3pt from WIs}

We define
\begin{align}
	\langle \mathcal{O}^{(1)}(x^-_1,x^i_1) \mathcal{O}^{(2)}(x^-_2,x^i_2) \mathcal{O}^{(3)}(x^-_3,x^i_3) \rangle \ = \ F(\tilde{x}_{12},\tilde{x}_{23},\tilde{x}_{13},|x_{12}|^2,|x_{23}|^2,|x_{13}|^2 ) \, ,
\end{align}
where we have used the relation $x^i_{12}+x^i_{23}+x^i_{31}=0$ to remove\footnote{Note that the presence of $\Omega_{ij}$ in the definition of $\tilde{x}$ means that they do not satisfy $\tilde{x}_{12}+\tilde{x}_{23}+\tilde{x}_{31}=0$.} all cross terms $x_{ab}^i x_{cd}^i$ with $ab \neq cd$. %To simplify the notation we introduce $(w_1,w_2,w_3)=(\tilde{x}_{12},\tilde{x}_{23},\tilde{x}_{31})$ and $(y_1,y_2,y_3)=(|x_{12}|^2,|x_{23}|^2,|x_{13}|^2)$.
To start we take as an ansatz
\begin{align}
	F \ = \ ( \tilde{x}_{12} )^{-\alpha_{12}} ( \tilde{x}_{23} )^{-\alpha_{23}} ( \tilde{x}_{13} )^{-\alpha_{13}} G( \tilde{x}_{12},\tilde{x}_{23},\tilde{x}_{13},|x_{12}|^2,|x_{23}|^2,|x_{13}|^2 ) \, . \label{3ptFnAnsatz}
\end{align}
Then the Lifshitz Ward identity \eqref{WIT} is solved if
\begin{align}
	0 \ =& \ \sum_{a<b}^3 \alpha_{ab} - \frac{1}{2} \Delta_a \, , \label{3ptsumalphas} \\
	0 \ =& \ \bigg( \sum_{a < b}^3 \tilde{x}_{ab} \frac{\partial}{\partial \tilde{x}_{ab}} + |x_{ab}|^2 \frac{\partial}{\partial |x_{ab}|^2} \bigg) G( \tilde{x}_{12},\tilde{x}_{23},\tilde{x}_{13},|x_{12}|^2,|x_{23}|^2,|x_{13}|^2 ) \, . \label{3ptDiffDil}
\end{align}
Substituting the ansatz \eqref{3ptFnAnsatz} into the Ward identity \eqref{WIMiplus} associated with the generator $M_{i+}$ and using \eqref{3ptsumalphas} fixes
\begin{align}
	\alpha_{ab} =  \Delta_a + \Delta_b - \tfrac{1}{2} \Delta_\text{T}
	%\alpha_{13} \ = \ - \tfrac{1}{2} \Delta + \Delta_1 + \Delta_2 \, , \quad \alpha_{13} \ = \ - \tfrac{1}{2} \Delta + \Delta_2 + \Delta_3 \, , \quad \alpha_{13} \ = \ - \tfrac{1}{2} \Delta +\Delta_1 + \Delta_3 \, ,
	%
\end{align}
where $\Delta_\text{T}$ denotes the total scaling dimension, $\Delta_\text{T} = \Delta_1+\Delta_2+\Delta_3$, together with constraining the functional form of $G$ to be
\begin{align}
	G( \tilde{x}_{12},\tilde{x}_{23},\tilde{x}_{13},|x_{12}|^2,|x_{23}|^2,|x_{13}|^2 ) \ = \ G( \xi_{12},\xi_{23},\xi_{13} ) \, .
\end{align}
This functional form for $G$ also solves \eqref{3ptDiffDil}. Using all this information, the $M_{i+}$ Ward identity can be reduced to
\begin{align}
	0 \ =& \ \sum_{a < b}^3 x_{ab}^i \bigg[ \bigg( \frac{1}{8R^2} \xi_{ab}^2 + 2 \bigg) \frac{\partial}{\partial \xi_{ab}} + \frac{1}{8R^2} ( - \tfrac{1}{2} \Delta_\text{T} + \Delta_a + \Delta_b ) \xi_{ab} - \frac{i}{3} p_{ab} \bigg] G( \xi_{12},\xi_{23},\xi_{13} ) \, , 
\end{align}
where we have also defined\footnote{Here we have chosen not to impose total $p_+$ conservation given by \eqref{momemtumconserv} in order to display the most symmetric form for $G$.}
\begin{align}
	p_{ab} \ \equiv \ p_a-p_b \, .
\end{align}
A solution to these coupled partial differential equations is
\begin{align}
	G( \xi_{12},\xi_{23},\xi_{13} ) \ =& \ C_{123} \bigg[ \prod_{a<b}^3 ( \xi_{ab}^2 + 16 R^2 )^{\frac{1}{4} \Delta_\text{T} - \frac{1}{2} \Delta_a - \frac{1}{2} \Delta_b} e^{\left( \frac{2iR}{3} p_{ab} \arctan\big( \tfrac{\xi_{ab}}{4R} \big) \right)} \bigg] \nn \\
	&\qquad \times H\bigg( \sum_{a < b}^3 \arctan\big( \tfrac{\xi_{ab}}{4R} \big) \bigg) \, ,
\end{align}  
where $H$ is an arbitrary function of its argument and $C_{123}$ is a constant which may depend on the $\Delta$'s and $p$'s. As with 2-points, the remaining Ward identity \eqref{WIKplus} is satisfied automatically. The full 3-point function is
\begin{align}
	\langle &\mathcal{O}^{(1)}(x^-_1,x^i_1) \mathcal{O}^{(2)}(x^-_2,x^i_2) \mathcal{O}^{(3)}(x^-_3,x^i_3) \rangle \nn \\
	=& \ \delta_{0,p_1+p_2+p_3} C_{123} \bigg[ \prod_{a<b}^3  [ \tilde{x}_{ab} ( \xi_{ab}^2 + 16 R^2 )^{\frac{1}{2}} ]^{\frac{1}{2}\Delta_\text{T} - \Delta_a - \Delta_b} e^{\left( \frac{2iR}{3} p_{ab} \arctan\big( \tfrac{\xi_{ab}}{4R} \big) \right)} \bigg] \nn\\
	& \qquad \qquad \qquad \times H\bigg(  \sum_{a < b}^3 \arctan\big( \tfrac{\xi_{ab}}{4R} \big) \bigg) \, .
\end{align}
Using the identity
\begin{equation}
\arctan\alpha+\arctan\beta+\arctan\gamma=\arctan\left(\frac{\alpha+\beta+\gamma-\alpha\beta\gamma}{1-\alpha\beta-\beta\gamma-\gamma\alpha}\right) \, ,
\end{equation}
the undetermined function can also be written as
\begin{equation}
H=H\left(\frac{16R^2\left( \xi_{12}+\xi_{23}+\xi_{31} \right)-\xi_{12}\xi_{23}\xi_{31}}{16R^2-\xi_{12}\xi_{23}-\xi_{23}\xi_{31}-\xi_{31}\xi_{12}}\right) \, .
\end{equation}
Thus in theories with $SU(1,3)$ conformal symmetry, 3-point functions of scalar operators are determined only up to an arbitrary function. This is in contrast to Lorentzian CFT's where 3-point functions are completely fixed by conformal symmetry. We will later show that dimensional reduction fixes the form of this function.

Written in terms of the complex variables \eqref{Defzonetwo}, we again see factorisation into holomorphic and anti-holomorphic pieces times the undetermined function $H$:
\begin{align}
&\langle \mathcal{O}^{(1)}(x^-_1,x^i_1) \mathcal{O}^{(2)}(x^-_2,x^i_2) \mathcal{O}^{(3)}(x^-_3,x^i_3) \rangle\nonumber\\
&= \ \delta_{0,p_1+p_2+p_3}\, C_{123} \prod_{a<b}^3\left(\frac{1}{z_{ab}}\right)^{-\frac{1}{4}\Delta_\text{T} + \frac{1}{2}\Delta_a + \frac{1}{2}\Delta_b -\frac{R}{3} p_{ab}} 
\left(\frac{1}{\bar z_{ab}}\right)^{-\frac{1}{4}\Delta_\text{T} + \frac{1}{2}\Delta_a + \frac{1}{2}\Delta_b + \frac{R}{3} p_{ab}}H\left( \frac{z_{12}z_{23}z_{31}}{\bar{z}_{12}\bar{z}_{23}\bar{z}_{31}} \right)\label{eq: 3pt Ward identity solution}\ .
\end{align}
 
\subsection{4-point Functions}\label{Four-point Functions}

For 4-point and higher-point correlation functions we expect to see the appearance of analogues of cross-ratios which are annihilated by the $SU(1,3)$ generators given in \eqref{eq: 5d algebra vector field rep}. Inspection of the 2- and 3-point functions found in the previous subsections suggests that there are two types of variables to consider constructing invariants from. These are 
\begin{align}
	x'_{ab} \ = \ \tilde{x}_{ab} (\xi_{ab}^2 + 16R^2)^{\frac{1}{2}} \, , \qquad {\xi}'_{ab} \ = \ 4R \arctan\bigg( \frac{\xi_{ab}}{4R} \bigg) \, .
\end{align}
We find that cross ratios of the $x'$'s are indeed annihilated by all the generators however because the action of the $M_{i+}$ and $K_+$ generators on the ${\xi}'_{ab}$'s are
\begin{align}
	M_{i+}({\xi}'_{ab}) \ =& \ 2\, x_{ab}^i \, , \qquad K_+({\xi}'_{ab}) \ = \ 16  \left( |x_a|^2 - |x_b|^2  \right) \, ,
\end{align}
we see that it is linear combinations which will be invariant rather than cross ratios.

Specialising to the case of 4-points, there are two independent cross ratios of the $x'$'s;
\begin{align}
	u \ =& \ \frac{{x}_{12}'{x}_{34}'}{{x}_{13}'{x}_{24}'} \, , \qquad  v \ = \ \frac{{x}_{14}'{x}_{23}'}{{x}_{13}'{x}_{24}'} \, ,
\end{align}
and three independent combinations of the ${\xi}'$'s;
\begin{align}
	{\xi}'_{12} + {\xi}'_{24} + {\xi}'_{41} \, , \qquad {\xi}'_{13} + {\xi}'_{34} + {\xi}'_{41} \, , \qquad {\xi}'_{23} + {\xi}'_{34} + {\xi}'_{42} \, .
\end{align}
Hence at 4-points under the $SU(1,3)$ symmetry group there are five independent conformal invariants compared to the more familiar case of two cross ratios under $SO(p,q)$.

Starting from a generic function of these five variables, we may solve each of the Ward identities following the same steps as for 2- and 3-point functions. We eventually find\footnote{Once again, we have not imposed total $p_+$ conservation in order to present a more symmetrical result.}
\begin{align}
	\langle \mathcal{O}^{(1)}(x^-_1,x^i_1) &\mathcal{O}^{(2)}(x^-_2,x^i_2) \mathcal{O}^{(3)}(x^-_3,x^i_3) \mathcal{O}^{(4)}(x^-_4,x^i_4) \rangle \ \nn \\
	=& \ \delta_{0,p_1+p_2+p_3+p_4}  \, \bigg[ \prod_{a<b}^4 (x'_{ab})^{\frac{1}{6} \Delta_\text{T} - \frac{1}{2} ( \Delta_a + \Delta_b)}  e^{ i p_{ab} {\xi}'_{ab}/8 } \bigg] \nn \\
	&\qquad \times H\big( u , v , {\xi}'_{12} + {\xi}'_{24} + {\xi}'_{41} \, , {\xi}'_{13} + {\xi}'_{34} + {\xi}'_{41} \, , {\xi}'_{23} + {\xi}'_{34} + {\xi}'_{42} \big) \, ,
\end{align}
where $H$ is an undetermined function and now $\Delta_\text{T} = \Delta_1+\Delta_2+\Delta_3+\Delta_4$. %\PR{I've used Rishi's note where he absorbs some of the $\hat{x}$'s into $H$ so that the exponents can be written nicely}\PR{Factorisation?}\PR{Crossing symmetry}\PR{Generating function approach?}

It is also covenient to define an alternative basis for the space of conformal invariants involving the $\xi'_{ab}$, 
\begin{align}
\lambda_{1}\ &= \ {\xi}'_{12}+{\xi}'_{24}+{\xi}'_{43}+{\xi}'_{31} \ = \ +\left( {\xi}'_{12} + {\xi}'_{24} + {\xi}'_{41} \right)  - \left( {\xi}'_{13} + {\xi}'_{34} + {\xi}'_{41} \right),\nn \\
\lambda_{2} \ &= \ {\xi}'_{13}+{\xi}'_{32}+{\xi}'_{24}+{\xi}'_{41} \ = \ +\left( {\xi}'_{13} + {\xi}'_{34} + {\xi}'_{41} \right)  - \left( {\xi}'_{23} + {\xi}'_{34} + {\xi}'_{42} \right),\\
\lambda_{3}\ &= \ {\xi}'_{14}+{\xi}'_{43}+{\xi}'_{32}+{\xi}'_{21} \ = \ -\left( {\xi}'_{23} + {\xi}'_{34} + {\xi}'_{42} \right)  - \left( {\xi}'_{12} + {\xi}'_{24} + {\xi}'_{41} \right)\nn\ .
\end{align}
Then, for instance under the interchange $x_1\leftrightarrow x_2$ they transform as:
\begin{equation}
\left(u,v,\lambda_{1},\lambda_{2},\lambda_{3}\right)\ \longrightarrow \ \left(-u/v,1/v,\lambda_3,-\lambda_2,\lambda_1\right)\ ,
\end{equation}
with similar rules for the other five permutations of $S_3$.

Generically, we may then consider the resulting crossing equations constraining the function $H$. However, the main focus of this paper is a particular class of $SU(1,3)$ theories, explored in section \ref{sect: Null Conformal Compactification}, which admit a six-dimensional interpretation. For these theories, we will see that almost all crossing symmetry is broken by the parameters $p_a$ even if all the scaling dimensions are the same.\\ 

We will however later consider a maximally symmetric case, in which all scaling dimensions equal, $\Delta_{a}=\Delta$, and all $p_a$ have equal magnitude $|p_a|=p> 0$. Then, after a convenient redefinition of $H$, we can write
\begin{align}
  \langle \pm \pm \pm \pm  \rangle = \langle  \mathcal{O}^{(1)}_{\pm p}\mathcal{O}^{(2)}_{\pm p}\mathcal{O}^{(3)}_{\pm p}\mathcal{O}^{(4)}_{\pm p}\rangle = \left({x}_{13}'{x}_{24}'\right)^{-\Delta}\exp\left(\frac{i}{8}\sum_{a<b}p_{ab}{\xi}'_{ab}\right)G_{\pm\pm\pm\pm}\ ,
\label{eq: 4pt G}
\end{align}
where the $G_{\pm\pm\pm\pm}=G_{\pm\pm\pm\pm}(u,v,\lambda_{1},\lambda_{2},\lambda_{3})$. Clearly by the overall conservation of $P_+$ charge, only the combinations with two pluses and two minuses can be non-zero. We will in fact find that theories with a six-dimensional interpretation have only $G_{++--}$ and $G_{+-+-}$ non-zero, with all other orderings vanishing.

We may then consider the crossing relations of, say, $G_{++--}$ with itself. For example, the interchange $x_1\leftrightarrow x_2$ corresponds to
\begin{align}
  G_{++--}(u,v,\lambda_{1},\lambda_{2},\lambda_{3})  \,\,\longrightarrow\,\,  v^{-\Delta} G_{++--}(-u/v,1/v,\lambda_3,-\lambda_2,\lambda_1) \, .
  \label{eq: G++-- 12 crossing}
\end{align}

\subsection{Higher-point Functions}

At $\Npt$-points, the general solution to the conformal Ward identities is
\begin{align}
	&\langle \mathcal{O}^{(1)}(x^-_1,x^i_1) \dots \mathcal{O}^{(N)}(x^-_N,x^i_N) \rangle \ \nn \\
	&\qquad= \ \delta_{0,p_1+\dots+p_N}  \, \bigg[ \prod_{a<b}^N (x'_{ab})^{-\alpha_{ab}}  e^{i p_{ab} {\xi}'_{ab}/2N } \bigg]  H\left( \frac{x'_{ab}x'_{cd}}{x'_{ac}x'_{bd}},\xi'_{ab} +\xi'_{bc} + \xi'_{ca}  \right) \, ,
\end{align}
where the $\alpha_{ab}=\alpha_{ba}$ satisfy $\sum_{b\neq a}\alpha_{ab}=\Delta_a$ for each $a=1,\dots, N$. Here, $H$ is a generic function of $N^2-3N+1$ variables, which fall into two classes:
\begin{itemize}
  \item The $N(N-3)/2$ independent cross-ratios of the form
  \begin{align}
  \frac{x'_{ab}x'_{cd}}{x'_{ac}x'_{bd}}\ .
\end{align}
  \item The $(N-1)(N-2)/2$ independent triplets of the form 
  \begin{align}
  \xi'_{ab} +\xi'_{bc} + \xi'_{ca}\ .
\end{align}\
\end{itemize}
To make explicit contact with our results at 3- and 4-points, note that one can always choose $H$ such that for $N\ge 3$ the $\alpha_{ab}$ can be taken as
\begin{align}
  \alpha_{ab} = \frac{1}{N-2}\left( \Delta_a+\Delta_b \right) - \frac{1}{(N-1)(N-2)}\Delta_\text{T} \ .
\end{align}
Where as before, $\Delta_\text{T}$ denotes the total scaling dimension, $\Delta_\text{T}=\sum_a \Delta_a$

As we are now used to, this form for the $\Npt$-point function is more naturally written in terms of the complex variables $z_{ab}$. We find then that the general $\Npt$-point function is written as
\begin{align}
	&\langle \mathcal{O}^{(1)}(x^-_1,x^i_1) \dots \mathcal{O}^{(N)}(x^-_N,x^i_N) \rangle \ \nn \\
	&\qquad= \ \delta_{0,p_1+\dots+p_N}  \, \left[ \prod_{a<b}^N (z_{ab} \bar{z}_{ab})^{-\alpha_{ab}/2}  \left( \frac{z_{ab}}{\bar{z}_{ab}} \right)^{p_{ab}R/N}\right]  H\left( \frac{|z_{ab}||z_{cd}|}{|z_{ac}||z_{bd}|}, \frac{z_{ab}z_{bc}z_{ca}}{\bar{z}_{ab}\bar{z}_{bc}\bar{z}_{ca}}  \right) \, .
	\label{eq: general N-point function}
\end{align} 

\section{Relation to Six-Dimensional CFTs}\label{sect: Null Conformal Compactification}

\subsection{Null Conformal Compactification}

We start with six-dimensional Minkowski spacetime in lightcone coordinates with metric 
\begin{align}
	ds^2_M = \hat g_{\mu\nu}d\hat x^\mu d\hat x^\nu =  -2d\hat x^+d\hat x^- + d\hat x^id \hat x^i\ ,
	\label{eq: 6d Minkowski metric}
\end{align}
where $\mu\in\{+,-,i\}$, and perform the coordinate transformation
\begin{align}\label{eq: coordinate transformation}
\hat x^+ &= 2R \tan (x^+/2R) \, , \nonumber\\ 
\hat x^- &= x^- +\frac{1}{4R}x^ix^i\tan (x^+/2R)\, ,\nonumber\\   
\hat x^i &=x^i - \tan(x^+/2R) R\Omega_{ij}x^j	\ .
\end{align}
Here $\Omega_{ij}$ is the same anti-self-dual constant spatial 2-form as appears in section \ref{sect: Ward Identities in Five-Dimensions}. 
%\PR{Should we put this much earlier in the draft when we first introduce $\Omega$?} $\Omega_{ij}$ is anti-symmetric, anti-self-dual and satisfies
%\begin{align}
%\Omega_{ij}\Omega_{jk} = -R^{-2}\delta_{ik}	\, .
%\end{align}
This transformation leads to the  metric
\begin{align}
ds^2_M = \frac{-2dx^+(dx^- - \frac12 \Omega_{ij}x^jdx^i) + dx^idx^i}{\cos^2(x^+/2R)}\ .	
\label{5dmetric}
\end{align}
Following this we perform a Weyl transformation $ds^2_\Omega = {\cos^2(x^+/2R)}ds^2_M$ to find
\begin{align}
ds^2_\Omega = g_{\mu\nu}x^\mu x^\nu =  -2dx^+\left(dx^- - \frac12 \Omega_{ij}x^jdx^i\right) + dx^idx^i \ .
\label{eq: conformally compactified metric}	
\end{align} 
Note the range of $x^+\in (-\pi R,\pi R)$ is finite. Thus we have conformally compactified the $x^+$ direction of six-dimensional Minkowski space. Note also that $\partial/\partial x^+$ is a Killing vector of $ds^2_\Omega$, while it was only a conformal Killing vector (with non-trivial conformal factor) of $ds^2_M$. This metric appears naturally as the conformal boundary to AdS$_7$ in the construction of \cite{Pope:1999xg}, arising from a $U(1)$ fibration of AdS$_7$ over a non-compact $\mathbb{C}P^3$.

\subsection{Mapping of Symmetries and Operators}

We first show how the five-dimensional symmetry algebra (\ref{eq: extended su(1,3) algebra}) is recovered from the full six-dimensional conformal algebra upon this null compactification. Let $\{P^\text{6d}_\mu,M^\text{6d}_{\mu\nu},D^\text{6d},K^\text{6d}_\mu\}$ denote the usual basis for $\frak{so}(6,2)$ in lightcone coordinates. These correspond to the conformal Killing vectors of the Minkowski metric (\ref{eq: 6d Minkowski metric}), which we take to be
\begin{align}
  \left( P^{\text{(6d)}}_\mu \right)_\partial &= \hat{\partial}_\mu \, , & \omega &= 0  \, , \nn\\
\left( M^{\text{(6d)}}_{\mu\nu} \right)_\partial &= \hat{x}_\mu \hat{\partial}_\nu - \hat{x}_\nu \hat{\partial}_\mu \, , & \omega &= 0 \, , \nn\\
\left( D^{\text{(6d)}} \right)_\partial &= \hat{x}^\mu \hat{\partial}_\mu \, , & \omega &= 1 \, , \nn\\
\left( K^{\text{(6d)}}_\mu \right)_\partial &= \hat{x}_\nu \hat{x}^\nu \hat{\partial}_\mu - 2 \hat{x}_\mu \hat{x}^\nu \hat{\partial}_\nu \, , & \omega &= -2 \hat{x}_\mu \, . 
\label{eq: 6d conformal Killing vectors}
\end{align}
These vector fields $V$ then satisfy $\mathcal{L}_V \hat{g} = 2\omega \hat{g}$ for the conformal factor $\omega$ as specified. Then, after the coordinate transformation (\ref{eq: coordinate transformation}) and Weyl transformation, these are still a basis for the space of conformal Killing vectors, just with shifted conformal factors; we have $\mathcal{L}_V g = 2\tilde{\omega} g$, where $\tilde{\omega} = \omega - \tfrac{1}{2R}\tan\left( \frac{x^+}{2R} \right)V^+$ for each vector field $V$.

The metric $g_{\mu\nu}$ as in (\ref{eq: conformally compactified metric}) has a null isometry along the $x^+$ direction. In terms of the full algebra of conformal Killing vectors (\ref{eq: 6d conformal Killing vectors}), this is realised by the combination $P_+:=P^{\text{(6d)}}_++\tfrac{1}{4}\Omega_{ij} M^{\text{(6d)}}_{ij} + \tfrac{1}{8R^2}K^{\text{(6d)}}_-$, which has vector field representation $(P_+)_\partial=\partial_+$.

Our next step is to reduce the theory into modes along this $x^+$ direction. At the level of the symmetry algebra, this amounts to choosing a basis for the space of local operators which diagonalises $P_+$. Equivalently, we expand all six-dimensional local operators in a Fourier series in the $x^+$ direction, as will be explored in more detail below.

A mode $\mathcal{O}_{p_+}(x)$ in such a decomposition satisfies $[\mathcal{O}_{p_+}(x),P_+]=-ip_+\mathcal{O}_{p_+}(x)$, and hence falls into a representation of the the maximal subalgebra $\frak{h}\subseteq\frak{so}(6,2)$ satisfying $[\frak{h},P_+]=0$. We find then that $\frak{h}$ is isomorphic to a centrally extended $\frak{su}(1,3)$, with basis
 \begin{align}
  	P_+ 	&= 	P^{\text{(6d)}}_+ + \tfrac{1}{4}\Omega_{ij} M^{\text{(6d)}}_{ij} + \tfrac{1}{8R^2} K^{\text{(6d)}}_-	\, , \nn\\
  	P_-		&=	P^{\text{(6d)}}_-	\, ,														\nn\\
  	P_i		&=	P^{\text{(6d)}}_i + \tfrac{1}{2}\Omega_{ij} M^{\text{(6d)}}_{j-}	 \, ,						\nn\\
  	B		&=	-\tfrac{1}{4}R\,\Omega_{ij} M^{\text{(6d)}}_{ij}			\, ,				\nn\\
  	C^I		&=	\tfrac{1}{4}\eta^I_{ij} M^{\text{(6d)}}_{ij}		\, ,						\nn\\
  	T		&=	D^{\text{(6d)}}-M^{\text{(6d)}}_{+-}		\, ,												\nn\\
  	M_{i+}	&=	M^{\text{(6d)}}_{i+}-\tfrac{1}{4}\Omega_{ij} K^{\text{(6d)}}_j	\, ,						\nn\\
  	K_+		&= 	K^{\text{(6d)}}_+ \, ,
  	\label{eq: 5d alg in terms of 6d}
\end{align}
which satisfies the algebra (\ref{eq: extended su(1,3) algebra}). The corresponding six-dimensional algebra of vector fields in terms of the $x^\mu$ is then
\begin{align}
	\left( P_+ \right)_\partial 	&= \partial_+ 																								\, ,&\omega = 0	\, ,					\nn\\
	\left( P_- \right)_\partial 	&= \partial_- 																								\, ,&\omega = 0 \, ,					\nn\\
	\left( P_i \right)_\partial 	&= \tfrac{1}{2}\Omega_{ij} x^j \partial_- + \partial_i														\, ,&\omega = 0 \, ,					\nn\\
	\left( B \right)_\partial 		&= -\tfrac{1}{2}R\,\Omega_{ij}x^i\partial_j 																	\, ,&\omega = 0\, , 					\nn\\
	\left( C^I \right)_\partial 	&= \eta^I_{ij}x^i\partial_j 																					\, ,&\omega = 0 \, ,					\nn\\
	\left( T \right)_\partial 		&= 2x^- \partial_- + x^i \partial_i																			\, ,&\omega = 1 \, ,					\nn\\
	\left( M_{i+} \right)_\partial 	&= x^i \partial_+ + \left( \tfrac{1}{2}\Omega_{ij} x^- x^j - \tfrac{1}{8}R^{-2} x^j x^j x^i \right)\partial_- + x^- \partial_i  				\nn\\
			&  \qquad +\tfrac{1}{4}( 2\Omega_{ik}x^k x^j + 2\Omega_{jk}x^k x^i - \Omega_{ij}x^k x^k )\partial_j	\, , &\omega = \tfrac{1}{2}\Omega_{ij}x^j \, , 	\nn\\
	\left( K_{+} \right)_\partial 	&= x^i x^i \partial_+ + ( 2 ( x^- )^2 - \tfrac{1}{8} R^{-2} ( x^i x^i )^2 )\partial_-															\nn\\			
			&  \qquad +( \tfrac{1}{2} \Omega_{ij} x^j x^k x^k + 2 x^- x^i )\partial_i							\, ,&\omega = 2x^- 	\, , 		
\end{align}
which satisfy $\mathcal{L}_V g = 2\omega g$ with $\omega$ as given. The five-dimensional vector field representation (\ref{eq: 5d algebra vector field rep}) is then recovered from the push-forward of these vector fields with respect to the projection map $(x^+, x^-, x^i)\to (x^-, x^i)$. \\
 
Next we consider the mapping of a local operator $\hat {\cal O}(\hat x^+,\hat x^-,\hat x^i)$ on six-dimensional Minkowski space.   Let us perform the coordinate transformation $\hat x^\mu\to x^\mu$ in (\ref{eq: coordinate transformation}) along with the Weyl transformation $d s^2 = \cos^2(x^+/2R) d\hat s^2$   which takes us to the $\Omega$-deformed space and maps $\hat {\cal O}(\hat x)$ to
\begin{align}
	 {\cal O}(x^+,x^-,x^i) = \cos^{-\Delta}(x^+/2R)\hat {\cal O}(\hat x^+(x),\hat x^-(x),\hat x^i(x))\ ,
	 \label{eq: 6d operator mapping}
\end{align}
where for simplicity we assumed that $\hat {\cal O}$ is a scalar operator with conformal dimension $\Delta$. 
Note that,  for operators that satisfy  $ \hat {\cal O}(\hat x)\to 0$ sufficiently quickly, {\it i.e.} faster than $1/|\hat x^+|^{\Delta}$,  as $\hat x^+\to \pm\infty$ then 
\begin{align}
	 {\cal O}(-\pi R,x^-,x^i)= {\cal O}(\pi R,x^-,x^i)= 0 \ . 
\end{align}
Such operators, as well as others,  can be expanded  in a Fourier series on $x^+\in (-\pi R,\pi R)$.   In particular, the Fourier mode $\mathcal{O}_\n$ satisfying $[\mathcal{O}_\n(x^-, x^i),P_+]=-ip_+\mathcal{O}_\n = -i \tfrac{\n}{R}\mathcal{O}_\n$, {\it i.e.} $p_+=\tfrac{\n}{R}$ , is given by  
\begin{align}
	{\cal O}_\n(x^-,x^i) &= \frac{1}{2\pi R}\int_{-\pi R}^{\pi R}dx^+ e^{i\n x^+/R}{\cal O}(x^+,x^-,x^i)\nonumber\\
	& = \frac{1}{2\pi R}\int_{-\pi R}^{\pi R}dx^+ e^{i\n x^+/R}\cos^{-\Delta}(x^+/2R) \hat {\cal O}(2R\tan( x^+/2R), \hat x^-(x),\hat x^i(x)) \nonumber\\
	%& = \frac{1}{2\pi R}\int_{-\infty}^{\infty}\frac{d \hat x^+}{1+(\hat x^+)^2/4R^2} e^{2i\n \arctan(\hat x^+/2R)}\left(\frac{1}{1+(\hat x^+)^2/4R^2}\right)^{-\Delta/2}\hat {\cal O}(\hat x^+,\hat x^-(x),\hat x^i(x)) \nonumber\\
	%& =  \frac{1}{2\pi R}\int_{-\infty}^{\infty} d\hat x^+ \left(\frac{2R+i\hat x^+}{2R-i\hat x^+}\right)^{\n }\left(\frac{4R^2}{4R^2+(\hat x^+)^2}\right)^{1-\Delta/2}\hat {\cal O}(\hat x^+,\hat x^-(x),\hat x^i(x)) \nonumber\\
	& =  \frac{(2R)^{1-\Delta}}{  \pi }\int_{-\infty}^{\infty} d \hat x^+  \frac{(2R+i\hat x^+)^{\n +\Delta/2-1}}{(2R-i \hat x^+)^{\n -\Delta/2+1}}  \hat {\cal O}( \hat x^+,\hat x^-(x),\hat x^i(x))\nonumber \\ 
	&= \frac{(-1)^n}{  \pi }\int_{-\infty}^{\infty} d u  \frac{(u-i)^{\n +\Delta/2-1}}{(u+i)^{\n -\Delta/2+1}}  \hat {\cal O}( 2Ru,\hat x^-(x),\hat x^i(x))\ ,
	\label{eq: Fourier expansion}
\end{align}
where we introduced  $u =\hat x^+/2R= \tan(x^+/2R)$ to simplify the integral.  This is still quite complicated however we can use translational invariance to set $\hat{x}^i=x^i=0$ and  $ \hat{x}^-=x^-=0$ so that we have, more simply, 
\begin{align}
	{\cal O}_\n (0,0) &=\frac{(-1)^n}{  \pi }\int_{-\infty}^{\infty}  du \left(\frac{u-i}{u+i}\right)^{\n }\left({1+u^2}\right)^{\Delta/2-1}\hat {\cal O}(2Ru,0,0) \ ,
%	\hat {\cal O}(\hat x^+,\hat x^-,0) &= \left(\frac{4R^2}{4R^2+(\hat x^+)^2}\right)^{\Delta/2}\sum_n \left(\frac{2R+i\hat x^+}{2R-i\hat x^+}\right)^{nR/R_+}{\cal O}_\n (\hat x^-,0)
\end{align}
whose inverse is
\begin{align}
%	{\cal O}_\n (0,0) &=\frac{1}{2\pi R_+}\int_{-\infty}^{\infty}  d\hat x^+ \left(\frac{2R-i\hat x^+}{2R+i\hat x^+}\right)^{nR/R_+}\left(\frac{4R^2+(\hat x^+)^2}{4R^2}\right)^{\Delta/2-1}\hat {\cal O}(\hat x^+,x^-,0)  \nonumber\\
	 \hat {\cal O}( \hat x^+,0,0) &= \left(\frac{4R^2}{4R^2+( \hat x^+)^2}\right)^{\Delta/2}\sum_{\n \in{\mathbb Z}} \left(\frac{2R-i\hat x^+}{2R+i\hat x^+}\right)^{\n }{\cal O}_\n (0,0) \, .
\end{align}

Using this map between operators, we can determine the relationship between the quantum numbers of an operator $\hat {\mathcal{O}}(\hat {x})$ in six-dimensions and those of its Fourier modes $\mathcal{O}_\n (x)$. We specialise to scalar operators for simplicity so that $\hat {\mathcal{O}}({x})$ is wholly characterised by a scaling dimension $\Delta_\text{6d}$ such that $[\hat {\mathcal{O}}(\hat {x}),D^\text{6d}]=(\hat  {x}^\mu \hat {\partial}_\mu + \Delta_\text{6d} )\hat {\mathcal{O}}(\hat {x})$, and has no spin, {\it  i.e.} $[\hat{\mathcal{O}}(\hat {x}),M^\text{6d}_{\mu\nu}]=( \hat {x}_\mu \hat {\partial}_\nu - \hat {x}_\nu \hat {\partial}_\mu )\hat {\mathcal{O}}(\hat {x})$. Then, using the explicit forms (\ref{eq: 5d alg in terms of 6d}) we determine that the Fourier mode $\mathcal{O}_\n $ as defined in (\ref{eq: Fourier expansion}) is a scalar operator in a five-dimensional sense:  $R_{\mathcal{O}_\n }[B]=R_{\mathcal{O}_\n }[C^I]=0$, with scaling dimension $\Delta$ as defined in (\ref{eq: 5d alg at origin}) given simply by $\Delta=\Delta_\text{6d}$.

 \subsection{2-point Functions}\label{subsec: 2pt dim red}

Let us consider a 2-point function in the six-dimensional theory of the form
\begin{equation}
	\big\langle  \hat {\cal O}^{(1)}({\hat x^+_1,\hat x_1^-,\hat x^i_1})\hat {\cal O}^{(2)}({\hat x^+_2,\hat x_2^-,\hat x^i_2}) \big\rangle = \frac{\hat C_{12}}{|\hat x_1-\hat x_2 |^{2\Delta}}\ .
	\label{eq: 6d 2pt}
\end{equation}
where $\hat{\mathcal{O}}^{(1,2)}$ both have scaling dimension $\Delta$. Using the reduction procedure described above, the five-dimensional 2-point function of the Fourier modes are given by
\begin{align}
	\left\langle \mathcal{O}^{(1)}_{n_1}\mathcal{O}^{(2)}_{n_2}\right\rangle &=\frac{\left( -1 \right)^{n_1+n_2}}{\pi^2}\int_{-\infty}^{\infty}\,d^2 u\prod_{a=1}^{2}{\left(u_{a}+i\right)^{-n_{a}+\Delta/2-1}\left(u_{a}-i\right)^{n_{a}+\Delta/2-1}}\nn\\
	&\hspace{40mm}\times \langle  \hat {\cal O}^{(1)}({2Ru_1,\hat x_1^-,\hat x^i_1})\hat {\cal O}^{(2)}({2Ru_2,\hat x_2^-,\hat x^i_2}) \rangle\nn\\[0.7em]
	&=\frac{\left( -1 \right)^{n_1+n_2}}{\pi^2}\hat C_{12}\int_{-\infty}^{\infty}\,d^2 u\,\frac{\Pi_{a=1}^{2}{\left(u_{a}+i\right)^{-n_{a}+\Delta/2-1}\left(u_{a}-i\right)^{n_{a}+\Delta/2-1}}}{(4R(u_2-u_1)(\hat x^-_1-\hat x^-_2)+(\hat x^i_1-\hat x_2^i)^2)^\Delta}\nn\\[0.7em]
&=\frac{\left( -1 \right)^{n_1+n_2}\hat C_{12}}{\pi^2(4 R)^{\Delta}}\frac{1}{\tilde{x}_{12}^{\Delta}}\int_{-\infty}^{\infty}\,\,d^2 u\,\frac{\Pi_{a=1}^{2}\left(u_{a}+i\right)^{-n_{a}+\Delta/2-1}\left(u_{a}-i\right)^{n_{a}+\Delta/2-1}}{\left(u_{2}-u_{1}+(1+u_{1}u_{2})\xi_{12}/4R\right)^{\Delta}}\ .	
	\label{eq: 2pt dim red integral}
\end{align} 
Here and for the rest of this paper, we focus on protected operators in six-dimensions with $\Delta\in2\mathbb{Z}$, and thus the integrand is free of branch points. Nonetheless, the integral is ill-defined; it has a whole curve of poles in the $u_1-u_2$ plane, in particular at
\begin{equation}
u_{2}=\frac{u_{1}-\xi_{12}/4R}{1+u_{1}\xi_{12}/4R}=v_{2}\ .
\end{equation}
This simply corresponds to the line of points at which the six-dimensional 2-point function diverges:\ when the two operators are light-like separated. Thankfully however the 2-point function $\langle \mathcal{O}^{(1)}_{n_1}\mathcal{O}^{(2)}_{n_2}\rangle$ is well-defined, and admits an integral representation given by a particular regularisation of (\ref{eq: 2pt dim red integral}). To see this, we note that the Lorentzian 2-point function (\ref{eq: 6d 2pt}) should more correctly be defined in terms of a Wick rotation of a six-dimensional Euclidean correlator. By considering this Wick rotation and its effect on the integral (\ref{eq: 2pt dim red integral}) more carefully, one determines this regularised integral, and hence a finite result for $\langle \mathcal{O}^{(1)}_{n_1}\mathcal{O}^{(2)}_{n_2}\rangle$.

The full details of this calculation can be found in appendix \ref{app: 2pt}, while here we present a heuristic description of its mechanism, and the answer it produces. At the computational level, the Wick rotation from Euclidean signature is neatly encapsulated by an $i\epsilon$ prescription, which has the familiar effect of shifting poles in the variables $u_a$ off the real line and into the complex plane. For instance, in (\ref{eq: 2pt dim red integral}) the effect of this prescription is to shift the pole at $u_2=v_2$ infinitesimally into the lower-half-plane, hence rendering the $u_2$ integral well-defined, and easily computable by closing the contour in the upper-half-plane and thus taking only the residue at $u_2=i$. The $u_1$ integral is then also finite and easily computed, to find
\begin{align}
	\langle \mathcal{O}^{(1)}_{n_1}\mathcal{O}^{(2)}_{n_2}\rangle &= \delta_{\n_1+\n_2,0}\, d\left( \Delta,\n_1 \right)\hat{C}_{12}\frac{1}{\tilde{x}_{12}^{\Delta}}\left(\frac{1+i\xi_{12}/4R}{1-i\xi_{12}/4R}\right)^{\n_1}\left(1+\xi_{12}^{2}/16R^2\right)^{-\Delta/2}\nn\\
	&=d(\Delta,\n_1)\hat{C}_{12}\frac{1}{\left(z_{12}\bar{z}_{12}\right)^{\Delta/2}}\left(\frac{z_{12}}{\bar{z}_{12}}\right)^{\n_1}\ .
	\label{eq: dim red 2pt from heuristics}
\end{align}
where the coefficient $d(\Delta,n)$ is found to be
\begin{align}
d(\Delta,n) = 	 \left( -2R\, i \right)^{-\Delta} {{\n+\tfrac{\Delta}{2}-1}\choose{\n-\tfrac{\Delta}{2}}}\ ,
\end{align} and we have  used the following convention for the generalised binomial coefficient here and throughout;
\begin{align}
  {\alpha\choose n} = \begin{cases}
 	\frac{\alpha(\alpha-1)\dots(\alpha-n+1)}{n!}	\qquad& n>0		\\
 	1	\qquad& n=0	\\
 	0	\qquad& n<0
 \end{cases}\ ,
\end{align}
for any $\alpha\in\mathbb{R}$, $n\in\mathbb{Z}$.

\subsubsection*{Fourier Resummation}

To verify that our choice of contour regularisation  is consistent we can perform the inverse procedure: resum the five-dimensional 2-point functions (\ref{eq: dim red 2pt from heuristics}) to get back to the six-dimensional 2-point function. So once again consider a scalar six-dimensional operator $\hat{\mathcal{O}}$ of scaling dimension $\Delta\in2\mathbb{Z}$. Then, writing $\hat{x}_{12}=\hat{x}_1-\hat{x}_2$ and $x_{12}=x_1-x_2$, we have
\begin{align}
  &\langle \hat{\mathcal{O}}^{(1)}(\hat{x}_1) \hat{\mathcal{O}}^{(2)}(\hat{x}_2) \rangle = \cos^\Delta\left( \tfrac{x_1^+}{2R} \right)\cos^\Delta\left( \tfrac{x_2^+}{2R} \right) \left\langle \mathcal{O}(x_1(\hat{x}_1)) \,\mathcal{O}(x_2(\hat{x}_2)) \right\rangle		\nn\\
  &\hspace{10mm}= \cos^\Delta\left( \tfrac{x_1^+}{2R} \right)\cos^\Delta\left( \tfrac{x_2^+}{2R} \right) \sum_{\n =\Delta/2}^\infty e^{-i\n x_{12}^+/R} \left\langle \mathcal{O}_{\n }(x_1^-, x_1^i) \mathcal{O}_{-\n }(x_2^-, x_2^i) \right\rangle		\nn\\
  &\hspace{10mm}= \hat{C}_{12}(-2R\, i)^{-\Delta}\cos^\Delta\left( \tfrac{x_1^+}{2R} \right)\cos^\Delta\left( \tfrac{x_2^+}{2R} \right) \left( z_{12}\bar{z}_{12} \right)^{-\tfrac{\Delta}{2}} \sum_{\n =\Delta/2}^\infty e^{-i\n x_{12}^+/R}\,\, {{\n +\tfrac{\Delta}{2}-1}\choose{\n -\tfrac{\Delta}{2}}}\left( \frac{z_{12}}{\bar{z}_{12}} \right)^{\n }	\nn\\
  &\hspace{10mm}= \hat{C}_{12} \left[ \frac{2R\, i \left(\bar{z}_{12} e^{ix_{12}^+/2R} - z_{12} e^{-i x_{12}^+/2R} \right)}{\,\cos\left( \tfrac{x_1^+}{2R} \right)\cos\left( \tfrac{x_2^+}{2R} \right)} \right]^{-\Delta} 	\nn\\
  &\hspace{10mm}= \hat{C}_{12} |-2 \hat{x}_{12}^+\hat{x}_{12}^- + \hat{x}_{12}^i \hat{x}_{12}^i|^{-\Delta}	\nn\\
  &\hspace{10mm}= \hat{C}_{12} |\hat{x}_{12}|^{-2\Delta}\ ,
\end{align}
as required. Note, as it is written, the infinite sum in this calculation does not converge. However the resummation is made precise by recalling the definition of the Lorentzian correlator $\langle \hat{\mathcal{O}}^{(1)}(\hat{x}_1) \hat{\mathcal{O}}(\hat{x}_2)^{(2)} \rangle$ in terms of the $i\epsilon$ prescription (\ref{eq: i epsilon}), encoding its Wick rotation from Euclidean signature. This in particular replaces $e^{-i\n x_{12}^+/R}\to e^{-\left( \epsilon_1-\epsilon_2 \right)n/R}e^{-i\n x_{12}^+/R}$, and thus $\epsilon_1>\epsilon_2$ ensures the convergence of the sum.

 This demonstrates how in principle the correlation functions of the six-dimensional CFT can be computed from the five-dimensional theory.
 
\subsection{3-point Functions}\label{subsec: 3pt dim red}

We can now pursue the dimensionally-reduced 3-point function in a similar manner.  Let us start with the six-dimensional 3-point function 
\begin{align}
  \big\langle \hat{\mathcal{O}}^{(1)}(\hat{x}_1) \hat{\mathcal{O}}^{(2)}(\hat{x}_2) \hat{\mathcal{O}}^{(3)}(\hat{x}_3) \big\rangle &= \frac{\hat{C}_{123}}{\left| \hat{x}_1-\hat{x}_2 \right|^{2\alpha_{12}}\left| \hat{x}_2-\hat{x}_3 \right|^{2\alpha_{23}}\left| \hat{x}_3-\hat{x}_1 \right|^{2\alpha_{31}}} \, ,
  \label{eq: 6d 3pt}
\end{align}
determined up to structure constants $\hat{C}_{123}$, and as in section \ref{subsec: 3pt from WIs},
\begin{align}
  \alpha_{ab} =  \Delta_a + \Delta_b - \tfrac{1}{2} \Delta_\text{T}\ .
\end{align}
We can now dimensionally reduce this to verify that it agrees with the solution to the five-dimensional Ward identity (\ref{eq: 3pt Ward identity solution})  and determine $H_{\n_1,\n_2,\n_3;\Delta_1,\Delta_2,\Delta_3} \equiv H$.

Once again, one must appeal to a more detailed treatment of Lorentzian correlators to perform this calculation, which is given in full detail in appendix \ref{app: 3pt}. For now, we simply state the result of this calculation,
\begin{align}
  &\big\langle \mathcal{O}^{(1)}_{n_1} \mathcal{O}^{(2)}_{n_2} \mathcal{O}^{(3)}_{n_3} \big\rangle\nn\\
  &\quad = \delta_{\n_1+\n_2+\n_3,0}\hat{C}_{123} \left( -2R\, i \right)^{-\tfrac{1}{2}\left( \Delta_1+\Delta_2+\Delta_3 \right)} \left( z_{12}\bar{z}_{12} \right)^{-\tfrac{1}{2}\alpha_{12}}\left( z_{23}\bar{z}_{23} \right)^{-\tfrac{1}{2}\alpha_{23}}\left( z_{31}\bar{z}_{31} \right)^{-\tfrac{1}{2}\alpha_{31}}\nn\\
  &\qquad\quad \times\sum_{m=0}^\infty {-\n_3-\tfrac{\Delta_3}{2}+\alpha_{23} - m -1\choose -\n_3-\tfrac{\Delta_3}{2}-m}{\n_1-\tfrac{\Delta_1}{2}+\alpha_{12}-m-1 \choose \n_1-\tfrac{\Delta_1}{2}-m}{\alpha_{31}+m-1 \choose m} \nn\\
  &\qquad\hspace{20mm}\times \left( \frac{z_{12}}{\bar{z}_{12}} \right)^{\n_1-m-\tfrac{1}{2}\alpha_{31}}\left( \frac{z_{23}}{\bar{z}_{23}} \right)^{-\n_3-m-\tfrac{1}{2}\alpha_{31}}\left( \frac{z_{31}}{\bar{z}_{31}} \right)^{-m-\tfrac{1}{2}\alpha_{31}}\ ,
  \label{eq: 3pt dim red from heuristics}
\end{align}
where, given $z=r e^{i\theta}$, we've chosen the branches $\left( z\bar{z} \right)^{1/2}=r$ and $\left( \tfrac{z}{\bar{z}} \right)^{1/2}=e^{i\theta}$. We note that the sum terminates at $\min\left( \n_1-\tfrac{\Delta_1}{2},-\n_3-\tfrac{\Delta_3}{2} \right)$, and thus as we have a finite, regularised result.

As with the 2-point function, the binomial coefficients encode the values of $\left( \Delta_a, \n_a \right)$ such that the 3-point function is non-vanishing. We immediately have that $\big\langle \mathcal{O}^{(1)}_{n_1} \mathcal{O}^{(2)}_{n_2} \mathcal{O}^{(3)}_{n_3} \big\rangle\neq 0$ requires $\n_1\ge \tfrac{\Delta_1}{2}$ and $\n_3\le -\tfrac{\Delta_3}{2}$. We find further constraints if either\footnote{At most one of the $\alpha_{ab}$'s can be non-positive.} $\alpha_{23}\le 0$ or $\alpha_{12}\le 0$. Writing these constraints back in terms of the conformal dimensions $\Delta_a$, we in particular have that if $\Delta_1\ge \Delta_2+\Delta_3$ we additionally need $\n_2=-\n_1-\n_3 \le -\tfrac{\Delta_2}{2}$, while if $\Delta_3\ge \Delta_2+\Delta_1$ we additionally need $\n_2=-\n_1-\n_3 \ge \tfrac{\Delta_2}{2}$.

We note also that the 3-point function admits a particularly compact form in terms of a contour integral of a generating function of two variables $w_1,w_2\in\mathbb{C}$,
\begin{align}
  &\hspace{-5mm}\big\langle \mathcal{O}^{(1)}_{n_1} \mathcal{O}^{(2)}_{n_2} \mathcal{O}^{(3)}_{n_3} \big\rangle\nn\\
   &= \delta_{\n_1+\n_2+\n_3,0}\hat{C}_{123} \left( -2R\, i \right)^{-\tfrac{1}{2}\left( \Delta_1+\Delta_2+\Delta_3 \right)}\nn\\
  &\quad \times \left( 2\pi i \right)^{-2}\oint_{|w_1|<1} dw_1 \oint_{|w_1|<1} dw_2\,\, \Big( w_1^{-n_1+\tfrac{\Delta_1}{2}-1}w_2^{n_3+\tfrac{\Delta_3}{2}-1} \left( \bar{z}_{23} - w_2 z_{23} \right)^{-\alpha_{23}}		\nn\\
  &\qquad \hspace{60mm} \times \left( \bar{z}_{12} - w_1 z_{12} \right)^{-\alpha_{12}}\left( z_{31} - w_1 w_2 \bar{z}_{31} \right)^{-\alpha_{31}} \Big)	\nn\\
  &= \delta_{\n_1+\n_2+\n_3,0}\hat{C}_{123} \left( -2R\, i \right)^{-\tfrac{1}{2}\left( \Delta_1+\Delta_2+\Delta_3 \right)}\nn\\
  &\quad \times \text{Res}_{\{w_1=0\}} \Big[ \text{Res}_{\{w_2=0\}}\,\, \Big[ w_1^{-n_1+\tfrac{\Delta_1}{2}-1}w_2^{n_3+\tfrac{\Delta_3}{2}-1} \left( \bar{z}_{23} - w_2 z_{23} \right)^{-\alpha_{23}}		\nn\\
  &\qquad \hspace{60mm} \times \left( \bar{z}_{12} - w_1 z_{12} \right)^{-\alpha_{12}}\left( z_{31} - w_1 w_2 \bar{z}_{31} \right)^{-\alpha_{31}} \Big]\Big]\ .
  \label{eq: 3pt gen func}
\end{align}

Lastly we can compare this result with the general solution (\ref{eq: 3pt Ward identity solution}) to ensure the two are consistent. This is indeed the case and  the function $H_{\n_1,\n_2,\n_3;\Delta_1,\Delta_2,\Delta_3}$ is determined to be
\begin{align}
  &\hspace{-5mm}H_{\n_1,\n_2,\n_3;\Delta_1,\Delta_2,\Delta_3} \nn\\
  &= \hat{C}_{123} \left( -2R\, i \right)^{-\tfrac{1}{2}\left( \Delta_1+\Delta_2+\Delta_3 \right)}\nn\\
  &\quad \times\sum_{m=0}^\infty {-\n_3-\tfrac{\Delta_3}{2}+\alpha_{23} - m -1\choose -\n_3-\tfrac{\Delta_3}{2}-m}{\n_1-\tfrac{\Delta_1}{2}+\alpha_{12}-m-1 \choose \n_1-\tfrac{\Delta_1}{2}-m}{\alpha_{31}+m-1 \choose m} \nn\\
  &\hspace{20mm}\times \left( \frac{z_{12}z_{23}z_{31}}{\bar{z}_{12}\bar{z}_{23}\bar{z}_{31}} \right)^{-\tfrac{1}{2}\alpha_{31}+\tfrac{1}{3}(\n_1-\n_3)-m}\ ,
  \label{eq: H at 3pts, finite k}
\end{align}
where, without loss of generality, we use the overall factor of $\delta_{\n_1+\n_2+\n_3,0}$ in $\big\langle \mathcal{O}^{(1)}_{n_1} \mathcal{O}^{(2)}_{n_2} \mathcal{O}^{(3)}_{n_3} \big\rangle$ to write $H_{\n_1,\n_2,\n_3;\Delta_1,\Delta_2,\Delta_3}$ in terms of only $\n_1$ and $\n_3$. We indeed see that as required, $H_{\n_1,\n_2,\n_3;\Delta_1,\Delta_2,\Delta_3}$ depends only on the argument of the product $z_{12}z_{23}z_{31}$.

\subsubsection*{Fourier resummation}

As we did for the 2-point function, we can verify the validity of our result (\ref{eq: 3pt dim red from heuristics}) by performing the inverse Fourier transform to recover the six-dimensional 3-point function (\ref{eq: 6d 3pt}). So, consider three six-dimensional operators $\hat{\mathcal{O}}^{(1)}, \hat{\mathcal{O}}^{(2)}, \hat{\mathcal{O}}^{(3)}$ with scaling dimensions $\Delta_1,\Delta_2,\Delta_3\in 2\mathbb{Z}$, respectively. Then, writing $\hat{x}_{ab}=\hat{x}_a-\hat{x}_b$ and $x_{ab}=x_a-x_b$, we have
\begin{align}
  &\langle \hat{\mathcal{O}}^{(1)}(\hat{x}_1) \hat{\mathcal{O}}^{(2)}(\hat{x}_2) \hat{\mathcal{O}}^{(3)}(\hat{x}_3) \rangle \nn\\
%  &\hspace{10mm}= \cos^{\Delta_1}\left( \tfrac{x_1^+}{2R} \right)\cos^{\Delta_2}\left( \tfrac{x_2^+}{2R} \right) \cos^{\Delta_3}\left( \tfrac{x_3^+}{2R} \right) \left\langle \mathcal{O}^{(1)}(x_1(\hat{x}_1)) \,\mathcal{O}^{(2)}(x_2(\hat{x}_2))\,\mathcal{O}^{(3)}(x_3(\hat{x}_3)) \right\rangle		\nn\\
%  &\hspace{10mm}= \cos^{\Delta_1}\left( \tfrac{x_1^+}{2R} \right)\cos^{\Delta_2}\left( \tfrac{x_2^+}{2R} \right) \cos^{\Delta_3}\left( \tfrac{x_3^+}{2R} \right) \nn\\
%   &\hspace{20mm}\times\sum_{\n_1=\Delta_1/2}^\infty \sum_{\n_3=\Delta_3/2}^\infty e^{-i\n_1x_{12}^+/R}e^{-i\n_3x_{23}^+/R} \left\langle \mathcal{O}^{(1)}_{\n_1}(x_1^-, x_1^i) \mathcal{O}^{(2)}_{\n_3-\n_1}(x_2^-, x_2^i) \mathcal{O}^{(2)}_{\n_3-\n_1}(x_2^-, x_2^i) \right\rangle		\nn\\
  &\hspace{10mm}= \hat{C}_{123} \left( -2R\, i \right)^{-\tfrac{1}{2}\left( \Delta_1+\Delta_2+\Delta_3 \right)} \left( \bar{z}_{12} \right)^{-\alpha_{12}}\left( \bar{z}_{23} \right)^{-\alpha_{23}}\left( z_{31} \right)^{-\alpha_{31}}\nn\\
  &\hspace{20mm}\times \cos^{\Delta_1}\left( \tfrac{x_1^+}{2R} \right)\cos^{\Delta_2}\left( \tfrac{x_2^+}{2R} \right) \cos^{\Delta_3}\left( \tfrac{x_3^+}{2R} \right)e^{-i\Delta_1 x_{12}^+/2R}e^{-i\Delta_3 x_{23}^+/2R} \mathcal{S}\ ,
\end{align}
where
\begin{align}
  \mathcal{S} &= \sum_{m=0}^\infty \sum_{\n_1=0}^\infty \sum_{\n_3=0}^\infty  e^{-i\n_1x_{12}^+/R}e^{-i\n_3x_{23}^+/R} \nn\\
  &\hspace{30mm}\times {\n_3+\alpha_{23} - m -1\choose \n_3-m}{\n_1+\alpha_{12}-m-1 \choose \n_1-m}{\alpha_{31}+m-1 \choose m}\nn\\
  &\hspace{30mm}\times \left( \frac{z_{12}}{\bar{z}_{12}} \right)^{\n_1-m}\left( \frac{z_{23}}{\bar{z}_{23}} \right)^{\n_3-m}\left( \frac{z_{31}}{\bar{z}_{31}} \right)^{-m}	\nn\\
%  &= \left( \sum_{m=0}^\infty {\alpha_{31}+m-1 \choose m} \left( \frac{\bar{z}_{31}}{z_{31}} e^{ix_{31}^+/R} \right)^m \right)	\nn\\
%  &\hspace{6mm}\times  \left( \sum_{\n_1=0}^\infty {\alpha_{12}+\n_1-1 \choose \n_1} \left( \frac{z_{12}}{\bar{z}_{12}} e^{-ix_{12}^+/R} \right)^{\n_1} \right)	\nn\\
%  &\hspace{6mm}\times  \left( \sum_{\n_3=0}^\infty {\alpha_{23}+\n_3-1 \choose \n_3} \left( \frac{z_{23}}{\bar{z}_{23}} e^{-ix_{23}^+/R} \right)^{\n_3} \right)	\nn\\
  &= \left( 1-\frac{\bar{z}_{31}}{z_{31}}e^{ix_{31}^+/R} \right)^{-\alpha_{31}}\left( 1-\frac{z_{12}}{\bar{z}_{12}}e^{-ix_{12}^+/R} \right)^{-\alpha_{12}}\left( 1-\frac{z_{23}}{\bar{z}_{23}}e^{-ix_{23}^+/R} \right)^{-\alpha_{23}}\ .
\end{align}
Hence, we find\footnote{As we saw at 2-point, the precise way to perform this resummation is to once again shift $x_a^+\to x_a^+-i\epsilon_a$ as defined in by the $i\epsilon$ prescription (\ref{eq: i epsilon}). It is then straightforwardly seen that the strict ordering $\epsilon_1>\epsilon_2>\epsilon_3>0$ ensures that all three infinite sums converge.} 
\begin{align}
  \left\langle \hat{\mathcal{O}}^{(1)}(\hat{x}_1) \hat{\mathcal{O}}^{(2)}(\hat{x}_2) \hat{\mathcal{O}}^{(3)}(\hat{x}_3) \right\rangle &= \hat{C}_{123} \prod_{a<b}^3 \left[ \frac{2R\, i \left(\bar{z}_{ab} e^{ix_{ab}^+/2R} - z_{ab} e^{-i x_{ab}^+/2R} \right)}{\,\cos\left( \tfrac{x_a^+}{2R} \right)\cos\left( \tfrac{x_b^+}{2R} \right)} \right]^{-\alpha_{ab}}\nn\\[1em]
  &= \hat{C}_{123}|\hat{x}_{12}|^{-2\alpha_{12}}|\hat{x}_{23}|^{-2\alpha_{23}}|\hat{x}_{31}|^{-2\alpha_{31}} \, .
\end{align}
as required.

\subsection{4-point Functions}

In the six-dimensional CFT, 4-point functions can be written in terms of general functions of two conformal cross ratios:
\begin{equation}
\hat{u}=\frac{\hat{x}_{12}^{2}\hat{x}_{34}^{2}}{\hat{x}_{13}^{2}\hat{x}_{24}^{2}},\,\,\,\hat{v}=\frac{\hat{x}_{14}^{2}\hat{x}_{23}^{2}}{\hat{x}_{13}^{2}\hat{x}_{24}^{2}}\ .
\end{equation}
Although 4-point functions are not fixed by conformal symmetry, they are heavily constrained by crossing symmetry. Under $1\leftrightarrow3$ exchange, they transform as follows:
\begin{equation}
(\hat{u},\hat{v})\rightarrow(\hat{v},\hat{u})\ .
\end{equation}
Under $1\leftrightarrow2$ exchange they transform as
\begin{equation}
(\hat{u},\hat{v})\rightarrow(\hat{u}/\hat{v},1/\hat{v})\ .
\end{equation}
For instance, consider a Lorentzian correlator of four scalar operators with identical scaling dimensions. In this case, the general solution to the conformal Ward identities is
\begin{equation}
\langle \hat{\mathcal{O}}^{(1)}(\hat{x}_1)\hat{\mathcal{O}}^{(2)}(\hat{x}_2)\hat{\mathcal{O}}^{(3)}(\hat{x}_3)\hat{\mathcal{O}}^{(4)}(\hat{x}_4)\rangle =\frac{G(\hat{u},\hat{v})}{\left(\hat{x}_{13}^{2}\hat{x}_{24}^{2}\right)^{\Delta}}\ ,
\end{equation}
where $G$ is an unspecified function. Invariance under crossing implies the following constraints on $G$:
\begin{align}
x_1\leftrightarrow x_3 &: \,\,\,G(\hat{u},\hat{v})=G(\hat{v},\hat{u})	\nn\\
x_1\leftrightarrow x_2 &: \,\,\,G(\hat{u}/\hat{v},1/\hat{v})=\hat{v}^{\Delta}G(\hat{u},\hat{v})\ .
\end{align}
\begin{comment}
\begin{center}
	\captionof{figure}{}\label{free}
	\includegraphics[width=12cm]{free_disc.jpg}
\end{center}
\end{comment}

Since 4-point functions are not fixed by symmetries, in order to proceed without making an assumption about dynamics we will consider disconnected correlators in a generalised free theory. Generalised free correlators are defined as those that decompose into products of 2-point correlators, and can be formally defined in theories like the six-dimensional $(2,0)$ theory \cite{Arutyunov:2002ff}.  Given a four six-dimensional scalar operators $\hat{\mathcal{O}}^{(a)}(\hat{x}_a)$ with scaling dimensions $\Delta_a$, the disconnected free 4-point function is 
\begin{align}
  \big\langle  \hat{\mathcal{O}}^{(1)}\hat{\mathcal{O}}^{(2)}\hat{\mathcal{O}}^{(3)}\hat{\mathcal{O}}^{(4)} \big\rangle = \sum_{(ab,cd)\in I} \big\langle \hat{\mathcal{O}}^{(a)}\hat{\mathcal{O}}^{(b)} \big\rangle \big\langle \hat{\mathcal{O}}^{(c)}\hat{\mathcal{O}}^{(d)} \big\rangle\ ,
  \label{eq: 4pt gen free}
\end{align}
where we have suppressed the dependence on the $\hat{x}_a$. Here, we denote by $I$ the index set $I=\{(13,24),(12,34),(14,23)\}$\footnote{Note that the ordering of operators in the 2-point functions is fixed by that of the operators in the 4-point function, which can be seen explicitly by defining Lorentzian correlators in terms of suitable $i\epsilon$-regulated Wick rotated Euclidean correlators as in (\ref{eq: i epsilon}).}. 

We can proceed to then calculate the corresponding 4-point functions of the Fourier modes of the $\hat{\mathcal{O}}^{(a)}$. As we did at 2- and 3-points, we can determine $\big\langle \mathcal{O}^{(1)}_{n_1}\mathcal{O}^{(2)}_{n_2}\mathcal{O}^{(3)}_{n_3}\mathcal{O}^{(4)}_{n_4} \big\rangle$ as an integral over variables $u_a$ of the $i\epsilon$-regulated six-dimensional 4-point function. One finds however that the integrand and ordering of the $\epsilon_a$ are such that the expression factorises into pairs of the integral expression for the 2-point function (\ref{eq: deformed 2pt integral}). All of this is to say, the factorisation persists in the same form at the level of Fourier modes in the five-dimensional theory; we have
\begin{align}
  \left\langle \mathcal{O}^{(1)}_{n_1}\mathcal{O}^{(2)}_{n_2}\mathcal{O}^{(3)}_{n_3}\mathcal{O}^{(4)}_{n_4} \right\rangle  = \sum_{(ab,cd)\in I} \left\langle \mathcal{O}^{(a)}_{n_a}\mathcal{O}^{(b)}_{n_b} \right\rangle \left\langle \mathcal{O}^{(c)}_{n_c}\mathcal{O}^{(d)}_{n_d} \right\rangle \ ,
  \label{eq: dim red 4pt}
\end{align}
in terms of the 2-point functions (\ref{eq: dim red 2pt from heuristics}) we have already determined. This in particular vanishes unless there exists two pairs of the $n_a$, say $(n_a, n_b)$ and $(n_c, n_d)$, each obeying $p_+$ momentum conservation independently: $n_a+n_b=0, n_c+n_d=0$.

We note that this result is consistent with the general Ward identity solution (\ref{eq: general N-point function}) for the 4-point function, and hence determines the function $H$, but we omit details of this calculation.\\
\begin{comment}
It is finally important to verify that this result is consistent with the general Ward identity solution (\ref{eq: general N-point function}) for the 4-point function. This is indeed the case, where 
In more detail, we identify the function $H$ as
\begin{align}
  H = \sum_{(ab,cd)\in I}& \delta_{\Delta_a,\Delta_b}\delta_{\Delta_c,\Delta_d} \delta_{n_a+n_b,0}\delta_{n_c+n_d,0} \hat{C}_{ab} \hat{C}_{cd}\left( -2R\,i \right)^{-(\Delta_a+\Delta_c)}		{{\n_a+\tfrac{\Delta_a}{2}-1}\choose{\n_a-\tfrac{\Delta_a}{2}}}{{\n_c+\tfrac{\Delta_c}{2}-1}\choose{\n_c-\tfrac{\Delta_c}{2}}}\nn\\
  &\quad\times 
  \left( \frac{|z_{ac}||z_{bd}|}{|z_{ab}||z_{cd}|} \right)^{\left( \Delta_a+\Delta_c \right)/6}
  \left( \frac{|z_{ad}||z_{bc}|}{|z_{ab}||z_{cd}|} \right)^{\left( \Delta_a+\Delta_c \right)/6}		\nn\\
  &\quad \times
  \left( \frac{z_{ab}z_{bd}z_{da}}{\bar{z}_{ab}\bar{z}_{bd}\bar{z}_{da}} \right)^{n_a/4} 
  \left( \frac{z_{ab}z_{bc}z_{ca}}{\bar{z}_{ab}\bar{z}_{bc}\bar{z}_{ca}} \right)^{n_a/4}
  \left( \frac{z_{bc}z_{cd}z_{da}}{\bar{z}_{bc}\bar{z}_{cd}\bar{z}_{da}} \right)^{n_c/4}
  \left( \frac{z_{ac}z_{cd}z_{da}}{\bar{z}_{ac}\bar{z}_{cd}\bar{z}_{da}} \right)^{n_c/4}
\end{align}
which is indeed a function only of $SU(1,3)$ invariants.
\end{comment}

For the sake of clarity, we finally consider a particular example of the generalised free 4-point function of Fourier modes, in which all of the scaling dimensions are equal, $\Delta_a=\Delta$, and all $P_+$ momenta are of equal magnitude, $|n_a|=n\ge \Delta/2$. Up to normalisation, this corresponds simply to
\begin{equation}\big\langle  \hat{\mathcal{O}}^{(1)}\hat{\mathcal{O}}^{(2)}\hat{\mathcal{O}}^{(3)}\hat{\mathcal{O}}^{(4)} \big\rangle=\frac{1}{\left(\hat{x}_{12}^{2}\hat{x}_{34}^{2}\right)^{\Delta}}+\frac{1}{\left(\hat{x}_{14}^{2}\hat{x}_{23}^{2}\right)^{\Delta}}+\frac{1}{\left(\hat{x}_{13}^{2}\hat{x}_{24}^{2}\right)^{\Delta}} \, .
\end{equation}
Using (\ref{eq: dim red 4pt}) as well as the vanishing conditions on 2-point functions, we find only two orderings of operators such that their 4-point functions are non-zero. Firstly, we have
\begin{align}
  \big\langle \mathcal{O}^{(1)}_{n}\mathcal{O}^{(2)}_{n}\mathcal{O}^{(3)}_{-n}\mathcal{O}^{(4)}_{-n} \big\rangle &= \big\langle \mathcal{O}^{(1)}_{n}\mathcal{O}^{(3)}_{-n} \big\rangle \big\langle \mathcal{O}^{(2)}_{n}\mathcal{O}^{(4)}_{-n} \big\rangle + \big\langle \mathcal{O}^{(1)}_{n}\mathcal{O}^{(4)}_{-n} \big\rangle \big\langle \mathcal{O}^{(2)}_{n}\mathcal{O}^{(3)}_{-n} \big\rangle\nn\\
  &\quad\propto\left(x_{13}'x_{24}'\right)^{-\Delta}e^{i \n\left({\xi}'_{13}+{\xi}'_{24}\right)/2R}	 + \left(x_{14}'x_{23}'\right)^{-\Delta}e^{i \n\left({\xi}'_{14}+{\xi}'_{23}\right)/2R}	\ ,
\end{align}
which has a manifest $\mathbb{Z}_2\times \mathbb{Z}_2$ crossing symmetry corresponding to $x_1\leftrightarrow x_2$ and $x_3\leftrightarrow x_4$. Secondly, we have
\begin{align}
  \big\langle \mathcal{O}^{(1)}_{n}\mathcal{O}^{(2)}_{-n}\mathcal{O}^{(3)}_{n}\mathcal{O}^{(4)}_{-n} \big\rangle &= \big\langle \mathcal{O}^{(1)}_{n}\mathcal{O}^{(2)}_{-n} \big\rangle \big\langle \mathcal{O}^{(3)}_{n}\mathcal{O}^{(4)}_{-n} \big\rangle 	\nn\\
  &\quad\propto\left(x_{12}'x_{34}'\right)^{-\Delta}e^{i \n\left({\xi}'_{12}+{\xi}'_{34}\right)/2R}	\ ,
\end{align}
which has a manifest $\mathbb{Z}_2$ crossing symmetry corresponding to the simultaneous swap $(x_1,x_2)\leftrightarrow(x_3,x_4)$. We further that see that there is no crossing relation between these two results, which is evident from their representation in terms of 2-point functions.

Using the parameterisation in \eqref{eq: 4pt G}, these non-zero results correspond to
\begin{align}
  G_{++--} & = \  e^{i\n\lambda_2/4R}\left( 1+e^{-in\lambda_2/2R} v^{-\Delta} \right)		\nn\\
  G_{+-+-} &= \ e^{-in\lambda_3/4R}u^{-\Delta}\ .
\end{align}
We indeed see that $G_{++--}$ is for instance invariant under the $x_1\leftrightarrow x_2$ crossing transformation \eqref{eq: G++-- 12 crossing}. 

\subsection{Comparison with the Five-Dimensional Theory}\label{subsec: comparison with 5d theory}

Earlier in this section we obtained the correlation functions between the Fourier modes of six-dimensional operators. We explicitly saw at 2- and 3-points that the 
resulting  correlation functions solved the five-dimensional $SU(1,3)$ Ward identities. Furthermore the reduction of the 3-point function determined the form of the function $H$ in (\ref{eq: 3pt Ward identity solution}). The reduction required a careful regularisation using an $i\epsilon$ prescription  of  various contour integrals. To confirm our analysis at 2- and 3- points we showed how the full six-dimensional correlation functions could be reconstructed from an infinite sum over five-dimensional correlation functions labelled by the Fourier mode number $n$. Conversely this gives hope that the five-dimensional gauge theory is sufficient to compute observables in the six-dimensional CFT, albeit non-perturbatively. With this in mind we would like to make some comments about this identification. 

In five-dimensional Lagrangian theories the action appears with a coupling constant
\begin{align}
	S =\frac{1}{g^2_{YM}}\int dx^-d^4x {\cal L}\ ,
\end{align}
here $g^2_{YM}$ has dimensions of length so that the theory is always strongly coupled at short distances. It is natural to  interpret the Fourier mode number as the instanton number
\begin{equation}
n = \frac{1}{8\pi^2}\int d^4x {\rm tr}(F\wedge F)	\ ,
\end{equation}
arising from the topological $U(1)$ current $J_I$ which must therefore be normalised to 
\begin{align}\label{JInorm}
	J_I =\frac{1}{8\pi^2 R}\star{\rm tr}(F\wedge F)\ .
\end{align} 
Thus we see that $g^2_{YM}\propto R$.\\

With this identification in place we note that the five-dimensional  $\Npt$-point correlators naturally spilt into a perturbative piece, with power-law decay determined by the scaling dimension, and an oscillating non-perturbative  contribution controlled by the instanton number. The observation that only modes with $n\geq\Delta/2$ contribute to the six-dimensional 2-point function implies that only anti-instantons can propagate and also that zero-modes do not contribute. This is consistent with the fact that the six-dimensional self-duality condition on the three-form restricts us to anti-self-dual gauge fields in the five-dimensional theory. In addition it would seem that the purely perturbative sector of the five-dimensional theory does not contribute to the 2-point functions of the six-dimensional theory. Furthermore if we construct the operators from scalar fields of naive scaling dimension 2 this suggests that we need at least one Fourier mode per scalar field.

 In appendix \ref{app: dimensional reduction from scratch}, we will carry out the analogous calculation
in the DLCQ limit and demonstrate a close analogy with the non-relativistic
limit of the Feynman propagator, for which anti-particles don't propagate.

\section{Recovering the DLCQ description}\label{sect: Recovering the DLCQ description}
 
We now investigate an interesting limit of our construction, in which our operators and their correlators become those of six-dimensional Minkowski space compactified on a null direction.

It is helpful to first reiterate our geometric set-up. We have coordinates $(x^+, x^-, x^i)$ which cover six-dimensional Minkowski space $\mathbb{R}^{1,5}$, as defined in (\ref{eq: coordinate transformation}). While $x^-,x^i$ are non-compact coordinates, $x^+$ runs over a finite interval $x^+\in(-\pi R, \pi R)$. In the limit $R\to\infty$, the transformation (\ref{eq: coordinate transformation}) degenerates and $(x^+, x^-, x^i)$ become normal lightcone coordinates on $\mathbb{R}^{1,5}$.

We now consider splitting the interval $x^+\in(-\pi R, \pi R)$ into $k\in\mathbb{N}$ subintervals each of length $2\pi R/k = 2\pi R_+$ where $R_+=R/k$, as depicted in Figure \ref{fig: orbifold}. We can then reduce our space of operators to only those which repeat on each subinterval; in other words, those satisfying $\mathcal{O}(x^++2\pi R_+)=\mathcal{O}(x^+)$. This defines a $\mathbb{Z}_k$ orbifold of our geometry, and the theory living on it.\\ \\ \\ \\ \\
\begin{center}
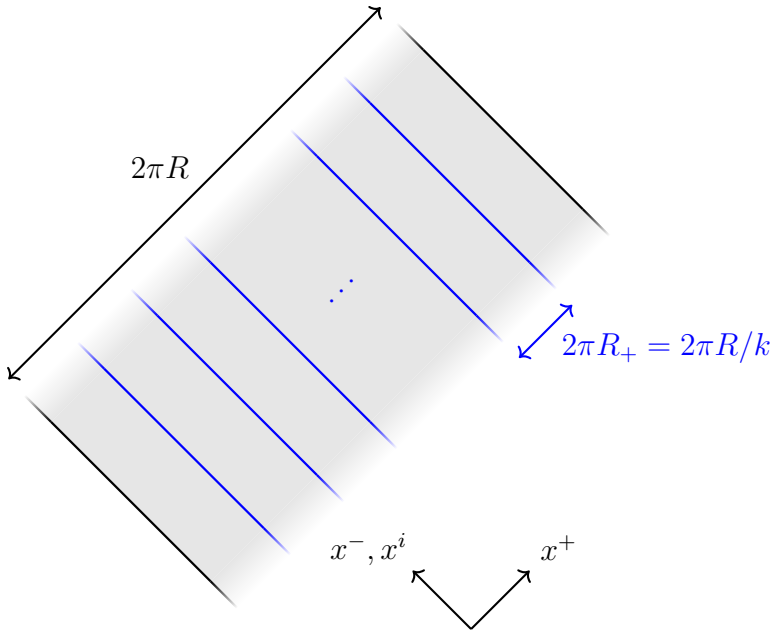

\captionof{figure}{The $\mathbb{Z}_k$ orbifold of the $x^+\in(-\pi R,\pi R)$ interval}\vspace{1em}\label{fig: orbifold}
\begin{minipage}[c][9cm]{\textwidth}\centering

\begin{tikzpicture}[transform canvas={rotate=45,shift={(-3.8,-0.8)}}]
	\definecolor{light-grey}{gray}{0.9}
	
		\fill[light-grey] (0,0) rectangle (7,3);
		\shade[top color=white, bottom color=light-grey] (0,3) rectangle (7,3.5);
		\shade[bottom color=white, top color=light-grey] (0,-0.5) rectangle (7,0);

		\fill[black] (-0.015,0) rectangle (0.015,3);
		\shade[top color=white, bottom color=black] (-0.015,3) rectangle (0.015,3.5);
		\shade[top color=black, bottom color=white] (-0.015,-0.5) rectangle (0.015,0);
		
		\fill[blue] (1-0.015,0) rectangle (1+0.015,3);
		\shade[top color=white, bottom color=blue] (1-0.015,3) rectangle (1+0.015,3.5);
		\shade[top color=blue, bottom color=white] (1-0.015,-0.5) rectangle (1+0.015,0);
		
		\fill[blue] (2-0.015,0) rectangle (2+0.015,3);
		\shade[top color=white, bottom color=blue] (2-0.015,3) rectangle (2+0.015,3.5);
		\shade[top color=blue, bottom color=white] (2-0.015,-0.5) rectangle (2+0.015,0);
		
		\fill[blue] (3-0.015,0) rectangle (3+0.015,3);
		\shade[top color=white, bottom color=blue] (3-0.015,3) rectangle (3+0.015,3.5);
		\shade[top color=blue, bottom color=white] (3-0.015,-0.5) rectangle (3+0.015,0);
		
		\node [blue] at (4,1.5) {$\dots$};
		
		\fill[blue] (5-0.015,0) rectangle (5+0.015,3);
		\shade[top color=white, bottom color=blue] (5-0.015,3) rectangle (5+0.015,3.5);
		\shade[top color=blue, bottom color=white] (5-0.015,-0.5) rectangle (5+0.015,0);
		
		\fill[blue] (6-0.015,0) rectangle (6+0.015,3);
		\shade[top color=white, bottom color=blue] (6-0.015,3) rectangle (6+0.015,3.5);
		\shade[top color=blue, bottom color=white] (6-0.015,-0.5) rectangle (6+0.015,0);
		
		\fill[black] (7-0.015,0) rectangle (7+0.015,3);
		\shade[top color=white, bottom color=black] (7-0.015,3) rectangle (7+0.015,3.5);
		\shade[top color=black, bottom color=white] (7-0.015,-0.5) rectangle (7+0.015,0);
		
		\draw [<->,thick] (0,3.8) -- (7,3.8);
		\draw [<->,thick,blue] (5,-0.8) -- (6,-0.8);
	
	\node [right,blue,rotate=-45] at (5.4,-1) { $2\pi R_+= 2\pi R/k$};	

	\node [left,rotate=-45] at (3.8,4) { $2\pi R$};
		
		\begin{scope}[shift={(0,0)}]
		\draw [->,thick] (2,-2.9) -- (3.1,-2.9);
		\draw [->,thick] (2,-2.9) -- (2,-1.8);
		\node [above right,rotate=-45] at (3.1,-2.9) {$x^+$};
		\node [above left,rotate=-45] at (2,-1.9) {$x^-,x^i$};
		\end{scope}

\end{tikzpicture}
\end{minipage}
\end{center}
Following our discussion in section \ref{subsec: comparison with 5d theory}, from the perspective of the five-dimensional theory, such an orbifold corresponds to taking coupling
\begin{align}
  g_\text{YM}^2 \propto R/k\ ,
\end{align}
such that the $\n$-instanton configuration is now identified with the $(kn)^\text{th}$ Fourier mode. Hence, taking $k>>1$ pushes the strong coupling regime to distances $R/\k<<R$, and so can be viewed as a sort of weak coupling limit.

Let us briefly comment on the holographic interpretation of this orbifold, in the case that the six-dimensional CFT is the $(2,0)$ superconformal theory, dual to M-theory on $\text{AdS}_7\times S^4$. Viewing $\text{AdS}_7$ as a circle fibration over a non-compact $\mathbb{C}P^3$, one identifies $x^+$ as the coordinate along the fibre \cite{Lambert:2019jwi}. Hence, the $\mathbb{Z}_k$ orbifold we have described, in which the space of operators is restricted to those with periodicity $2\pi R_+$, corresponds to the same $\mathbb{Z}_k$ orbifold of this fibration. This is how the holograhic duality is defined for $k>1$. We note that this is analogous to what was done in the context of M2-brane theories \cite{Aharony:1997an}, where the dual geometry is $\text{AdS}_4 \times S^7$ and the $S^7$ is thought of as a circle fibration of a compact $\mathbb{C}P^3$, which is then subjected to a $\mathbb{Z}_k$ orbifold. In that case however, the orbifold was implemented by rescaling the fibre coordinate by $1/k$, and in the limit $k \rightarrow \infty$, the bulk theory reduced to IIA string theory on AdS$_4 \times \mathbb{C}P^3$. \\

Given this $\mathbb{Z}_k$ orbifold, we now consider the combined limit in which we take $R\to\infty$ and $k\to\infty$ while holding $R_+=R/k$ fixed. In this limit, $\left( x^+, x^-, x^i \right)$ become standard lightcone coordinates on $\mathbb{R}^{1,5}$, but all operators must be periodic with period $2\pi R_+$ along the null direction $x^+$. In other words, we arrive at a null compactification $x^+\sim x^+ + 2\pi R_+$ of Minkowski space, a background first considered for the M5-brane in the DLCQ proposal of \cite{Aharony:1997an,Aharony:1997th}. For this reason, we will refer to this combined limit as the \textit{DLCQ limit}.

 We note that the maximally supersymmetric Lagrangian $SU(1,3)$ theory \cite{Lambert:2019jwi,Lambert:2018lgt} simplifies in the DLCQ limit. The resulting theory's dynamics are constrained to the moduli space of anti-self-dual gauge fields in $\mathbb{R}^4$ \cite{Mouland:2019zjr}, in line with the original DLCQ proposal for the M5-brane.\\

In the remainder of this section, we will investigate the behaviour of correlators in the DLCQ limit. Mirroring our analysis of the finite $R$ theory, we will first explore the constraining power of (bosonic) symmetries on the correlators of Fourier modes on a null compactification, recovering and extending the results of \cite{Aharony:1997an}.

We will then seek the precise form of these five-dimensional Fourier mode correlators by dimensionally reducing known six-dimensional correlators. Equivalently, this will determine necessary conditions on a five-dimensional theory with the correct symmetries to admit a six-dimensional interpretation. Although this calculation can in principle be performed in a way analogous to the dimensional reduction performed at finite $R$ in section \ref{sect: Null Conformal Compactification}, in practice one encounters divergences essentially due to the infinite range of $x^+$ as we approach $R\to\infty$. We will therefore pursue the five-dimensional DLCQ correlators by considering the DLCQ limit of our results at finite $R$, determining the leading order asymptotics of the 2-point, 3-point and some special 4-point functions.
  
  In appendix \ref{app: dimensional reduction from scratch}, we present an alternative derivation of the DLCQ asymptotics of correlators, in which the dimensional reduction from six dimensions is performed from scratch. In doing so, we point out some analogies between the $i\epsilon$ prescription used at finite $R$ and the standard non-relativistic limit of Lorentzian theories.

\subsection{Ward Identity Constraints on Correlators}

We first suppose that we lie exactly at the DLCQ limit. Then, $\left( x^+, x^-, x^i \right)$ are standard lightcone coordinates on $\mathbb{R}^{1,5}$, but with periodic null coordinate $x^+ \sim x^+ + 2\pi R_+$. We first determine and solve the Ward identities which constrain the correlators of Fourier modes on this compact null direction $x^+$. At the level of symmetries, this once again simply corresponds to choosing a basis of operators which diagonalise the translation $\left( P_+ \right)_\partial = \partial_+$. Such operators then fall into representations of the maximal subalgebra of $\frak{so}(2,6)$ that commutes with $P_+$.

This subalgebra is guaranteed to include 16 generators obtained by simply taking the $R\to\infty$ of our $\frak{su}(1,3)$ generators at finite $R$. In fact  the algebra is enhanced as $R\to\infty$ by an additional 2 generators, corresponding to the subalgebra of spatial rotations $\frak{u}(1)\oplus \frak{su}(2) = \text{span}\left\{ B, C^I \right\}$ of rotations which preserve $\Omega_{ij}$ becoming  the full $\frak{su}(2)\oplus \frak{su}(2)\cong \frak{so}(4)$ \cite{Lambert:2019fne}. Hence, the correlators of Fourier modes on the compact null direction $x^+$ are constrained by the Ward identities corresponding to 18 symmetry generators in total. We now present the general solution to these Ward identities.\\
% Thus we can readily take the $R\to\infty$ limit, with $R_+$ fixed,  of our results in section \ref{sect: Ward Identities in Five-Dimensions} (although we need to rescale the coefficients to obtain non-vanishing answers, we will return to this issue later).

An operator $\mathcal{O}^\text{DCLQ}$ on our null compactified geometry can be written as a sum of Fourier modes $\mathcal{O}^\text{DCLQ}_n$, with $P_+$ eigenvalues $[\mathcal{O}^\text{DCLQ}_n,P_+]=-i\left( n/R_+ \right)\mathcal{O}^\text{DCLQ}_n$. The 2-point function of such operators is then completely fixed by the Ward identities to be
\begin{align}
  \left\langle \mathcal{O}^{(1),\text{DCLQ}}_{\n_1}\left( x_1^-, x_1^i \right) \mathcal{O}^{(2),\text{DCLQ}}_{\n_2}\left( x_2^-, x_2^i \right) \right\rangle \,\,\propto\,\, \delta_{n_1+n_2,0} \delta_{\Delta_1,\Delta_2} \left( x_{12}^- \right)^{-\Delta_1} \exp\left( \frac{i\n_1}{2R_+}\frac{|x_{12}^i|^2}{x_{12}^-} \right)\ .
  \label{eq: 2pt DLCQ WI solution}
\end{align}
as was first found in \cite{Aharony:1997an}. We note in particular that in contrast to the result at finite $R$, there is no longer any spatial power-law decay.

At 3-points, the general solution is
\begin{align}
  &\left\langle \mathcal{O}^{(1),\text{DCLQ}}_{\n_1}\left( x_1^-, x_1^i \right) \mathcal{O}^{(2),\text{DCLQ}}_{\n_2}\left( x_2^-, x_2^i \right) \mathcal{O}^{(3),\text{DCLQ}}_{\n_3}\left( x_3^-, x_3^i \right) \right\rangle\nn\\ &\quad\propto  \left[ \prod_{a<b}^3 \left( x^-_{ab} \right)^{\tfrac{1}{2}\Delta_{\text{T}} -\Delta_a-\Delta_b} e^{i\left( n_a-n_b \right)\xi_{ab}/6R_+} \right] H\left( \xi_{12}+\xi_{23}+\xi_{31} \right)\, .
  \label{eq: 3pt DLCQ WI solution}
\end{align}
where $H$ is a general function of one variable, and we abuse notation slightly by using $\xi_{ab}$ to denote the DLCQ limit of $\xi_{ab}$ at finite $R$,
\begin{align}
  \xi_{ab} = \lim_{R\to\infty} \left( \tfrac{|x_{ab}^i|^2}{x_{ab}^-+\tfrac{1}{2}\Omega_{ij} x_a^i x_b^j} \right) =  \frac{|x_{ab}^i|^2}{x_{ab}^-}\ .
  \label{eq: xi DLCQ limit}
\end{align}
More generally, at $\Npt$-points we have
\begin{align}
  &\left\langle \mathcal{O}^{(1),\text{DCLQ}}_{\n_1}\left( x_1^-, x_1^i \right) \dots \mathcal{O}^{(N),\text{DCLQ}}_{\n_N}\left( x_N^-, x_N^i \right) \right\rangle\nn\\ &\quad\propto  \left[ \prod_{a<b}^N \left( x^-_{ab} \right)^{-\alpha_{ab}} e^{i\left( n_a-n_b \right)\xi_{ab}/2NR_+} \right] H\left( \frac{x_{ab}^- x_{cd}^-}{x_{ac}^- x_{bd}^-},\xi_{ab}+\xi_{bc}+\xi_{ca} \right)\ .
  \label{eq: DLCQ Npt WI sol}
\end{align}
where the $\alpha_{ab}=\alpha_{ba}$ satisfy $\sum_{b\neq a}\alpha_{ab} = \Delta_a$ for each $a=1,\dots, N$.  

\subsection{Correlators from Six Dimensions}

We now  repeat our analysis from section \ref{sect: Null Conformal Compactification} and consider the dimensional reduction of known six-dimensional correlators to determine the correlators of Fourier modes on a null compactification exactly

Suppose first that we lie exactly at the DLCQ point $k,R\to\infty$. Given an operator $\mathcal{O}$ on the non-compact spacetime, the naive construction for an operator ${\cal O}^\text{DLCQ}$ on the null-compactified space $x^+\sim x^+ + 2\pi R_+$ is then a sum over images,
\begin{align}
  \mathcal{O}^\text{DLCQ}\left( x^+, x^-, x^i \right) = \sum_{\s\in\mathbb{Z}}  \mathcal{O} \left( x^+ + 2\pi R_+ \s, x^-, x^i \right)\ .
\end{align}
%is periodic but leads to divergences in correlation functions due to an infinite over-counting of repeated terms. 
%Given a six-dimensional operator $\mathcal{O}^\text{n.c.}$ on \textit{non-compact} Minkowski space, we can define
%\begin{align}
%  \mathcal{O}\left( x^+, x^-, x^i \right) = \sum_{\s\in\mathbb{Z}}  \mathcal{O}^\text{n.c.} \left( x^+ + 2\pi R_+ \s, x^-, x^i \right)
%\end{align}
%which is manifestly periodic on $x^+$ with period $2\pi R_+$. From the perspective of the original six-dimensional space, $\mathcal{O}$ is non-local. But if we now view the spacetime as null compactified $x^+\sim x^+ + 2\pi R_+$, $\mathcal{O}$ defines a local operator. Said another way, $\mathcal{O}$ is simply a sum of images of $\mathcal{O}^\text{n.c.}$ on the null circle.
The utility of this formulation is that we know the correlation functions of the operators $\mathcal{O}$. However, we soon run into issues if we try to use them to write down the correlation functions of our new compactified operators. In particular, for a generic $\Npt$-point function we have
\begin{align}
  &\left\langle\mathcal{O}^{(1),\text{DLCQ}} \left( x_1^+\right) \mathcal{O}^{(2),\text{DLCQ}} \left( x_2^+ \right) \dots \mathcal{O}^{(\Npt),\text{DLCQ}} \left( x_\Npt^+ \right)		\right\rangle		\nn\\
  & \quad= \sum_{\s_a\in \mathbb{Z}}  \left\langle\mathcal{O}^{(1) } \left( x_1^+ +2\pi R_+ \s_1 \right) \mathcal{O}^{(2) } \left( x_2^++2\pi R_+ \s_2\right) \dots \mathcal{O}^{(\Npt)} \left( x_\Npt^++2\pi R_+ \s_N\right)		\right\rangle		\nn\\
  & \quad= \sum_{\tilde{\s}_a\in \mathbb{Z}}  \left\langle\mathcal{O}^{(1) } \left( x_1^+ \right) \mathcal{O}^{(2) } \left( x_2^++2\pi R_+ \tilde{\s}_2\right) \dots \mathcal{O}^{(\Npt)} \left( x_\Npt^++2\pi R_+ \tilde{\s}_N\right)		\right\rangle\ ,
 \label{eq: DLCQ Npt divergent}
\end{align}
where $a=1,\dots,\Npt$, and we have suppressed dependence on the coordinates $x_a^-, x_a^i$. Here, we have used the translational symmetry of the six-dimensional correlators in the $x^+$ direction to move the first operators to $x_1^+$ for every term in the multiple sums. Hence, the terms being summed over have no dependence on $\tilde{s}_1$, yet we still sum over all $\tilde{s}_1$, introducing a divergence. Hence, the decomposition is ill-defined, and requires regularisation. Crucially, we do \textit{not} encounter this divergence in the finite $k,R$ theory, simply due to the finite range of $x^+\in(-\pi R, \pi R)$. We can therefore regularise the DLCQ $\Npt$-point function by approaching from finite $k,R$ more carefully. \\

So consider the behaviour of correlators at finite $k,R$ as we approach the DLCQ limit. We begin with the theory at finite $R$, and perform the $\mathbb{Z}_k$ orbifold as defined at the beginning of section \ref{sect: Recovering the DLCQ description}. This amounts to restricting to operators ${\cal O}^\text{orb}$ on our interval $x^+\in (-\pi R, \pi R)$ which have periodicity $2\pi R_+=2\pi R/k$. Such an operator can be written in terms of an operator $\mathcal{O}$ on the un-orbifolded spacetime by once again summing over images; but now, only $k$ images need to be summed over. As a formal device to simplify notation, we let $x^+$ be a periodic coordinate $x^+\sim x^+ + 2\pi R$. Then, this sum of images is written simply as
\begin{align}
  \mathcal{O}^\text{orb}\left( x^+, x^-, x^i \right) = \sum_{s=0}^{k-1} \mathcal{O} \left( x^++2\pi R_+ s, x^-, x^i \right)\ .
  \label{eq: orbifolded operator}
\end{align}
We can now once again seek the correlators of these compactified operators in terms of the known correlations of the $\mathcal{O}$. We have in particular,
\begin{align}
  &\left\langle\mathcal{O}^{(1),\text{orb}} \left( x_1^+\right) \mathcal{O}^{(2),\text{orb}} \left( x_2^+ \right) \dots \mathcal{O}^{(\Npt),\text{orb}} \left( x_\Npt^+ \right)		\right\rangle		\nn\\
  & \quad= \sum_{\s_a=0}^{k-1}  \left\langle\mathcal{O}^{(1) } \left( x_1^+ +2\pi R_+ \s_1 \right) \mathcal{O}^{(2) } \left( x_2^++2\pi R_+ \s_2\right) \dots \mathcal{O}^{(\Npt) } \left( x_\Npt^++2\pi R_+ \s_N\right)		\right\rangle		\nn\\
  & \quad= \sum_{\tilde{\s}_a=0}^{k-1}  \left\langle\mathcal{O}^{(1) } \left( x_1^+ \right) \mathcal{O}^{(2) } \left( x_2^++2\pi R_+ \tilde{\s}_2\right) \dots \mathcal{O}^{(\Npt) } \left( x_\Npt^++2\pi R_+ \tilde{\s}_N\right)		\right\rangle		\nn\\
  & \quad= k\sum_{\substack{\tilde{\s}_a=0\\a\neq 1}}^{k-1}  \left\langle\mathcal{O}^{(1) } \left( x_1^+ \right) \mathcal{O}^{(2) } \left( x_2^++2\pi R_+ \tilde{\s}_2\right) \dots \mathcal{O}^{(\Npt) } \left( x_\Npt^++2\pi R_+ \tilde{\s}_N\right)		\right\rangle	\ .
  \label{eq: DLCQ reg sum}
\end{align}
We have   utilised translational invariance, so that the summation variable that drops out contributes a multiplicative factor of $k$. Note, this use of translational invariance, in particular the $x^+$ translations invariance of the $\Npt$-point function of the $\mathcal{O}$'s, is slightly subtle. While $\partial/\partial x^+$ is not an isometry of six-dimensional Minkowski space, it \textit{is} an isometry of the Weyl rescaled metric $ds_\Omega$, on which the $\mathcal{O}$ are defined. Hence, this use of translational invariance is valid, as can be verified by assessing the functional forms of our known results.

Unsurprisingly, the $\Npt$-point function diverges at large $k$ due to the overall factor of $k$. In this way, $k$ provides a regulator for the divergence encountered in (\ref{eq: DLCQ Npt divergent}). It does however make sense to consider an asymptotic expansion of the $N$-point function of orbifolded operators as $k\to\infty$. In particular, we find that the 2-point function has leading order behaviour
\begin{align}
  \left\langle\mathcal{O}^{(1),\text{orb}} \left( x_1^+\right) \mathcal{O}^{(2),\text{orb}} \left( x_2^+ \right) 	\right\rangle \sim k \sum_{\s\in\mathbb{Z}} \left\langle\mathcal{O}^{(1) } \left( x_1^+ \right) \mathcal{O}^{(2) } \left( x_2^++2\pi R_+ \s\right)\right\rangle\ ,
  \label{eq: 2pt leading k}
\end{align}
where the coefficient of this leading order term is straightforwardly seen to converge by the known functional form of the 2-point function. Similarly, at 3-points we have
\begin{align}
  &\left\langle\mathcal{O}^{(1),\text{orb}} \left( x_1^+\right) \mathcal{O}^{(2),\text{orb}} \left( x_2^+ \right)\mathcal{O}^{(3),\text{orb}} \left( x_3^+ \right) 	\right\rangle \nn\\
  &\qquad \sim k \sum_{\s_a\in\mathbb{Z}} \left\langle\mathcal{O}^{(1) } \left( x_1^+ \right) \mathcal{O}^{(2) } \left( x_2^++2\pi R_+ \s_1\right)\mathcal{O}^{(3) } \left( x_3^++2\pi R_+ \s_2\right)\right\rangle\ ,
  \label{eq: 3pt leading k}
\end{align}
where once again, the function form of the 3-point function guarantees convergence of this leading order coefficient.

At higher points, we generically encounter yet more divergent behaviour, corresponding to the orbifolding of correlators with additional degrees of translational symmetry. This is explored in more detail in section \ref{subsubsec: DLCQ 4pt and beyond}. \\

We can finally consider the behaviour of the five-dimensional correlators of Fourier modes as we approach the DLCQ limit. Using the definition (\ref{eq: orbifolded operator}) of $\mathcal{O}^\text{orb}$ in terms of images of the operator $\mathcal{O}$, we find the expansion
\begin{align}
  \mathcal{O}^\text{orb}\left( x^+, x^-, x^i \right) = \sum_\n e^{-inx^+/R_+} \,\mathcal{O}^\text{orb}_{n}( x^-, x^i ) = \sum_\n e^{-inx^+/R_+}\Big( k\,\mathcal{O}_{kn}( x^-, x^i )\Big)\ ,
\end{align}
in terms of the Fourier modes of $\mathcal{O}$, i.e. $\mathcal{O}^\text{orb}_{n}=k\mathcal{O}_{kn}$. Then, the six-dimensional orbifolded $\Npt$-point function is reconstructed from Fourier mode correlators by
\begin{align}
  &\left\langle\mathcal{O}^{(1),\text{orb}} \left( x_1^+\right) \mathcal{O}^{(2),\text{orb}} \left( x_2^+ \right) \dots \mathcal{O}^{(\Npt),\text{orb}} \left( x_\Npt^+ \right)		\right\rangle		\nn\\
  & \quad= \sum_{n_a\in\mathbb{Z}} \exp\left( -\tfrac{i}{R_+} \textstyle{\sum_a} n_a x_a^+ \right)  \left\langle \mathcal{O}^{(1),\text{orb}}_{n_1}\mathcal{O}^{(2),\text{orb}}_{n_2}\dots \mathcal{O}^{(N),\text{orb}}_{n_N} \right\rangle 	\nn\\
  & \quad= \sum_{n_a\in\mathbb{Z}} \exp\left( -\tfrac{i}{R_+} \textstyle{\sum_a} n_a x_a^+ \right) \Big( k^N \left\langle \mathcal{O}^{(1)}_{kn_1}\mathcal{O}^{(2)}_{kn_2}\dots \mathcal{O}^{(N)}_{kn_N} \right\rangle \Big)\ ,
  \label{eq: orbifolded Npt decomposition}
\end{align}
where the sum over Fourier modes is regulated by the $i\epsilon$ prescription (\ref{eq: i epsilon}). We therefore have a formula by which to reconstruct six-dimensional correlators on the orbifolded spacetime from the five-dimensional correlators of the $(kn)^\text{th}$ Fourier modes of $\mathcal{O}$, which we have already calculated at 2-points (\ref{eq: dim red 2pt from heuristics}), 3-points (\ref{eq: 3pt dim red from heuristics}), and for a particular example at 4-points (\ref{eq: dim red 4pt}). We now present the leading order asymptotics of these results at large $k$, which can be used to define the DLCQ limit of the theory.\\

Note, appendix \ref{app: dimensional reduction from scratch} presents an alternative derivation of the leading order asymptotics of correlators as we approach the DLCQ limit, by considering the null dimensional reduction for six-dimensions from scratch.

\subsubsection{2-point Functions}

The six-dimensional 2-point function $\left\langle\mathcal{O}^{(1),\text{orb}} \mathcal{O}^{(2),\text{orb}} \right\rangle$ is decomposed into a sum over Fourier mode correlators $\big\langle\mathcal{O}^{(1),\text{orb}}_{n_1} \mathcal{O}^{(2),\text{orb}}_{n_2} \big\rangle$, which are then in turn determined in terms of Fourier mode correlators in the un-orbifolded theory. We are concerned with the leading order asymptotics of these correlators at large $k$, which from (\ref{eq: 2pt leading k}) are determined to appear at order $k$.

We find that $\big\langle\mathcal{O}^{(1),\text{orb}}_{n_1} \mathcal{O}^{(2),\text{orb}}_{n_2} \big\rangle$ vanishes, unless $\Delta_1=\Delta_2=\Delta$ and $n_1=-n_2=n>0$. In this case, we find the leading order asymptotics

\begin{align}
  \big\langle\mathcal{O}^{(1),\text{orb}}_{n} \mathcal{O}^{(2),\text{orb}}_{-n} \big\rangle &=k^2\big\langle \mathcal{O}^{(1)}_{\k \n}(x_1) \mathcal{O}^{(2)}_{-\k \n}(x_2)  \big\rangle \nn\\
 	  &\quad\sim k\,\frac{\hat{C}_\Delta \left( 2R_+ i \right)^{-\Delta} \n^{\Delta-1}}{\Gamma(\Delta)} \left( x_{12}^- \right)^{-\Delta} \exp \left( \frac{i\n}{2R_+} \frac{|x_{12}^i|^2}{x_{12}^-} \right) \ ,
 \label{eq: DLCQ 2pt from limit}
\end{align}
which is indeed consistent with the general solution (\ref{eq: 2pt DLCQ WI solution}) to the DLCQ Ward identities.

\subsubsection{3-point Functions}

The result at 3-points is most straightforwardly approached from the generating function representation (\ref{eq: 3pt gen func}) for the 3-point function at finite $k$. The resulting 3-point functions $\langle \mathcal{O}^{(1),\text{orb}}_{\n_1}(x_1) \mathcal{O}^{(2),\text{orb}}_{\n_2}(x_2) \mathcal{O}^{(3),\text{orb}}_{\n_3}(x_3)  \rangle=\k^3\langle \mathcal{O}^{(1)}_{\k\n_1}(x_1) \mathcal{O}^{(2)}_{\k\n_2}(x_2) \mathcal{O}^{(3)}_{\k\n_3}(x_3)  \rangle$ vanish, unless $n_1+n_2+n_3=0$, $n_1>0$ and $n_3<0$. In this case, we find the large $k$ asymptotics,
\begin{align}
  &\langle \mathcal{O}^{(1),\text{orb}}_{\n_1}(x_1) \mathcal{O}^{(2)}_{-\n_1-\n_3}(x_2) \mathcal{O}^{(3)}_{\n_3}(x_3)  \rangle\nn\\
  &\,  \sim k\, \hat{C}_{123} \left( 2R_+ i \right)^{-\tfrac{1}{2}\left( \Delta_1+\Delta_2+\Delta_3 \right)} \left( x_{12}^- \right)^{-\alpha_{12}}\left( x_{23}^- \right)^{-\alpha_{23}}\left( x_{31}^- \right)^{-\alpha_{31}} \exp \left( \tfrac{in_1}{2R_+}\xi_{12}-\tfrac{in_3}{2R_+}\xi_{23} \right)	\nn\\
  &\,\hspace{6mm} \times \text{Res}_{\{w_1=0\}}\text{Res}_{\{w_2=0\}}\Bigg[ \left( e^{n_3w_2-n_1w_1} w_1^{-\alpha_{12}}w_2^{-\alpha_{23}}\left( w_1+w_2 - \tfrac{i}{2R_+}\left( \xi_{12}+\xi_{23}+\xi_{31} \right) \right)^{-\alpha_{31}} \right)	\nn\\
  &\, \hspace{44mm} +\left\{\begin{aligned}
	\exp\left( \tfrac{in_3}{2R_+} \left( \xi_{12}+\xi_{23}+\xi_{31} \right)\right) e^{n_3w_2-\left( n_1+n_3 \right)w_1} w_1^{-\alpha_{12}} w_2^{-\alpha_{31}}\qquad\\
	\times\left( w_2-w_1 + \tfrac{i}{2R_+}\left( \xi_{12}+\xi_{23}+\xi_{31} \right) \right)^{-\alpha_{23}} \text{ if } n_1+n_3\ge 0 \\[2em]
	\exp\left( -\tfrac{in_1}{2R_+} \left( \xi_{12}+\xi_{23}+\xi_{31} \right)\right) e^{\left( n_1+n_3 \right)w_1 -n_1 w_2} w_1^{-\alpha_{23}} w_2^{-\alpha_{31}}\qquad\\
	\times\left( w_2-w_1 + \tfrac{i}{2R_+}\left( \xi_{12}+\xi_{23}+\xi_{31} \right) \right)^{-\alpha_{12}} \text{ if } n_1+n_3< 0
\end{aligned}\,\right]\ ,
\label{eq: DLCQ 3pt from limit}
\end{align}
where once again the $\mathcal{O}(k)$ behaviour is as expected from (\ref{eq: 3pt leading k}). This result is easily shown to be consistent with the general solution (\ref{eq: 3pt DLCQ WI solution}) for the 3-point function to the DLCQ Ward identities, and hence determines the function $H$, although we omit the details of this calculation.

As we saw at finite $k,R$, the 3-point function takes on special forms if any of the $\alpha_a\le 0$. In this DLCQ limit, we see this simply as the generating function reducing to a single term. Further, it is straightforward to see that if $\Delta_1\ge \Delta_2+\Delta_3$, then a non-zero 3-point function requires $n_2=-n_1-n_3<0$, while if $\Delta_3\ge \Delta_2+\Delta_1$, then a non-zero 3-point function requires $n_2=-n_1-n_3>0$. Otherwise, the 3-point function is non-zero for all $n_1>0, n_3<0$.

\subsubsection{4-point and Higher-point Functions}\label{subsubsec: DLCQ 4pt and beyond}

We can finally investigate the behaviour of the generalised free 4-point function (\ref{eq: 4pt gen free}) as we approach the DLCQ limit of large $k$. The form of the 4-point function, and in particular it's representation in terms of disconnected diagrams, leads to an overall $\mathcal{O}(k^2)$ asymptotic behaviour at large $k$, rather than the $\mathcal{O}(k)$ behaviour encountered at 2- and 3-points.

To see this, consider the 4-point function of orbifolded operators,
\begin{align}
  &\left\langle\mathcal{O}^{(1),\text{orb}} \left( x_1^+\right) \mathcal{O}^{(2),\text{orb}} \left( x_2^+ \right)\mathcal{O}^{(3),\text{orb}} \left( x_3^+ \right)\mathcal{O}^{(4),\text{orb}} \left( x_4^+ \right) 	\right\rangle \nn\\
  &\qquad = k \sum_{\substack{\s_a=0\\ a=1,2,3}}^{k-1} \left\langle\mathcal{O}^{(1) } \left( x_1^+ \right) \mathcal{O}^{(2) } \left( x_2^++2\pi R_+ \s_1\right)\mathcal{O}^{(3) } \left( x_3^++2\pi R_+ \s_2\right)\mathcal{O}^{(4) } \left( x_4^++2\pi R_+ \s_3\right)\right\rangle\ .
\end{align}
However, the form of $ \big\langle  \hat{\mathcal{O}}^{(1)}\hat{\mathcal{O}}^{(2)}\hat{\mathcal{O}}^{(3)}\hat{\mathcal{O}}^{(4)} \big\rangle$ implies the same factorisation for the (coordinate and Weyl) transformed operators $\mathcal{O}^{(a)}$, namely
\begin{align}
  \big\langle  \mathcal{O}^{(1)}\mathcal{O}^{(2)}\mathcal{O}^{(3)}\mathcal{O}^{(4)} \big\rangle = \sum_{(ab,cd)\in I} \big\langle \mathcal{O}^{(a)}\mathcal{O}^{(b)} \big\rangle \big\langle \mathcal{O}^{(c)}\mathcal{O}^{(d)} \big\rangle\ ,
\end{align}
which also persists for the orbifolded operators,
\begin{align}
  &\left\langle\mathcal{O}^{(1),\text{orb}} \left( x_1^+\right) \mathcal{O}^{(2),\text{orb}} \left( x_2^+ \right)\mathcal{O}^{(3),\text{orb}} \left( x_3^+ \right)\mathcal{O}^{(4),\text{orb}} \left( x_4^+ \right) 	\right\rangle \nn\\
  &\qquad = k^2 \sum_{(ab,cd)\in I} \sum_{\substack{\s_a=0\\ a=1,2}}^{k-1} \left\langle\mathcal{O}^{(a) } \left( x_a^+ \right) \mathcal{O}^{(b) } \left( x_b^++2\pi R_+ \s_1\right)\right\rangle \left\langle\mathcal{O}^{(c) } \left( x_c^+\right)\mathcal{O}^{(d) } \left( x_d^++2\pi R_+ \s_2\right)\right\rangle			\nn\\
  &\qquad= \sum_{(ab,cd)\in I} \left\langle\mathcal{O}^{(a),\text{orb}} \left( x_a^+\right) \mathcal{O}^{(b),\text{orb}} \left( x_b^+ \right)\right\rangle \left\langle\mathcal{O}^{(c),\text{orb}} \left( x_c^+ \right)\mathcal{O}^{(d),\text{orb}} \left( x_d^+ \right) 	\right\rangle\ .
  \label{eq: extra divergence for 4pt}
  \end{align}
Since we have $\left\langle\mathcal{O}^{(a),\text{orb}} \left( x_a^+\right) \mathcal{O}^{(b),\text{orb}} \left( x_b^+ \right)\right\rangle = \mathcal{O}(k)$ as $k\to \infty$, we see that the 4-point function goes as $k^2$. In other words, the additional degree of $x^+$ translational symmetry enjoyed by the terms of the generalised free correlator gives rise to an additional degree of divergence as we approach the DLCQ limit $k\to \infty$.\\

We can similarly consider the large $k$ asymptotics of the Fourier modes of the 4-point function, as determined in (\ref{eq: orbifolded Npt decomposition}). Making use of the factorisation (\ref{eq: dim red 4pt}) of the Fourier mode 4-point function at finite $k$, have simply
\begin{align}
  \big\langle \mathcal{O}^{(1),\text{orb}}_{\n_1} \mathcal{O}^{(2),\text{orb}}_{ \n_2} \mathcal{O}^{(3),\text{orb}}_{\n_3}\mathcal{O}^{(4),\text{orb}}_{\n_4}  \big\rangle &= k^4\big\langle \mathcal{O}^{(1)}_{\k \n_1} \mathcal{O}^{(2)}_{\k \n_2} \mathcal{O}^{(3)}_{\k \n_3}\mathcal{O}^{(4)}_{\k \n_4}  \big\rangle \nn\\
 	  &= \sum_{(ab,cd)\in I} \Big( k^2\big\langle \mathcal{O}^{(a)}_{\k\n_a}\mathcal{O}^{(b)}_{\k\n_b} \big\rangle \Big) \Big( k^2\big\langle \mathcal{O}^{(c)}_{\k\n_c}\mathcal{O}^{(d)}_{\k\n_d}\big\rangle  \Big)\ ,
 	  \label{eq: 4pt DLCQ final}
\end{align}
where the large $k$ asymptotics of the right-hand-side are given by (\ref{eq: DLCQ 2pt from limit}). One can further check that this result is consistent with the general solution (\ref{eq: DLCQ Npt WI sol}) to the DLCQ Ward identities, and hence determines the function $H$, although we omit the details of this calculation.\\

We finally make some comments on the situation at higher points. Following (\ref{eq: DLCQ reg sum}), we see that the general $\Npt$-point function goes at least as $k$ in the DLCQ limit. But in a generalisation of our analysis of the generalised free 4-point function, we find that if a $\Npt$-point function factorises into $A$ connected sub-correlators, then the leading order behaviour is in fact at order $k^A$.

\section{Conclusion}\label{sect: Conclusion}

In this paper we studied scalar correlators of five-dimensional non-Lorentzian theories with a centrally extended $SU(1,3)$ conformal symmetry. Examples of such theories have recently been constructed in \cite{Lambert:2019jwi,Lambert:2019fne,Lambert:2020jjm} as $\Omega$-deformed Yang-Mills Lagrangian theories. In particular we solved the Ward identities for general $\Npt$-point functions. Although the $SU(1,3)$ symmetry is weaker than the full Lorentzian conformal group in five dimensions it nevertheless imposes significant constraints on the correlation functions. For example we saw that there was a natural holomorphic structure similar to two-dimensional CFT's. In particular the generic $\Npt$-point function factorises into  holomorphic and anti-holomorphic  parts, which are determined by the conformal dimension and central extension, along with a single undetermined function of conformal invariants. At 2-points there are no such conformal invariants, there is one at 3-points and five at 4-points.  Furthermore the correlators have a power-law decay as well as an  oscillating contribution. From the point of view of a Lagrangian description these appear to have  perturbative and  non-perturbative origins respectively.  

Another  aim of this paper has been to show how, at least in principle, features of a six-dimensional CFT can be computed from a five-dimensional  Lagrangian theory. Indeed the theories obtained in \cite{Lambert:2019jwi,Lambert:2019fne,Lambert:2020jjm} arose from considering $(2,0)$ and   $(1,0)$ theories on the conformally compactified spacetime (\ref{eq: conformally compactified metric}). 
 In particular we explicitly showed that correlation functions of the the Fourier components of six-dimensional operators, as computed in the six-dimensional theory,  solved the five-dimensional Ward identities. Conversely one   can therefore use an infinite sum over correlators of the five-dimensional theory to reconstruct a six-dimensional correlator. Thus there is a reasonable hope that the Lagrangians of \cite{Lambert:2019jwi,Lambert:2019fne,Lambert:2020jjm} define six-dimensional conformal field theories with $(2,0)$ and $(1,0)$ supersymmetry. 
 
 On the other hand  one need not expect that all five-dimensional theories with $SU(1,3)$ spacetime symmetry arise from reduction of a six-dimensional conformal field theory (for example in the above constructions there is no restriction on the choice of gauge group which is expected for six-dimensional CFT's). Thus it is interesting to see what the conditions are for such a five-dimensional theory to lift to six dimensions.  One such restriction comes from the form of the correlation functions which are only determined in the five-dimensional theory up to constant $C_{\Delta,\n}$, for 2-point functions,  and a function $H_{\n_1,\n_2,\n_3;\Delta_1,\Delta_2,\Delta_3}$ in the case of 3-point functions. However, the reduction of the six-dimensional 2-point and 3-point correlation functions leads to precise predictions for these, determined entirely by  two constants $\hat C_{12}$ and $\hat C_{123}$ which appear in the six dimensional 2-point and 3-point functions, respectively. We have also shown that 4-point disonnected free correlators in six-dimensions reduce to disconnected free correlators in five-dimensions, defined by decomposition into 2-point functions, although we note that the five dimensional 2-point functions contain instanton contributions so are not necessarily free from the point of view of the Lagrangian theory. %Our analysis makes no assumptions about the 6d theory, so to place further constraints on 4-point functions we need to specify what 6d theory we are talking about. 

 In order to perform the 
 dimensional reduction, we invoked an $i \epsilon$ prescription to regulate the Fourier mode integrals. This  encodes the six-dimensional time ordering and implies intricate constraints on the five-dimensional correlators. For example the sum of  Fourier momentum modes from right to left can never go positive, which implies that only anti-instantons can propagate in the gauge theory. Morever at 2-points, each operator must satisfy $\left|\n\right|\geq\Delta/2$ so the perturbative  zero instanton number sector does contribute. At higher points, we find more complicated constraints. 
 %At three points, the bound $|n| \ge \Delta$  does not apply to middle operator if scaling dimensions satisfy triangle inequality. 

Our construction also allows for a  ${\mathbb Z}_k$ orbifold which leads to weak coupling in the five-dimensional gauge theory. This is analogous to the weak coupling limit in ABJM. Furthermore combining this with a decompactification limit $R\to\infty $ such that $R/k=R_+$ is fixed sets $\Omega_{ij}=0$ and leads to a DLCQ picture,  where the $x^+$ direction of ordinary six-dimensional Minkowski space is periodically identified. 
  In this scenario our correlation functions reduce to DLCQ correlators.  We reproduced the form of the 2-point function given in \cite{Aharony:1997an}  and  generalised it to $\Npt$-points (see also \cite{Henkel:1993sg} for 2- and 3-points). Furthermore setting $\Omega_{ij}=0$  in the five-dimensional Lagrangian field theories \cite{Lambert:2019jwi,Lambert:2019fne,Lambert:2020jjm} leads to a constraint that  localises the gauge fields to anti-self-dual connections on ${\mathbb R}^4$. Therefore the dynamics reduces to quantum mechanics on instanton moduli space \cite{Mouland:2019zjr}, in line with the proposals of \cite{Aharony:1997an,Aharony:1997th}.

Let us now close with some additional comments on future work. In this paper we restricted our attention to scalar operators where the scaling dimension was an even integer, {\it i.e.} protected operators. A related point is that these theories have a large amount of supersymmetry which we have not exploited here (apart from ensuring that such protected operators exist). It would be interesting to extend our analysis to include the constraints of supersymmetry and consider more general operators where one must deal with the associated branch cuts. 
Furthermore our discussion of  4-point correlators was rather general and so less specific. It would be interesting to consider a particular theory, such as $(2,0)$ expanded in large-$N$, and examine its 4-point correlators in greater detail. In addition one might try to match with Witten diagrams in AdS$_7\times S^4$, where AdS$_7$ is thought of as a fibration over non-compact $\mathbb{C}P^3$ and external states have mode number $k n$, $n \in \mathbb{Z}$ along the fibre.

 It may also be interesting to examine defects from the construction presented here. In particular one could look for solutions to the conformal Ward identities  for 2-point functions on semi-infinite space as was done for $\Omega_{ij}=0$ in \cite{Henkel:1993sg} within the context of condensed matter physics. From the M5-brane point of view we could try to make contact with the defects considered in \cite{Drukker:2020swu,Drukker:2020atp}.

In order to reconstruct six-dimensional correlators from the five-dimensional theory we need to better understand the non-perturbative structure of the theory and in particular how to construct operators with non-vanishing instanton number and compute their correlation functions. To this end it would interesting to examine the role that instanton operators \cite{Lambert:2014jna,Tachikawa:2015mha} or related objects play. In addition for $\Omega_{ij}\ne 0$ the constraints impose a novel anti-self-duality condition on the gauge fields which allows for dependence on $x^-$ and it would be interesting to explore the solutions \cite{Lambert:2021}.

 More generally, it would be interesting to explore theories with $SU(1,d/2)$ symmetry obtained from dimensional reduction of $d$-dimensional CFT's, where $d=2,4$.  The origin of this symmetry group can be understood holographically by considering complex embedding coordinates for AdS$_{d+1}$ 
\begin{equation}
-\left|Z^{0}\right|^{2}+\left|Z^{1}\right|^{2}+...+\left|Z^{d/2}\right|^{2}=-1\ ,		
\end{equation}
and noting that it can be written as a Hopf fibration of a non-compact $\mathbb{C}P^{d/2}$ \cite{Pope:1999xg}. After reducing along the fibre, the isometry group is broken from $SO(2,d)$ to $SU(1,d/2)$. Denoting the fibre coordinate at $x^+$, the generators of $SU(1,d/2)$ then correspond to the subset of $SO(2,d)$ generators which commute with $\partial_+$, which contain in particular a Lifshitz scaling. 
%Recalling that $d$-dimensional dilatations and boosts are given by
%\[
%D=x^{+}\partial_{+}+x^{-}\partial_{-}+x^{i}\partial_{i},\,\,\,M_{+-}=x^{+}\partial_{+}-x^{-}\partial_{-},
%\]
%we see that the linear combination which has no $x^+$ dependence is precisely the generator of Lifshitz scaling:
%\[
%D-M_{+-}=2x^{-}\partial_{-}+x^{i}\partial_{i}.
%\]
Hence, $SU(1,d/2)$ can be thought of as the non-relativistic analogue of the conformal group in $(d-1)$ dimensions.  It would  be interesting to derive the conformal blocks for this symmetry group and develop the non-relativistic conformal bootstrap, which may have applications to condensed matter physics \cite{Moroz:2019qdw,Seiberg:2020bhn,Orlando:2020idm, Hellerman:2020eff,Chen:2020vvn}. 

Lastly we might apply our construction to four-dimensional superconformal field theories which admit a Lagrangian. As a result we would obtain an explicit three-dimensional Lagrangian with a Kaluza-Klein tower operators. The resulting theory appears to share similar features with the works \cite{Beem:2013sza,Baiguera:2020jgy}. Indeed already in six-dimensions there might be a a natural relation to the the chiral algebra studied in \cite{Beem:2014kka,Bobev:2020vhe}

\section*{Acknowledgement}

N.~Lambert and P.~Richmond were supported   by STFC grant ST/L000326/1, A.~Lipstein by the Royal Society as a Royal Society University Research Fellowship holder and  R.~Mouland by the STFC studentship ST10837. 

\appendix

\section{Derivation of Dimensionally Reduced Correlators}

\subsection{2-point Functions}\label{app: 2pt} 
%\NL{Maybe move this to an appendix?}
Let us now derive the result of section \ref{subsec: 2pt dim red}, and in particular show how it arises from a six-dimensional $i\epsilon$ prescription. We follow the familiar routine of defining Lorentzian correlation functions by Wick rotating their Euclidean counterparts \cite{Hartman:2015lfa}. In doing this, one encounters ambiguities corresponding to how the branch points in the complex time plane are navigated. The resulting family of Lorentzian results are naturally captured through a Wightman function, inside which operators no longer commute. In this way, the Wick rotation induces a natural operator ordering, which one usually chooses to coincide with time ordering. We instead choose to order operators with respect to the coordinate $x^+$, which can be seen as a deformation of the more familiar ordering by the regular lightcone coordinate $\hat{x}^+$ \cite{Fitzpatrick:2018ttk}. This then defines the Lorentzian $n$-point function to be
\begin{align}
  \left\langle \hat{\mathcal{O}}^{(1)}(\hat{x}_1^+,\hat{x}_1^-,\hat{x}_1^i)\dots \hat{\mathcal{O}}^{(n)}(\hat{x}_N^+,\hat{x}_N^-,\hat{x}_N^i) \right\rangle &=\nn\\ 
  &\hspace{-60mm}\lim_{\substack{\epsilon_a\to 0 \\ \epsilon_1>\dots >\epsilon_N>0}} \left\langle \hat{\mathcal{O}}^{(1)}(\hat{x}_1^+(\epsilon_1),\hat{x}_1^-(\epsilon_1),\hat{x}_1^i(\epsilon_1))\dots \hat{\mathcal{O}}^{(n)}(\hat{x}_N^+(\epsilon_N),\hat{x}_N^-(\epsilon_N),\hat{x}_N^i(\epsilon_n)) \right\rangle_\text{Wick}\ ,
  \label{eq: i epsilon}
\end{align}
where the correlation function on the right hand side is the naive result of Wick rotating the Euclidean correlation function, and we define
\begin{align}
  	\hat{x}^+(\epsilon) &= \hat{x}^+\left(x^+-i\epsilon,x^-, x^i\right) = 2R\left( \frac{\left( \tfrac{\hat{x}^+}{2R} \right)-i\tanh\left( \tfrac{\epsilon}{2R} \right)}{1+i\tanh\left( \tfrac{\epsilon}{2R} \right)\left( \tfrac{\hat{x}^+}{2R} \right)} \right)  \, ,	\nn\\
  	\hat{x}^-(\epsilon)	&= \hat{x}^-\left(x^+-i\epsilon,x^-, x^i\right) 	\, ,\nn\\
  	\hat{x}^i(\epsilon)	&= \hat{x}^i\left(x^+-i\epsilon,x^-, x^i\right) \, .
\end{align}
It is important here to keep track of this deformation for \textit{finite} $\epsilon$, since the action of the prescription in $\hat{x}^+$ space is inhomogeneous. Note, the effect of this $i\epsilon$ prescription on the integral (\ref{eq: 2pt dim red integral}) is to shift any potential poles in $\hat{x}_1^+$ and $\hat{x}_2^+$ on the real line infinitesimally up or down into the complex plane, thus regularising the integral.  

Using this, we dimensionally reduce the six-dimensional 2-point function (\ref{eq: 6d 2pt}), finding five-dimensional 2-point function
%\begin{align}
%  F^{\left( \text{5d} \right)}_{12} &= \frac{\hat{C}_\Delta}{\pi^2}\left( 4R \right)^{-\Delta}(-1)^{\n_1+\n_2} \nn\\
%  &\qquad\times\lim_{\substack{\epsilon_a\to 0\\ \epsilon_1>\epsilon_2>0}}\int_{-\infty}^{\infty} du_1\int_{-\infty}^{\infty} du_2\,\, \left( u_1+i \right)^{-\n_1-1}\left( u_1-i \right)^{\n_1-1} \left( u^\epsilon_1+i \right)^{\tfrac{\Delta}{2}}\left( u^\epsilon_1-i \right)^{\tfrac{\Delta}{2}}		\nn\\
%  &\hspace{53.5mm}\times \left( u_2+i \right)^{-\n_2-1}\left( u_2-i \right)^{\n_2-1}\left( u^\epsilon_2+i \right)^{\tfrac{\Delta}{2}}\left( u^\epsilon_2-i \right)^{\tfrac{\Delta}{2}}	\nn\\
%  &\hspace{53.5mm}\times \Big( \left( \,\tilde{x}_{12} \right) \left( u^\epsilon_2-u^\epsilon_1 +\left( 1+ u^\epsilon_1 u^\epsilon_2 \right)\tfrac{\xi_{12}}{4R} \right) \Big)^{-\Delta}
%  \label{eq: deformed 2pt integral}
%\end{align}
\begin{align}
  F_2:&= \left\langle \mathcal{O}_{n_1}\mathcal{O}_{n_2}\right\rangle\nn\\
   &= \frac{\hat{C}_\Delta}{\pi^2}\left( 4R \right)^{-\Delta}(-1)^{\n_1+\n_2} \nn\\
  &\qquad\times\lim_{\substack{\epsilon_a\to 0\\ \epsilon_1>\epsilon_2>0}}\int_{-\infty}^{\infty} du_1\int_{-\infty}^{\infty} du_2\,\, \prod_{a=1}^2 \left( u_a+i \right)^{-\n_a-1}\left( u_a-i \right)^{\n_a-1} \left( u^\epsilon_a+i \right)^{\frac{\Delta}{2}}\left( u^\epsilon_a-i \right)^{\frac{\Delta}{2}}		\nn\\
  &\hspace{58.5mm}\times \Big( \tilde{x}_{12} \left( u^\epsilon_2-u^\epsilon_1 +\left( 1+ u^\epsilon_1 u^\epsilon_2 \right)\tfrac{\xi_{12}}{4R} \right) \Big)^{-\Delta} \, ,
  \label{eq: deformed 2pt integral}
\end{align}
where $a=1,2$, and we define
\begin{align}
  u^\epsilon_a = \frac{u_a-i\epsilon_a}{1+i\epsilon_a u_a} \, .
  \label{eq: u epsilon def}
\end{align}

We have importantly made sure to apply the $i\epsilon$ prescription also to the Weyl rescaling factor (\ref{eq: 6d operator mapping}). Also, in arriving at this expression we have exploited the strict monotonicity of $\tanh$ to replace $\tanh\left( \tfrac{\epsilon_a}{2R} \right)$ with simply $\epsilon_a$ while preserving the ordering of the $\epsilon_a$. In doing this, we have also shortened the range of $\epsilon_a$ to $\epsilon_a\in(0,1)$.

We now find that the integrand is strictly non-zero for all $u_1,u_2\in\mathbb{R}$, and goes like $u_a^{-2}$ as $|u_a|\to\infty$. Hence, it is convergent.

Although the $\epsilon_a$ enter in $u_a^\epsilon$ in a somewhat complicated way, their effect on the total integrand is simplified  by noting the identity
\begin{align}
  \tilde{x}_{12} \Big( u^\epsilon_2-u^\epsilon_1 +\left( 1+ u^\epsilon_1 u^\epsilon_2 \right)\tfrac{\xi_{12}}{4R} \Big) &= \frac{i}{2}\Big[ \left( u_1^\epsilon - i \right)\left( u_2^\epsilon+i \right)\bar{z}_{12} -  \left( u_1^\epsilon + i \right)\left( u_2^\epsilon-i \right)z_{12} \Big]\ ,
\end{align}
and hence
\begin{align}
  &\frac{\left( \tilde{x}_{12} \left( u^\epsilon_2-u^\epsilon_1 +\left( 1+ u^\epsilon_1 u^\epsilon_2 \right)\tfrac{\xi_{12}}{4R} \right) \right)^2}{\left( u_1^\epsilon +i \right)\left( u_1^\epsilon -i \right)\left( u_2^\epsilon +i \right)\left( u_2^\epsilon -i \right)} \nn\\
  &\qquad= -\frac{1}{4} \frac{\Big[ \left( 1+\epsilon_1 \right)\left( 1-\epsilon_2 \right)\left( u_1 - i \right)\left( u_2+i \right)\bar{z}_{12} -  \left( 1-\epsilon_1 \right)\left( 1+\epsilon_2 \right)\left( u_1 + i \right)\left( u_2-i \right)z_{12} \Big]^2}{\left( 1-\epsilon_1^2 \right)\left( 1-\epsilon_2^2 \right)\left( u_1 +i \right)\left( u_1 -i \right)\left( u_2 +i \right)\left( u_2 -i \right)}\ .
  \label{eq: fraction identity}
\end{align}
We can therefore write
%\begin{align}
%  F^{\left( \text{5d} \right)}_{12} = \frac{\hat{C}_\Delta}{\pi^2}\left( 2R\, i \right)^{-\Delta}(-1)^{\n_1+\n_2}\hspace{100mm}  \nn\\
%  \qquad\times\lim_{\substack{\epsilon_a\to 0\\ \epsilon_1>\epsilon_2>0}}\left( 1-\epsilon_1^2 \right)^{\tfrac{\Delta}{2}}\left( 1-\epsilon_2^2 \right)^{\tfrac{\Delta}{2}}\int_{-\infty}^{\infty} du_1\int_{-\infty}^{\infty} du_2\,\, \left( u_1+i \right)^{-\n_1+\tfrac{\Delta}{2}-1}\left( u_1-i \right)^{\n_1+\tfrac{\Delta}{2}-1} 	\nn\\
% \times \left( u_2+i \right)^{-\n_2+\tfrac{\Delta}{2}-1}\left( u_2-i \right)^{\n_2+\tfrac{\Delta}{2}-1}	\nn\\
%  \times \Big( \left( 1+\epsilon_1 \right)\left( 1-\epsilon_2 \right)\left( u_1 - i \right)\left( u_2+i \right)\bar{z}_{12} -  \left( 1-\epsilon_1 \right)\left( 1+\epsilon_2 \right)\left( u_1 + i \right)\left( u_2-i \right)z_{12} \Big)^{-\Delta}
%  \label{eq: deformed 2pt integral 2}
%\end{align}
\begin{align}
  F_2 &= \frac{\hat{C}_\Delta}{\pi^2}\left( 2R\, i \right)^{-\Delta}(-1)^{\n_1+\n_2}\label{eq: deformed 2pt integral 2}\\
  &\times\lim_{\substack{\epsilon_a\to 0\\ \epsilon_1>\epsilon_2>0}}\left( 1-\epsilon_1^2 \right)^{\frac{\Delta}{2}}\left( 1-\epsilon_2^2 \right)^{\frac{\Delta}{2}}\int_{-\infty}^{\infty} du_1\int_{-\infty}^{\infty} du_2\,\, \prod_{a=1}^2 \left( u_a+i \right)^{-\n_a+\frac{\Delta}{2}-1}\left( u_a-i \right)^{\n_a+\frac{\Delta}{2}-1} 	\nn\\
% \times \left( u_2+i \right)^{-\n_2+\tfrac{\Delta}{2}-1}\left( u_2-i \right)^{\n_2+\tfrac{\Delta}{2}-1}	\nn\\
  &\times \Big( \left( 1+\epsilon_1 \right)\left( 1-\epsilon_2 \right)\left( u_1 - i \right)\left( u_2+i \right)\bar{z}_{12} -  \left( 1-\epsilon_1 \right)\left( 1+\epsilon_2 \right)\left( u_1 + i \right)\left( u_2-i \right)z_{12} \Big)^{-\Delta} \nn \, .
\end{align}
There are now many ways to proceed to calculate $F_{12}^{(\text{5d})}$ explicitly. Here, we follow a particularly streamlined approach. For a more general discussion of the evaluation of integrals of this type, including an alternative contour derivation of their explicit values, see appendix \ref{app: contours}.\\

To proceed, note that we can write the final part of the integrand as
\begin{align}
  &\Big( \left( 1+\epsilon_1 \right)\left( 1-\epsilon_2 \right)\left( u_1 - i \right)\left( u_2+i \right)\bar{z}_{12} -  \left( 1-\epsilon_1 \right)\left( 1+\epsilon_2 \right)\left( u_1 + i \right)\left( u_2-i \right)z_{12} \Big)^{-\Delta}\nn\\
  &\hspace{5mm}= \left( \left( 1+\epsilon_1 \right)\left( 1-\epsilon_2 \right)\left( u_1 - i \right)\left( u_2+i \right)\bar{z}_{12} \right)^{-\Delta}\left( 1 -  \frac{\left( 1-\epsilon_1 \right)\left( 1+\epsilon_2 \right)\left( u_1 + i \right)\left( u_2-i \right)z_{12}}{\left( 1+\epsilon_1 \right)\left( 1-\epsilon_2 \right)\left( u_1 - i \right)\left( u_2+i \right)\bar{z}_{12}} \right)^{-\Delta}\ .
  \label{eq: Taylor expn 1}
\end{align}
Then, we have
\begin{align}
  \left|\frac{\left( 1-\epsilon_1 \right)\left( 1+\epsilon_2 \right)\left( u_1 + i \right)\left( u_2-i \right)z_{12}}{\left( 1+\epsilon_1 \right)\left( 1-\epsilon_2 \right)\left( u_1 - i \right)\left( u_2+i \right)\bar{z}_{12}}\right| = \frac{\left( 1-\epsilon_1 \right)\left( 1+\epsilon_2 \right)}{\left( 1+\epsilon_1 \right)\left( 1-\epsilon_2 \right)}<1\quad \text{ since }\quad \epsilon_1>\epsilon_2\ ,
\end{align}
and hence we can use the Taylor expansion for $(1-w)^{-\Delta}$, as the argument falls just within the radius of convergence. Indeed, the partial sums converge uniformly to functions both of $u_1$ and $u_2$, allowing us to integrate term-wise. Hence, we see in this way how our integral is regularised: we have a convergent series expansion which would have otherwise been indeterminate. Further, if we had instead  $\epsilon_1<\epsilon_2$, we would have instead written $(1-w)^{-\Delta}=\left( -w \right)^{-\Delta}\left( 1-w^{-1} \right)^{\Delta}$ and used the series expansion for the latter factor; in this way, we can see how the ordering prescription manifests in our calculation.

So we now simply substitute this series expansion into (\ref{eq: deformed 2pt integral}) to find
\begin{align}
  F_2 &= \frac{\hat{C}_\Delta}{\pi^2}\left( -2R\, i \right)^{-\Delta}(-1)^{\n_1+\n_2}   \nn\\
  &\qquad\times\lim_{\substack{\epsilon_a\to 0\\ \epsilon_1>\epsilon_2>0}}\sum_{m=0}^\infty \left( \tfrac{\left( 1-\epsilon_1 \right)\left( 1+\epsilon_2 \right)}{\left( 1+\epsilon_1 \right)\left( 1-\epsilon_2 \right)} \right)^{\tfrac{\Delta}{2}+m}{\Delta+m-1\choose m}\left( z_{12} \right)^m \left( \bar{z}_{12} \right)^{-\Delta-m}\nn\\
  &\qquad\hspace{30mm}\times\left( \int_{-\infty}^{\infty} \frac{du_1}{1+u_1^2} \left( u_1+i \right)^{-\n_1+\tfrac{\Delta}{2}+m}\left( u_1-i \right)^{\n_1-\tfrac{\Delta}{2}-m}	 \right)	\nn\\
  &\qquad\hspace{30mm}\times \left( \int_{-\infty}^{\infty} \frac{du_2}{1+u_2^2} \left( u_2+i \right)^{-\n_2-\tfrac{\Delta}{2}-m}\left( u_2-i \right)^{\n_2+\tfrac{\Delta}{2}+m} \right)	\, .
  \label{eq: 2pt integral expanded}
\end{align}
We can finally perform the integrals explicitly, using the identity
\begin{align}
  \int_{-\infty}^\infty \frac{du}{1+u^2} \left( u+i \right)^m\left( u-i \right)^{-m} = \pi \delta_{m,0} \, .
  \label{eq: orthogonality relation}
\end{align}
We in particular find that every term in the sum vanishes, \textit{unless} $\n_1=-\n_2=\tfrac{\Delta}{2}+m$ for some $m\in\{0,1,\dots\}$. If this is the case, then there is a single non-zero term in the sum. Computing this term and then safely taking the limits $\epsilon_a\to 0$, we finally arrive at the dimensionally reduced 2-point function,
\begin{align}
  \left\langle \mathcal{O}^{(1)}_{\n_1}(x_1^-, x_1^i) \mathcal{O}^{(2)}_{\n_2}(x_2^-,x_2^i) \right\rangle=
	\delta_{\n_1+\n_2,0}\hat{C}_\Delta \left( -2R\, i \right)^{-\Delta} {{\n_1+\tfrac{\Delta}{2}-1}\choose{\n_1-\tfrac{\Delta}{2}}} \left( z_{12}\bar{z}_{12} \right)^{-\tfrac{\Delta}{2}} \left( \frac{z_{12}}{\bar{z}_{12}} \right)^{\n_1} 
	\label{eq: dim red 2pt final} \, .
\end{align}

\begin{comment}
This calculation additionally tells us that the correct continuation of (\ref{eq: dim red 2pt final}) to odd $\Delta$ such that the lift to six-dimensions still works is given by
\begin{align}
  \left\langle \mathcal{O}^{(1)}_{\n }(x_1^-, x_1^i), \mathcal{O}^{(2)}_{-\n }(x_2^-,x_2^i) \right\rangle= \frac{\hat{C}_\Delta}{(2R)^\Delta}(-i)^{\Delta}{{\n +\tfrac{\Delta}{2}-1}\choose{\n -\tfrac{\Delta}{2}}} \left( z\bar{z} \right)^{-\tfrac{\Delta}{2}} \left( \frac{z}{\bar{z}} \right)^{\n }
  \label{eq: 5d 2pt from dim red}
\end{align}
where the binomial coefficient is continued in the usual way using the gamma function, and we take the positive square root of $z\bar{z}$.
\end{comment}

\subsection{Derivation of the Five-dimensional 3-point Functions}\label{app: 3pt}

We also present the full derivation of the dimensionally reduced 3-point function of section \ref{subsec: 3pt dim red}. Using the $i\epsilon$ prescription (\ref{eq: i epsilon}), we find that the Fourier modes of the 3-point function (\ref{eq: 6d 3pt}) are given by the regularised integral
%\begin{align}
%  F^{\left( \text{5d} \right)}_{123} = \frac{\hat{C}_{123}}{\pi^3} \left( 4R \right)^{-\frac{1}{2}\left( \Delta_1+\Delta_2+\Delta_3 \right)}\left( -1 \right)^{\n_1+\n_2+\n_3}\hspace{75mm} \nn\\
%  \times\lim_{\substack{\epsilon_a\to 0\\ \epsilon_1>\epsilon_2>\epsilon_3>0}}\int_{-\infty}^{\infty} du_1\int_{-\infty}^{\infty} du_2\int_{-\infty}^{\infty} du_3\,\, \left( u_1+i \right)^{-\n_1-1}\left( u_1-i \right)^{\n_1-1}\left( u^\epsilon_1+i \right)^{\tfrac{\Delta_1}{2}}\left( u^\epsilon_1-i \right)^{\tfrac{\Delta_1}{2}}		\nn\\
%  \times \left( u_2+i \right)^{-\n_2-1}\left( u_2-i \right)^{\n_2-1}\left( u^\epsilon_2+i \right)^{\tfrac{\Delta_2}{2}}\left( u^\epsilon_2-i \right)^{\tfrac{\Delta_2}{2}}	\nn\\
%  \times \left( u_3+i \right)^{-\n_3-1}\left( u_3-i \right)^{\n_3-1}\left( u^\epsilon_3+i \right)^{\tfrac{\Delta_3}{2}}\left( u^\epsilon_3-i \right)^{\tfrac{\Delta_3}{2}}	\nn\\
%  \times \Big( \left( \tilde{x}_{12} \right)\left( u^\epsilon_2-u^\epsilon_1 +\left( 1+ u^\epsilon_1 u^\epsilon_2 \right)\tfrac{\xi_{12}}{4R} \right) \Big)^{-\alpha_{12}}		\nn\\
%  \times \Big( \left( \tilde{x}_{23} \right)\left( u^\epsilon_3-u^\epsilon_2 +\left( 1+ u^\epsilon_2 u^\epsilon_3 \right)\tfrac{\xi_{23}}{4R} \right) \Big)^{-\alpha_{23}}		\nn\\
%  \times \Big( \left( \tilde{x}_{31} \right)\left( u^\epsilon_1-u^\epsilon_3 +\left( 1+ u^\epsilon_3 u^\epsilon_1 \right)\tfrac{\xi_{31}}{4R} \right) \Big)^{-\alpha_{31}}
%\end{align}
\begin{align}
  F_3:&=\left\langle \mathcal{O}^{(1)}_{\n_1}(x_1^-, x_1^i) \mathcal{O}^{(2)}_{\n_2}(x_2^-,x_2^i)\mathcal{O}^{(3)}_{\n_3}(x_3^-,x_3^i) \right\rangle 	
\nn\\
& =\lim_{\substack{\epsilon_a\to 0\\ \epsilon_1>\epsilon_2>\epsilon_3>0}}\int_{-\infty}^{\infty} d^3 u\, \frac{\hat{C}_{123}}{\pi^3} \left( 4R \right)^{-\frac{1}{2}\left( \Delta_1+\Delta_2+\Delta_3 \right)}\left( -1 \right)^{\n_1+\n_2+\n_3} \nn\\
  &\qquad\times\prod_{a=1}^3 \frac{ \left( u_a-i \right)^{\n_a-1} }{\left( u_a+i \right)^{\n_a+1}} \left( \left( u^\epsilon_a+i \right)\left( u^\epsilon_a-i \right) \right)^{\frac{\Delta_a}{2}}\prod_{a<b}^3\Big( \left( \tilde{x}_{ab} \right)\left( -u^\epsilon_a+u^\epsilon_b +\left( 1+ u^\epsilon_a u^\epsilon_b \right)\tfrac{\xi_{ab}}{4R} \right) \Big)^{-\alpha_{ab}}\ ,
\end{align}
The $u^\epsilon_a$ are as defined in (\ref{eq: u epsilon def}), and we have again assumed $\Delta_1,\Delta_2,\Delta_3\in 2\mathbb{Z}$ to avoid the issue of branch points.

As we saw for the 2-point function, the role of the $i\epsilon$ prescription is made clearer by rewriting this as
\begin{align}
  F_3 &= \frac{\hat{C}_{123}}{\pi^3} \left( 2R\, i \right)^{-\tfrac{1}{2}\left( \Delta_1+\Delta_2+\Delta_3 \right)}\left( -1 \right)^{\n_1+\n_2+\n_3} \nn\\
  &\quad\times\lim_{\substack{\epsilon_a\to 0\\ \epsilon_1>\epsilon_2>\epsilon_3>0}} \prod_{a=1}^3 \left( 1-\epsilon_a^2 \right)^{\frac{\Delta_a}{2}}\int_{-\infty}^{\infty} d^3 u \, \prod_{a=1}^3 \frac{ \left( u_a-i \right)^{\n_a+\frac{\Delta_a}{2}-1} }{ \left( u_a+i \right)^{\n_a-\frac{\Delta_a}{2}+1} } \nn\\
  &\quad\times \prod_{a<b}^3 \Big( \left( 1+\epsilon_a \right)\left( 1-\epsilon_b \right)\left( u_a - i \right)\left( u_b+i \right)\bar{z}_{ab} -  \left( 1-\epsilon_a \right)\left( 1+\epsilon_b \right)\left( u_a + i \right)\left( u_b-i \right)z_{ab} \Big)^{-\alpha_{ab}}	\, .
  \label{eq: 3pt ready for series expn}
\end{align}
To proceed to calculate $F_{123}^{(\text{5d})}$ explicitly, we follow the same procedure as we did at 2-points. For a more general discussion of the evaluation of integrals of this type, including an alternative contour derivation of their explicit values, see appendix \ref{app: contours}.\\

As we saw at 2-points, the final three terms can be expanded in convergent series expansions, making use of $\epsilon_1>\epsilon_2$, $\epsilon_2>\epsilon_3$ and $\epsilon_1>\epsilon_3$ respectively. Doing so, we arrive at
\begin{align}
  F_3&= \frac{\hat{C}_{123}}{\pi^3} \left( -2R\, i \right)^{-\tfrac{1}{2}\left( \Delta_1+\Delta_2+\Delta_3 \right)}\left( -1 \right)^{\n_1+\n_2+\n_3}\nn\\
  &\quad \times\sum_{m_1,m_2,m_3=0}^\infty {\alpha_{23}+m_1-1 \choose m_1}{\alpha_{31}+m_2-1 \choose m_2}{\alpha_{12}+m_3-1 \choose m_3} \nn\\
  &\hspace{20mm}\times \left( z_{12} \right)^{m_3}\left( \bar{z}_{12} \right)^{-\alpha_{12}-m_3}\left( z_{23} \right)^{m_1}\left( \bar{z}_{23} \right)^{-\alpha_{23}-m_1}\left( z_{31} \right)^{-\alpha_{31}-m_2}\left( \bar{z}_{31} \right)^{m_2}\nn\\
  &\hspace{20mm}\times\left( \int_{-\infty}^\infty \frac{du_1}{1+u_1^2}\,\, \left( u_1+i \right)^{-\n_1+\tfrac{\Delta_1}{2}+m_2+m_3}\left( u_1-i \right)^{\n_1-\tfrac{\Delta_1}{2}-m_2-m_3} \right)		\nn\\
  &\hspace{20mm}\times\left( \int_{-\infty}^\infty \frac{du_2}{1+u_2^2}\,\, \left( u_2+i \right)^{-\n_2+\tfrac{1}{2}\left( \Delta_3-\Delta_1 \right)+m_1-m_3}\left( u_2-i \right)^{\n_2-\tfrac{1}{2}\left( \Delta_3-\Delta_1 \right)-m_1+m_3} \right)		\nn\\
  &\hspace{20mm}\times\left( \int_{-\infty}^\infty \frac{du_3}{1+u_3^2}\,\, \left( u_3+i \right)^{-\n_3-\tfrac{\Delta_3}{2}-m_1-m_2}\left( u_3-i \right)^{\n_3+\tfrac{\Delta_3}{2}+m_1+m_2} \right)		\ ,
\end{align}
which again is finite due to the identity (\ref{eq: orthogonality relation}). Hence, we finally find the dimensionally reduced 3-point function
\begin{align}
  F_3= \delta_{\n_1+\n_2+\n_3,0}&\hat{C}_{123} \left( -2R\, i \right)^{-\tfrac{1}{2}\left( \Delta_1+\Delta_2+\Delta_3 \right)} \left( z_{12}\bar{z}_{12} \right)^{-\tfrac{1}{2}\alpha_{12}}\left( z_{23}\bar{z}_{23} \right)^{-\tfrac{1}{2}\alpha_{23}}\left( z_{31}\bar{z}_{31} \right)^{-\tfrac{1}{2}\alpha_{31}}\nn\\
  \qquad \times\sum_{m=0}^\infty &{-\n_3-\tfrac{\Delta_3}{2}+\alpha_{23} - m -1\choose -\n_3-\tfrac{\Delta_3}{2}-m}{\n_1-\tfrac{\Delta_1}{2}+\alpha_{12}-m-1 \choose \n_1-\tfrac{\Delta_1}{2}-m}{\alpha_{31}+m-1 \choose m} \nn\\
  &\times \left( \frac{z_{12}}{\bar{z}_{12}} \right)^{\n_1-m-\tfrac{1}{2}\alpha_{31}}\left( \frac{z_{23}}{\bar{z}_{23}} \right)^{-\n_3-m-\tfrac{1}{2}\alpha_{31}}\left( \frac{z_{31}}{\bar{z}_{31}} \right)^{-m-\tfrac{1}{2}\alpha_{31}}\ ,
  \label{eq: final dim red 3pt}
\end{align}
where, given $z=r e^{i\theta}$, we've chosen the branches $\left( z\bar{z} \right)^{1/2}=r$ and $\left( \tfrac{z}{\bar{z}} \right)^{1/2}=e^{i\theta}$. We note that the sum terminates at $\min\left( \n_1-\tfrac{\Delta_1}{2},-\n_3-\tfrac{\Delta_3}{2} \right)$, and thus as promised we have a finite, regularised result.

\section{$\Npt$-point Integrals and Their Residue Representation}\label{app: contours}

In section \ref{sect: Null Conformal Compactification}, we performed the dimensional reduction of six-dimensional 2-point, 3-point, and some special 4-point functions to five-dimensions. Here, we present a more general disucssion of integrals that would appear at $\Npt$-points. This will in particular include an alternative route to explicitly calculating their values, {\it via} contour integrals.\\

In dimensionally reducing an $\Npt$-point function in six dimensions, we will encounter integrals of the form
\begin{align}
  \mathcal{I}^{(\Npt)} := \prod_{a=1}^\Npt\int_{-\infty}^{\infty}  du_a \left( u_a + i \right)^{-\n_a+\tfrac{\Delta_a}{2}-1}\left( u_a - i \right)^{\n_a+\tfrac{\Delta_a}{2}-1} \prod_{b<c} \left( \tilde{x}_{bc}\left( u_c-u_b +\left( 1+ u_b u_c \right)\tfrac{\xi_{bc}}{4R} \right) \right)^{-\alpha_{bc}}\ .
\end{align}
where the $\alpha_{ab}=\alpha_{ba}$ are integers, which from six-dimensional scale invariance satisfy $\sum_{b\neq a}\alpha_{ab}=\Delta_a$. These integrals are however generically ill-defined; although the integrand has integrable behaviour at large $|u_a|$ ({\it i.e.} $u_a\sim u_a^{-2}$ as $|u_a|\to\infty$), there are poles at finite points which render it divergent.

To see this more explicitly, consider trying to perform the $u_\Npt$ integral. Then, the integrand generically has $(\Npt-1)$ poles at the points
\begin{align}
  u_\Npt = v_{\Npt,a}:=\frac{4Ru_a-\xi_{a\Npt}}{4R+\xi_{a\Npt}u_a},\quad\text{ for each }\quad a\neq \Npt\ .
\end{align}
and hence the integral is not well defined. \\

We can regularise $\mathcal{I}^{(\Npt)}$ by redefining it as the limit of a well-defined $\epsilon$-deformed integral,
\begin{align}
  \mathcal{I}^{(\Npt)} := \prod_{a=1}^\Npt\int_{-\infty}^{\infty}  du_a \left( u_a + i \right)^{-\n_a-1}\left( u_a - i \right)^{\n_a-1}\left( u^\epsilon_a + i \right)^{\tfrac{\Delta_a}{2}}\left( u^\epsilon_a - i \right)^{\tfrac{\Delta_a}{2}} \nn\\
  \times\prod_{b<c} \left( \tilde{x}_{bc}\left( u^\epsilon_c-u^\epsilon_b +\left( 1+ u^\epsilon_b u^\epsilon_c \right)\tfrac{\xi_{bc}}{4R} \right) \right)^{-\alpha_{bc}}\ .
\end{align}
Utilising a generalisation of the identity (\ref{eq: fraction identity}), we can rewrite this as
\begin{align}
  \mathcal{I}^{(\Npt)} = \lim_{\substack{\epsilon_a\to 0\\ \epsilon_1>\dots\epsilon_\Npt>0}}\prod_{a=1}^\Npt\left( 2i\left( \epsilon_a^2-1 \right) \right)^{\tfrac{\Delta_a}{2}}\int_{-\infty}^{\infty}  du_a \left( u_a + i \right)^{-\n_a+\tfrac{\Delta_a}{2}-1}\left( u_a - i \right)^{\n_a+\tfrac{\Delta_a}{2}-1} \hspace{25mm} \nn\\
  \times\prod_{b<c} \Big( \left( 1+\epsilon_b \right)\left( 1-\epsilon_c \right)\left( u_b-i \right)\left( u_c+i \right) \bar{z}_{bc} - \left( 1-\epsilon_b \right)\left( 1+\epsilon_c \right)\left( u_b+i \right)\left( u_c-i \right) z_{bc} \Big)^{-\alpha_{bc}}\ ,
\end{align}
which is indeed a generalisation of (\ref{eq: deformed 2pt integral 2}) at 2-points and (\ref{eq: 3pt ready for series expn}) at 3-points. We could then proceed to calculate this explicitly by series expanding the factors involving the $z_{bc}$, and using the relation (\ref{eq: orthogonality relation}) to pick out the non-zero terms in the resulting sums.\\

We will now  explore an alternative way in which we could proceed, which will in particular demonstrate that the action of $i\epsilon$ prescription is to shift the integrand's poles off the real line and into the complex $u_\Npt$-plane. Further, this occurs in a controlled way, which allows for relatively simple expression for $\mathcal{I}^{(\Npt)}$ in terms of iterated residues. 

First, we perform the $u_\Npt$ integral by continuing $u_\Npt$ to $\mathbb{C}$ and seeking a related contour integral. In addition to the integrand's obvious potential poles at $u_\Npt=\pm i$, we have up to $(\Npt-1)$ additional poles at
\begin{align}
  u_\Npt = v_{\Npt,a}^\epsilon &:=  -i\, \frac{\left( 1+\epsilon_a \right)\left( 1-\epsilon_\Npt \right)\left( u_a-i \right)\bar{z}_{a\Npt}+\left( 1-\epsilon_a \right)\left( 1+\epsilon_\Npt \right)\left( u_a+i \right)z_{a\Npt}}{\left( 1+\epsilon_a \right)\left( 1-\epsilon_\Npt \right)\left( u_a-i \right)\bar{z}_{a\Npt}-\left( 1-\epsilon_a \right)\left( 1+\epsilon_\Npt \right)\left( u_a+i \right)z_{a\Npt}}\nn\\
  &\phantom{:}= v_{n,i} + \mathcal{O}\left( \epsilon_1,\epsilon_2 \right)\ .
\end{align}
We find in particular the imaginary part,
\begin{align}
  \Im\left( v_{n,i}^\epsilon \right) = -\frac{\left( 1+u_a^2 \right)\left( 16R^2+\xi_{a\Npt}^2 \right)\left( 1-\epsilon_a\epsilon_\Npt \right)\left( \epsilon_a-\epsilon_\Npt \right)}{\left( \epsilon_a-\epsilon_\Npt \right)^2\left( 4R u_a-\xi_{a\Npt} \right)^2+\left( 1-\epsilon_a\epsilon_\Npt \right)^2\left( 4R+u_a\xi_{a\Npt} \right)^2}\ ,
  \label{eq: imaginary part}
\end{align}
and hence because we have $\epsilon_a>\epsilon_\Npt$ for all $a\neq \Npt$, we see that all $(\Npt-1)$ of the $v_{\Npt,a}^\epsilon$ lie strictly in the lower-half-plane. Hence, we choose to complete our contour with a large semi-circle in the upper-half-plane, as shown in Figure \ref{fig: n-pt contours}.
\begin{center}
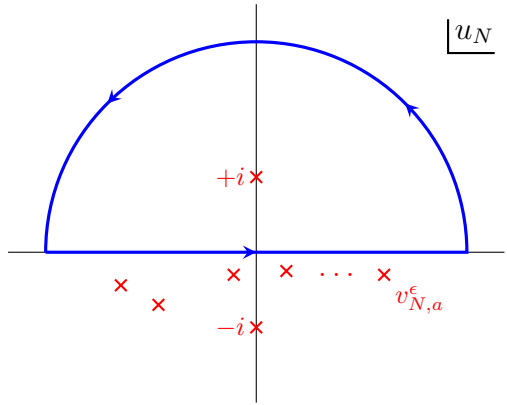
\captionof{figure}{The $u_n$ contour integral, with $(N-1)$ poles shifted into the lower-half-plane}\vspace{1em}\label{fig: n-pt contours}.
\begin{tikzpicture}
\draw  (-8,0) -- (-1.4,0) ;

\draw  (-4.7,-2) -- (-4.7,3.3) ;

\node  at (-1.8,2.9) {$u_N$};
\draw [thick] (-2.15,3.1) -- (-2.15,2.65) -- (-1.53,2.65) ;

\node [cross=3,thick,red ] at (-4.7,1) {};
\node [left,red] at (-4.7,1) {\footnotesize\,$+i$};

\node [cross=3,thick,red ] at (-4.7,-1) {};
\node [left,red] at (-4.7,-1) {\footnotesize\,$-i$};

\node [cross=3,thick,red ] at (-3,-0.3) {};
\node [cross=3,thick,red ] at (-4.3,-0.25) {};
\node [cross=3,thick,red ] at (-6.5,-0.44) {};
\node [cross=3,thick,red ] at (-5,-0.3) {};
\node [cross=3,thick,red ] at (-6,-0.7) {};
\node [red] at (-3.6,-0.3) {$\dots$};
\node [below right, red] at (-3,-0.3) {\footnotesize$v_{\Npt,a}^\epsilon$};

\path [draw=blue,very thick,postaction={on each segment={mid arrow=blue}}] (-7.5,0) -- (-1.9,0) arc(0:180:2.8) ;

\end{tikzpicture}
\end{center}

We omit the explicit form of this residue, but note that it is a meromorphic function of the remaining variables $u_1,\dots, u_{\Npt-1}$ with possible poles only at $u_a= \pm i$. This is seen by noting

\begin{align}
&\Big( \left( 1+\epsilon_a \right)\left( 1-\epsilon_N \right)\left( u_a-i \right)\left( u_N+i \right)\bar{z}_{aN}\nn\\
&\qquad -\left( 1-\epsilon_a \right)\left( 1+\epsilon_N \right)\left( u_a+i \right)\left( u_N-i \right)z_{aN} \Big)\Big|_{u_N=i} = 2i\left( 1+\epsilon_a \right)\left( 1-\epsilon_N \right)\left( u_a-i \right)\bar{z}_{aN}\ .
\end{align}
Thus, its only effect of evaluating the residue at $u_N=i$ on the singularity structure of the resulting integrand is to shift the degree of the poles at $u_a= i$.

So we now do the $u_{\Npt-1}$ integral. There are now $(\Npt-2)$ potential poles other than $u_{\Npt-1}=\pm i$ in the integrand, but since $\epsilon_a>\epsilon_{\Npt-1}$ for all $a<\Npt-1$, once again we see that all of these poles lie strictly in the lower-half-plane. Hence, the integral is once again given entirely by the residue at $u_{\Npt-1}=i$. Iterating this procedure, we finally find
\begin{align}
 \mathcal{I}^{(\Npt)} =(2\pi i)^N\,\text{Res}_{\{u_1=i\}}\bigg[ \,\text{Res}_{\{u_2=i\}}\bigg[ \, \dots \, \text{Res}_{\{u_\Npt=i\}}\bigg[ \hspace{70mm}\nn\\
  \prod_{a=1}^\Npt \left( u_a + i \right)^{-\n_a+\tfrac{\Delta_a}{2}-1}\left( u_a - i \right)^{\n_a+\tfrac{\Delta_a}{2}-1} \prod_{b<c} \left( \tilde{x}_{bc}\left( u_c-u_b +\left( 1+ u_b u_c \right)\tfrac{\xi_{bc}}{4R} \right) \right)^{-\alpha_{bc}}\bigg]\dots \bigg] \bigg]\ .
\end{align}
Computing this directly for the 2-point and 3-point functions, we indeed recover the results (\ref{eq: dim red 2pt from heuristics}) and (\ref{eq: 3pt dim red from heuristics}), respectively.

\section{DLCQ from Direct Dimensional Reduction}
\label{app: dimensional reduction from scratch}

In this section, we pursue an alternative but equiavelent approach to determine the leading order asymptotics of Fourier mode correlators as $k\to\infty$, which involves performing the dimensional reduction from six-dimensions from scratch.

\paragraph{2-point}\ \label{2ptreduction}
\\
\\
Our aim is to first calculate the leading order behaviour of the orbifolded 2-point function at large $k$. As we have seen in (\ref{eq: 2pt leading k}), the leading order term is of order $k$, with coefficient given by

\begin{align}
F_2:=\sum_{\s\in\mathbb{Z}}\left\langle \mathcal{O}\left(x_{1}^{+}\right)\mathcal{O}\left(x_{2}^{+}+2\pi R_+ \s\right)\right\rangle =\sum_{\s\in\mathbb{Z}}\frac{1}{\left(\vec{x}_{12}^{2}-2\left(x_{12}^{+}+2\pi R_+ \s\right)x_{12}^{-}\right)^{\Delta}}\ .
\label{eq: F2 def}
\end{align}
where the operator $\mathcal{O}$ has scaling dimension $\Delta$.

We can now Fourier decompose this expression, and in doing so recover the coefficient in (\ref{eq: DLCQ 2pt from limit}). We write
\begin{equation}
F_2=\sum_{\n}e^{-i \n x_{12}^{+}/R_+}\Phi_{\n}\ ,
\label{2ptdimred}
\end{equation}
where
\begin{align}
	\Phi_{\n}&=\frac{1}{2\pi}\int_{0}^{2\pi}dx_{12}^{+}e^{i \n x_{12}^{+}/R_+}\sum_{\s\in\mathbb{Z}}\left(\vec{x}_{12}^{2}-2\left(x_{12}^{+}+2\pi R_+ \s\right)x_{12}^{-}\right)^{-\Delta}
\nn\\ &
\propto\left(x_{12}^{-}\right)^{-\Delta}\int_{-\infty}^{\infty}dx_{12}^+\,\,e^{ i \n x_{12}^{+}/R_+}\left(x_{12}^{+}-\xi_{12}/2\right)^{-\Delta}\ ,
\end{align}
where we once again abuse notation slightly by using $\xi_{ab}$ to denote the $R\to\infty$ limit of the correspondonding variable in the finite $R$ theory. In other words, $\xi_{ab}$ is as defined in (\ref{eq: xi DLCQ limit}).

%\[
%\xi_{ij}=\frac{\vec{x}_{ij}^{2}}{2x_{ij}^{-}}.
%\]
On the other hand, using (\ref{eq: F2 def}) we identify
\begin{equation}
F_2 =\sum_{\n}e^{-i \n x_{12}^{+}/R_+}\left(\, \lim_{k\to\infty}k^{-1} \left\langle \mathcal{O}^\text{orb}_{\n}\mathcal{O}^\text{orb}_{-\n}\right\rangle \right), 
\label{2ptfourier}
\end{equation}
where by the results of section \ref{sect: Recovering the DLCQ description} we know that this limit exists.

We now focus only on the functional form of the resulting 2-point function. Combining equations \eqref{2ptdimred} and \eqref{2ptfourier} then gives the following integral formula for five-dimensional correlators in terms of scaling dimensions and Fourier momentum modes as we approach the DLCQ limit,
\begin{equation}
\left\langle \mathcal{O}^\text{orb}_{\n}\mathcal{O}^\text{orb}_{-\n}\right\rangle \propto\left(x_{12}^{-}\right)^{-\Delta}\int_{-\infty}^{\infty}dx_{12}^+\,\,e^{i \n x_{12}^{+}/R_+}\left(x_{12}^{+}-\xi_{12}/2\right)^{-\Delta}\ .
\end{equation}
Note however that this integral is not well defined because of the pole on the real axis. As we will see below this is very similar to the poles encountered in standard Feynman propagators so we will remedy it using a standard $i \epsilon$ prescription:
\begin{equation}
x_{i}^{+}\rightarrow x_{i}^{+}-i\epsilon_{i}\ ,
\end{equation}
where $\epsilon_{1}>\epsilon_{2}>...>\epsilon_{n}>0$. Note that this is the same $i \epsilon$ prescription introduced in section \ref{sect: Null Conformal Compactification}, in the limit that $R\to\infty$. After implementing this prescription, the real pole gets shifted into the upper half plane and we obtain 
\begin{equation}
\left\langle \mathcal{O}^\text{orb}_{\n}\mathcal{O}^\text{orb}_{-\n}\right\rangle \propto\left(x_{12}^{-}\right)^{-\Delta}\int_{-\infty}^{\infty}dx_{12}^+\,\,e^{i \n x_{12}^{+}/R_+}\left(x_{12}^{+}-\left(\xi_{12}/2+i\epsilon_{12}\right)\right)^{-\Delta}\ .
\label{2ptreg}
\end{equation}
If $\n>0$ can close in upper half plane picking up the residue of the pole, but if $\n<0$ we close in lower half plane and get zero:
\begin{align}
  \left\langle \mathcal{O}^\text{orb}_{\n}\mathcal{O}^\text{orb}_{-\n}\right\rangle \propto \left\{\begin{aligned}
 	\left(x_{12}^{-}\right)^{-\Delta}e^{i \n \zeta_{12}/2R_+},& \quad &n>0		\\
 	0,& &n<0
 \end{aligned}\right.\ ,
\end{align}
which is consistent with the result (\ref{eq: DLCQ 2pt from limit}) as found by taking the limit of the 2-point function at finite $k,R$. We see that only anti-instantons can propagate. The zero mode case $\n=0$ is slightly ambiguous, but is determined to be zero, again by the result (\ref{eq: DLCQ 2pt from limit}).

\begin{figure}
\centering
\includegraphics[width=10cm]{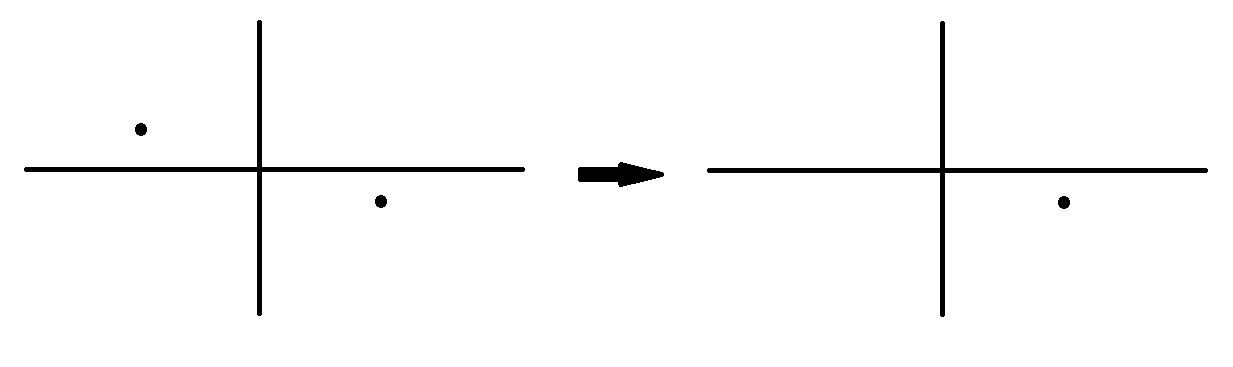}
\caption{On the left, we depict the poles of the Feynman propagator which encode the propagation of particles and antiparticles. In the non-relativistic limit, only one pole survives corresponding to the propagation of particles.}
\label{nonrelfeynman}
\end{figure}

Let us briefly pause to compare this to the non-relativistic limit of the Feynman propagator:
\begin{equation}
\frac{1}{-\left(k^{0}\right)^{2}+\vec{k}^{2}+m^{2}+i\epsilon}=-\frac{1}{\left(k^{0}+\sqrt{\vec{k}^{2}+m^{2}}+i\epsilon\right)\left(k^{0}-\sqrt{\vec{k}^{2}+m^{2}}-i\epsilon\right)}\ .
\end{equation}
Using the standard $i \epsilon$ prescription for time-ordered propagators, the poles are shifted into the complex plane as depicted in Figure \ref{nonrelfeynman}. Note that the pole on the left corresponds to propagation of anti-particles while the pole on the right corresponds to particle propagation. If we take the $k^{0}=m+E$, in the limit $E \ll m$, this reduces to
\begin{equation}
\frac{1}{-\left(k^{0}\right)^{2}+\vec{k}^{2}+m^{2}+i\epsilon}\rightarrow-\frac{1}{E-\vec{k}^{2}/2m+i\epsilon}\ ,
\end{equation}
and we see that only one pole remains so only particles can propagate, analogous to the pole structure in \eqref{2ptreg}. Indeed, compactifying along a null direction can be thought of as taking a non-relativistic limit since the resulting five-dimensional theory has a Lifshitz scaling symmetry.

\paragraph{3-point}\
\\
\\
Following similar steps to the ones outlined in the previous subsection, we obtain the following integral formula for 3-point functions:
\begin{align}
&\big\langle \mathcal{O}^{(1),\text{orb}}_{\n_{1}}\mathcal{O}^{(2),\text{orb}}_{-\n_1-\n_3}\mathcal{O}^{(3),\text{orb}}_{\n_3}\big\rangle\nn\\
 &\quad \propto\Pi_{i<j,k\neq i,j}\left(x_{ij}^{-}\right)^{-\left(\Delta_{i}+\Delta_{j}-\Delta_{k}\right)/2}\int_{-\infty}^{\infty} dx_{13}^{+}dx_{23}^{+} e^{i\left(\n_{1}x_{12}^{+}-\n_{3}x_{23}^{+}\right)/R_+}\nn\\
&\hspace{70mm}\times\left(x_{12}^{+}-\left(\xi_{12}/2+i\epsilon_{12}\right)\right)^{-\left(\Delta_{1}+\Delta_{2}-\Delta_{3}\right)/2}\nn\\
&\hspace{70mm}\times\left(x_{12}^{+}+x_{23}^{+}-\left(\xi_{13}/2+i\epsilon_{13}\right)\right)^{-\left(\Delta_{1}+\Delta_{3}-\Delta_{2}\right)/2}\nn\\
&\hspace{70mm}\times\left(x_{23}^{+}-\left(\xi_{23}/2+i\epsilon_{23}\right)\right)^{-\left(\Delta_{2}+\Delta_{3}-\Delta_{1}\right)/2}\ .
\label{3ptintegral}
\end{align}
Notice that all three poles are once again shifted into the upper half of the complex plane. Hence, if $\n_{1}<0$ we can close the $x_{12}^{+}$ contour in lower half plane obtaining a zero
residue. Similarly, if $\n_{3}>0$ we can close the $x_{23}^{+}$ contour in lower half
place getting a zero residue. Hence, $\n_{1}\geq0$ and $\n_{3}\leq0$ and
$k_{2}$ only constrained by momentum conservation. This is consistent with the results found for $\Omega \neq 0$, in particular the sum over  Fourier momentum modes from right to left should never go positive so instantons cannot be produced.

We now proceed to calculate the integral by residues, restricting to cases in which  $\Delta_{a}\geq\Delta_{b}+\Delta_{c}$ for some $a$ for simplicity. The more general result is given by (\ref{eq: DLCQ 3pt from limit}). Suppose $\Delta_{2}\geq\Delta_{1}+\Delta_{3}$. Then there is no pole in the middle term in \eqref{3ptintegral}. Evaluating residues of the poles in first and third term gives
\begin{equation}
\big\langle \mathcal{O}^{(1),\text{orb}}_{\n_{1}}\mathcal{O}^{(2),\text{orb}}_{-\n_1-\n_3}\mathcal{O}^{(3),\text{orb}}_{\n_3}\big\rangle \propto\Pi_{i<j,k\neq i,j}\left(x_{ij}^{-}\right)^{-\left(\Delta_{i}+\Delta_{j}-\Delta_{k}\right)/2}e^{i\left(\n_{1}\xi_{12}-\n_{3}\xi_{23}\right)/2R_+}\psi\ ,
\end{equation}
where
\begin{align}
\psi &=e^{-i\left(\n_{1}\xi_{12}-\n_{3}\xi_{23}\right)/2R_+}\partial_{x_{12}^{+}}^{\left(\Delta_{1}+\Delta_{2}-\Delta_{3}\right)/2-1}\partial_{x_{23}^{+}}^{\left(\Delta_{2}+\Delta_{3}-\Delta_{1}\right)/2-1}\nn\\
&\left.\left(e^{i\left(\n_{1}x_{12}^{+}-\n_{3}x_{23}^{+}\right)/R_+}\left(x_{12}^{+}+x_{23}^{+}-\left(\xi_{13}/2-i\epsilon_{13}\right)\right)^{-\left(\Delta_{1}+\Delta_{3}-\Delta_{2}\right)/2}\right)\right|_{x_{12}^{+}=\xi_{12}/2,x_{23}^{+}=\xi_{23}/2}\ .
\end{align}
Note that $\psi$ is a function of 
\begin{equation}
\rho=\xi_{12}+\xi_{23}+\xi_{31}=-\frac{\left(x_{13}^{i}x_{23}^{-}-x_{23}^{i}x_{13}^{-}\right)^{2}}{x_{12}^{-}x_{23}^{-}x_{31}^{-}}\ ,
\end{equation}
in agreement with the general solution in \cite{Henkel:1993sg}. Some manipulation shows that this result is indeed consistent with the exact result (\ref{eq: DLCQ 3pt from limit}) obtained from the theory at finite $R$.

For $\Delta_{3}\geq\Delta_{1}+\Delta_{2}$ we can also obtain a generating function by re-writing the integral as 
\begin{align}
&\big\langle \mathcal{O}^{(1),\text{orb}}_{\n_{1}}\mathcal{O}^{(2),\text{orb}}_{-\n_1-\n_3}\mathcal{O}^{(3),\text{orb}}_{\n_3}\big\rangle \nn\\
&\quad\propto\Pi_{i<j,k\neq i,j}\left(x_{ij}^{-}\right)^{-\left(\Delta_{i}+\Delta_{j}-\Delta_{k}\right)/2}\int_{-\infty}^{\infty}dx_{13}^{+}dx_{23}^{+}e^{i\left(\n_{1}x_{13}^{+}+\n_{2}x_{23}^{+}\right)/R_+}\nn\\
&\hspace{70mm}\times\left(x_{13}^{+}-x_{23}^{+}-\left(\xi_{12}/2+i\epsilon_{12}\right)\right)^{-\left(\Delta_{1}+\Delta_{2}-\Delta_{3}\right)/2}\nn\\
&\hspace{70mm}\times\left(x_{13}^{+}-\left(\xi_{13}/2+i\epsilon_{13}\right)\right)^{-\left(\Delta_{1}+\Delta_{3}-\Delta_{2}\right)/2}\nn\\
&\hspace{70mm}\times\left(x_{23}^{+}-\left(\xi_{23}/2+i\epsilon_{23}\right)\right)^{-\left(\Delta_{2}+\Delta_{3}-\Delta_{1}\right)/2}\ ,
\end{align}
and evaluating the residue of the pole in the second and third term.

\paragraph{4-point}\
\\
\\
Finally, let us consider the reduction of 4-point functions. Using arguments similar to those at two points, we find that a general five-dimensional 4-point function is given by
\begin{align}
&\left\langle \mathcal{O}^{(1),\text{orb}}_{\n_{1}}\mathcal{O}^{(2),\text{orb}}_{\n_{2}}\mathcal{O}^{(3),\text{orb}}_{\n_{3}}\mathcal{O}^{(4),\text{orb}}_{-\n_{1}-\n_{2}-\n_{3}-\n_{4}}\right\rangle\nn\\
 &\qquad\propto\frac{1}{(2\pi)^{3}}\int_{-\infty}^{\infty}dx_{14}^{+}dx_{24}^{+}dx_{34}^{+}e^{i\left(\n_{1}x_{14}^{+}+\n_{2}x_{24}^{+}+\n_{3}x_{34}^{+}\right)/R_+}\left\langle \mathcal{O}^{(1)}\mathcal{O}^{(2)}\mathcal{O}^{(3)}\mathcal{O}^{(4)}\right\rangle\ .\label{eq:4ptreduction}
\end{align}
Unlike at 2 and 3 point functions, the six-dimensional 4-point functions are not fixed by symmetry, so we will consider a disconnected free correlator with $\Delta_{i}=\Delta$:
\begin{align}
\left\langle \mathcal{O}^{(1)}\mathcal{O}^{(2)}\mathcal{O}^{(3)}\mathcal{O}^{(4)}\right\rangle =&\left(x_{12}^{-}x_{34}^{-}\right)^{-\Delta}\left(x_{14}^{+}-x_{24}^{+}-\left(\xi_{12}/2+i\epsilon_{12}\right)\right)^{-\Delta}\left(x_{34}^{+}-\left(\xi_{34}/2+i\epsilon_{34}\right)\right)^{-\Delta}
\nn\\
&+\left(x_{13}^{-}x_{24}^{-}\right)^{-\Delta}\left(x_{14}^{+}-x_{34}^{+}-\left(\xi_{13}/2+i\epsilon_{13}\right)\right)^{-\Delta}\left(x_{24}^{+}-\left(\xi_{24}/2+i\epsilon_{24}\right)\right)^{-\Delta}
\nn\\
&+\left(x_{14}^{-}x_{23}^{-}\right)^{-\Delta}\left(x_{14}^{+}-\left(\xi_{14}/2+i\epsilon_{14}\right)\right)^{-\Delta}\left(x_{24}^{+}-x_{34}^{+}-\left(\xi_{23}/2+i\epsilon_{23}\right)\right)^{-\Delta}\ ,\label{eq:4ptdisc1}
\end{align}
where we have implemented the $i \epsilon$ prescription to make the integral in \eqref{eq:4ptreduction} well-defined. 

Consider $\n_{1}=\n_{2}=-\n_{3}=-\n_{4}=\n>0$. Looking at the exponential in \eqref{eq:4ptreduction}, we must then close the
$x_{14}^{+}$ and $x_{24}^{+}$ contours in upper half plane while closing the $x_{34}^{+}$ contour
in the lower half. All poles are shifted into the upper half plane so the first term in
the disconnected free correlator in \eqref{eq:4ptdisc1} doesn't contribute
since the contour does not contain the $x_{34}^{+}$ pole. Computing
the residues of the poles in $x_{14}^{+}$ and $x_{24}^{+}$ in the
second and third terms leaves no remaining $x_{34}^{+}$ dependence,
so the integral over $x_{34}^{+}$ just gives a divergent constant corresponding to the additional factor of $k$ encountered in (\ref{eq: extra divergence for 4pt}). We
then find that
\begin{align}
\left\langle ++--\right\rangle &=\left(x_{13}^{-}x_{24}^{-}\right)^{-\Delta}e^{i \n \left(\xi_{13}+\xi_{24}\right)/2R_+}+\left(x_{14}^{-}x_{23}^{-}\right)^{-\Delta}e^{i \n \left(\xi_{14}+\xi_{23}\right)/2R_+}\ ,
\end{align}
hence recovering the result (\ref{eq: 4pt DLCQ final}) of taking the DLCQ limit of the finite $R$ result.
Similarly, we find that 
\begin{equation}
\left\langle +-+-\right\rangle =\left(x_{12}^{-}x_{34}^{-}\right)^{-\Delta}e^{i \n \left(\xi_{12}+\xi_{34}\right)/2R_+}\ ,
\end{equation}
again recovering the result (\ref{eq: 4pt DLCQ final}). All other correlators vanish.

\end{document}